\documentclass[showpacs,aps,prd,reprint,superscriptaddress,nofootinbib]{revtex4-1}
\usepackage[colorlinks=true, pdfstartview=FitV, linkcolor=red,citecolor=blue, urlcolor=blue,
bookmarks=true, bookmarksnumbered=true, breaklinks]{hyperref}
\usepackage[dvipdfmx]{graphicx}
\usepackage{amsmath,amssymb,bm,ascmac,color,longtable,mathrsfs}

\usepackage{slashed}

\allowdisplaybreaks[1]

\usepackage{color}

\usepackage[normalem]{ulem}

\def\simge{\mathrel{
   \rlap{\raise 0.511ex \hbox{$>$}}{\lower 0.511ex \hbox{$\sim$}}}}
\def\simle{\mathrel{
   \rlap{\raise 0.511ex \hbox{$<$}}{\lower 0.511ex \hbox{$\sim$}}}}
\def\bigs{\mathrel{
   \rlap{\raise 0.531ex \hbox{$>$}}{\lower 0.531ex \hbox{$<$}}}}

\allowdisplaybreaks[3]
\allowdisplaybreaks

\usepackage[dvipdfmx]{graphicx}

\begin{document}

%\preprint{KEK-TH-1960}
\begin{flushright}
KEK-TH-1960
\end{flushright}

\title{Topology and stability of the Kondo phase in quark matter
}

\author{Shigehiro~Yasui}
\email[]{{\tt yasuis@th.phys.titech.ac.jp} (corresponding author)}
\affiliation{Department of Physics, Tokyo Institute of Technology, Tokyo 152-8551, Japan}
\author{Kei~Suzuki}
\email[]{{\tt kei.suzuki@kek.jp}}
\affiliation{KEK Theory Center, Institute of Particle and Nuclear
Studies, High Energy Accelerator Research Organization, 1-1 Oho, Tsukuba, 
Ibaraki, 305-0801, Japan}
\author{Kazunori~Itakura}
\email[]{{\tt kazunori.itakura@kek.jp}}
\affiliation{KEK Theory Center, Institute of Particle and Nuclear
Studies, High Energy Accelerator Research Organization, 1-1 Oho, Tsukuba, 
Ibaraki, 305-0801, Japan}
\affiliation{Graduate University for Advanced Studies (SOKENDAI),
1-1 Oho, Tsukuba, Ibaraki, 305-0801, Japan}

\begin{abstract}

We investigate properties of the ground state of a light quark matter with heavy quark impurities. This system exhibits the ``QCD Kondo effect" where the interaction strength between a light quark near the Fermi surface and a heavy quark increases with decreasing energy of the light quark towards the Fermi energy, and diverges at some scale near the Fermi energy, called the Kondo scale. Around and below the Kondo scale, we must treat the dynamics nonperturbatively. As a typical nonperturbative method to treat the strong coupling regime, we adopt a mean-field approach where we introduce a condensate, the Kondo condensate, representing a mixing between a light quark and a heavy quark, and determine the ground state in the presence of the Kondo condensate. We show that the ground state is a topologically non-trivial state and the heavy quark spin forms the hedgehog configuration in the momentum space. We can define the Berry phase for the ground-state wavefunction in the momentum space which is associated with a monopole at the position of a heavy quark. We also investigate fluctuations around the mean field in the random-phase approximation, and show the existence of (exciton-like) collective excitations made of a hole $h$ of a light quark and a heavy quark $Q$. 
\end{abstract}


\maketitle

%\tableofcontents

%% main text
% !TEX root = 0_main.tex
\section{Introduction}

The Kondo effect has been studied in a variety of fermionic systems containing heavy particles as impurities. It is the phenomenon that the interaction between a light fermion near the Fermi surface and a heavy impurity particle becomes stronger due to quantum fluctuations at low temperatures, and it drastically affects the transportation and thermodynamic properties of the bulk matter~\cite{Kondo:1964}. The essential conditions for the Kondo effect to occur are summarized as the existence of the following three ingredients: (i) Fermi surface (degenerate state), (ii) loop effects (particle-hole creation) and (iii) non-Abelian interaction between a light fermion and a heavy impurity~\cite{Hewson,Yosida,Yamada}. Historically, the Kondo effect was observed in metals with impurity atoms having a finite spin~\cite{Kondo:1964}. There, the non-Abelian interaction is played by the spin $\mathrm{SU}(2)_{\mathrm{spin}}$ symmetry supplied by the spin-exchange between the electron and the impurity atom, and it was demonstrated by J.~Kondo that the second order perturbation for the scattering amplitude between an electron and the heavy spin yields the enhancement of the amplitude due to the three ingredients shown above. Since then, the Kondo effect has been widely applied to various systems including artificial materials, such as quantum dots, and also to the multi-band electron systems which are regarded to have the $\mathrm{SU}(n)$  symmetry with $n$ being the number of electron bands~\cite{Goldhaber-Gordon:1998,Cronenwett:1998,Wiel:2000,Jeong:2001,Park:2002,Hur:2015}. Hence, the Kondo effect is now recognized as a typical example of the strongly correlated condensed matter systems.

Recently, the Kondo effect has expanded its region of applicability to the world of strong interaction which is the fundamental force of nuclei, hadrons, quarks and antiquarks~\cite{Yasui:2013xr,Hattori:2015hka,Ozaki:2015sya,Yasui:2016ngy,Yasui:2016svc,Yasui:2016hlz,Yasui:2016yet,Kanazawa:2016ihl,Kimura:2016zyv}. There are several sources of non-Abelian interaction in strong interaction: spin, isospin (flavor) and color. We can study different types of the Kondo effect depending on what type of the non-Abelian interaction we choose. The first application \cite{Yasui:2013xr} of the Kondo effect to the strong interaction was done in a nuclear matter containing $\bar{D}$ or $B$ mesons as impurities~\footnote{For a review on current status of heavy hadrons in nuclear matter, see Ref.~\cite{Hosaka:2016ypm}.}. In fact, $\bar{D}$ and $B$ mesons are heavier than a nucleon in a nucleus, $m_{\bar{D}}=1.87$ GeV and $m_{B}=5.28$ GeV~\cite{Agashe:2014kda}, and thus can be treated as heavy impurities. The non-Abelian interaction is provided by the isospin-exchange interaction between a $\bar{D}$ or $B$ meson and a nucleon, whose symmetry is given by $\mathrm{SU}(2)_{\mathrm{isospin}}$~\cite{Yasui:2013xr,Yasui:2016ngy}. When a $\bar{D}^{\ast}$ ($B^{\ast}$) meson is regarded as the spin partner of a $\bar{D}$ ($B$) meson, we can also consider the spin-exchange interaction as the non-Abelian interaction governed by the $\mathrm{SU}(2)_{\mathrm{spin}}$ symmetry~\cite{Yasui:2016hlz}. The same paper also discussed the Kondo effect in a light flavor quark matter containing $c$ ($b$) quarks as impurities \cite{Yasui:2013xr}. In addition to the fact that $c,$ $b$ quarks are much heavier than the light quarks, $m_{c}=1.27$ GeV, $m_{b}=4.18$ GeV~\cite{Agashe:2014kda}, the heaviness of $c$ and $b$ quarks also ensures the existence of a light quark matter which will be realized at a relatively large chemical potential $\mu$ so that there is a hierarchy $\mu \ll m_c, m_b$. The non-Abelian interaction is given by the color-exchange interaction with $\mathrm{SU}(3)_{\mathrm{c}}$ symmetry. This type of the Kondo effect working in a light quark matter is in particular called the QCD Kondo effect, and this is the main subject of the present paper.

Since the first publication on the QCD Kondo effect \cite{Yasui:2013xr}, we have already seen a considerable progress in understanding unique and intriguing aspects of the QCD Kondo effect. Let us briefly give an overview on the previous works~\cite{Hattori:2015hka,Ozaki:2015sya,Yasui:2016svc,Yasui:2016yet,Kimura:2016zyv}.
The first analysis on the QCD Kondo effect \cite{Yasui:2013xr} was performed with the four-Fermi interaction between a light quark and a heavy quark which minimally represents the non-Abelian property of the color-exchange interaction. Namely, the interaction is proportional to a product of two Gell-Mann matrices $\lambda^{a}\!\cdot\!\lambda^{a}$ ($a=1,\dots,N_{c}^{2}-1$) for the $\mathrm{SU}(N_{c})$ color symmetry. For the analysis of scattering between a light quark near the Fermi energy and a heavy impurity, we are able to take the QCD coupling $\alpha_s(\mu)$ small enough, which implies that the four-Fermi interaction strength is also taken to be small enough. However, a one-loop perturbative calculation for the scattering between a light quark and a heavy impurity turns out to logarithmically increase as the energy of the light quark decreases towards the Fermi energy. This means that the effective interaction becomes stronger with decreasing energy scale and that perturbative analysis would become invalid at some lower energy scale. This is nothing but the appearance of the Kondo effect, and the lower energy scale where the perturbative calculation breaks down corresponds to the Kondo scale.

Later, a more precise calculation was performed based on the finite-density QCD perturbation theory~\cite{Hattori:2015hka}. While the four-Fermi interaction in the previous work is a contact interaction with a zero interaction range, the actual interaction between a light quark and a heavy quark must be described by the exchange of a gluon and thus has a finite interaction range. Indeed, at finite densities, gluon propagation is screened by the medium effects: the electric component is screened to acquire the Debye mass, while the magnetic component is dynamically screened. Still, we are able to define the effective coupling similar to the one in the four-Fermi interaction through the s-wave projection of the scattering amplitude.  With a small QCD coupling $\alpha_s(\mu)\ll 1$ for a large chemical potential $\mu$, we can perform perturbative analysis for the calculation of scattering amplitudes of a light quark near the Fermi surface off a heavy quark impurity. In order to study how the effective coupling strength varies with decreasing energy of a light quark, the renormalization group analysis was done up to the one-loop order and it was shown that the effective coupling increases with decreasing energy scales of a light quark, and that there exists the Kondo scale at which the effective coupling diverges. Notice that these results are all qualitatively consistent with those of the previous work in Ref.~\cite{Yasui:2013xr}, which justifies the use of the four-Fermi interaction in the study of the QCD Kondo effect. 

Incidentally, what the first condition (i) truly implies is that there must be a finite degeneracy at the lowest energy state (without a mass gap). Then, we expect that the QCD Kondo effect will also take place in a strong magnetic field which induces the formation of the Landau levels with a nonzero degeneracy in the lowest Landau level and allows for a linear dispersion in the direction parallel to the magnetic field. This idea was explicitly demonstrated in Ref.~\cite{Ozaki:2015sya} and the effect is called magnetically induced QCD Kondo effect. This is a unique phenomena in QCD which is not seen in the ordinary Kondo effect in condensed matter physics: while imposing a magnetic field inhibits the ordinary Kondo effect with the spin-flip interaction, the non-Abelian interaction in the QCD Kondo effect does not directly feel the magnetic field (only through the magnetic screening of gluon propagation) and it makes sense to impose a strong magnetic field.

As repeatedly mentioned, the perturbative analysis (or improved analysis with the renormalization group) of the Kondo effect indicates that there exists a strongly coupled regime. It corresponds to a shell outside of the Fermi surface whose thickness is typically specified by the Kondo scale. The perturbative analysis works only outside of this shell, and we have to resort to some nonperturbative method to go into the shell. For example, if one is exactly on the Fermi sphere, one can utilize the (boundary) conformal field theory as a nonperturbative technique as was recently done in Ref.~\cite{Kimura:2016zyv}. It was suggested that the quark matter with two flavors ($u$ and $d$) and three colors ($N_c=3$) will be a non-Fermi liquid. While the conformal field theory is a powerful nonperturbative technique, the connection with the perturbative region (i.e., the region outside the shell) is not clear. It should also be noticed that both the perturbative analysis of the scattering amplitudes and the conformal field theory treat only a single heavy impurity. In the actual situation, however, we expect impurities are randomly distributed in a quark matter with a small averaged density. In order to find a ground state of such a system, we need to introduce a nonperturbative and field-theoretical technique as was done in Ref.~\cite{Yasui:2016svc}.

It is instructive to recall the mechanism of superconductivity and chiral symmetry breaking. In both cases, interactions in the relevant channels (electron-electron scattering for superconductivity and quark-antiquark scattering for chiral symmetry breaking) increase with decreasing scattering energies, which leads to  formation of bound states accompanied by nonzero condensates. We expect that similar phenomena occur in the QCD Kondo effect when we treat it in a field-theoretical way. Namely, enhancement of the light and heavy quark interaction will entail the formation of a bound state and the generation of a nonzero condensate. As is well-known in the superconductivity and chiral symmetry breaking, we can nonperturbatively study these phenomena by the mean-field approach. Indeed, the mean-field approach is found to be useful in the Kondo effect in condensed matter physics~\cite{Takano:1966,Yoshimori:1970,Lacroix:1979,Eto:2001,Yanagisawa:2015conf,Yanagisawa:2015}. Motivated by these observations, the mean-field approach was applied to the QCD Kondo effect by us in Ref.~\cite{Yasui:2016svc}. The mean-field was defined as the expectation value of a product of a light quark field and a heavy quark field, and we called it the Kondo condensate. The mean-field approximation for the four-Fermi interaction generates a mixing term between a light quark and a heavy quark, with the strength of mixing determined by the Kondo condensate. By solving the gap equation, we found that a finite value of the mean-field is favored as the stable solution. We can map the region of nonzero Kondo condensate on the $\lambda$-$\mu$ plane ($\lambda$ is the Lagrange multiplier for the heavy-quark density and $\mu$ is the chemical potential for the light quark) and called it the Kondo phase. In Ref.~\cite{Yasui:2016svc}, a uniform number density of heavy quarks was assumed for simplicity so that the mean-field is also homogeneous. Although this setting is far from the actual situation with a small number of randomly distributed impurities, we can extract physical information from this simple calculation by estimating the energy gain per a heavy quark. Later, in Ref.~\cite{Yasui:2016yet}, the opposite situation with a single heavy quark was also analyzed in the mean-field approach. In this case, one has to treat the mean field which is not homogeneous in space. It was shown that the energy gain of a single heavy quark in the Kondo condensate is  comparable with that obtained from uniform heavy quark density. This comparison suggests that the actual value would be close to these values.

In the present article, we continue the mean-field analysis of the previous work~\cite{Yasui:2016svc} and perform a detailed study on the properties of the ground state in the presence of the Kondo condensate. We also investigate the stability of the mean-field by including quantum fluctuations around it. The Kondo condensate found in Ref.~\cite{Yasui:2016svc} has a unique structure: it is made of a combination of the scalar-type condensate and the vector-type condensate (see also Ref.~\cite{Yasui:2016yet}). We show that such a coexistence of the scalar and vector condensates leads to the locking of chiral symmetry and heavy quark spin symmetry as well as to topologically non-trivial properties in the ground state. These are new properties which have not been known in a quark matter. As for the topological properties, we show that the heavy quark spin forms a hedgehog configuration in momentum space. By using the quasi-quark wave function, which is a mixed state of the light and heavy quarks in the presence of the Kondo condensate, we can define the Berry phase in momentum space which is associated with a monopole. As for the quantum fluctuations around the Kondo condensate, we consider the collective excitations of the light quark $q$ (hole $h$) and the heavy quark $Q$ within the random-phase approximation, and find that $hQ$ bound states, which are analog of excitons, appear as collective modes. As already mentioned, one of the important properties of the Kondo effect is the enhancement of the interaction strength between a light quark and a heavy quark. We show this enhancement in the effective Lagrangian including the mean-field and the quantum fluctuations.

The contents of the present article are the following. In Sec.~\ref{sec:Kondo_interaction}, we define the Lagrangian describing the system with light quarks and heavy quarks which are interacting with each other via the color-current-current interaction. This interaction has the $\mathrm{SU}(N_{c})$ color symmetry which is responsible for the non-Abelian interaction necessary for the Kondo effect, and $\mathrm{U}(N_{f})_{\mathrm{V}}\times\mathrm{U}(N_{f})_{\mathrm{A}}$ chiral symmetry for $N_{f}$ light flavors. In Sec.~\ref{sec:ground_state}, we study the ground state with the Kondo effect by applying the mean-field approach. We show that the Kondo condensate is realized in the ground state, and that the wavefunction of the ground state has topologically non-trivial properties: hedgehog configurations of heavy quark spin and monopoles associated with the Berry phase, both of which appear in momentum space. In Sec.~\ref{sec:Kondo_excitations}, we discuss the collective excitations of a hole of a light quark $h$ and a heavy quark $Q$ beyond the mean-field using the random-phase approximation. The final section is devoted to conclusion.
% !TEX root = 0_main.tex
\section{Color-current interaction}
\label{sec:Kondo_interaction}

In this section, we define the model we analyze, in particular, the interaction between a light quark and a heavy quark. As we commented in the Introduction, the QCD Kondo effect is well studied by the contact four-Fermi interaction which reproduces qualitatively the same results as the one-gluon exchange interaction. Thus, for the purpose of developing new nonperturbative method, we adopt the contact four-Fermi interaction. We construct the model so that it possesses the chiral symmetry in the light quark sector, and the heavy quark (spin) symmetry (HQS) in the heavy quark sector.

As for the heavy quark sector, we use the framework of the heavy quark effective theory~\cite{Manohar:2000dt,Neubert:1993mb}. In this formalism, we focus on the dynamics of a heavy quark associated with deviation from the on-mass-shell motion. Namely, we separate the heavy quark momentum $P^{\mu}$ into the on-mass-shell part ($m_{Q}v^{\mu}$ with $m_Q$ being the mass and $v^\mu$ being the four velocity of the heavy quark) and the off-mass-shell part ($p^{\mu}$), $P^{\mu}=m_{Q}v^{\mu}+p^{\mu}$, and consider the dynamics with respect to the off-mass-shell momentum $p^\mu$. The four velocity $v^{\mu}=(v^{0},\vec{v}\,)$ is defined so as to satisfy the on-mass-shell condition, $v^{\mu}v_{\mu}=1$, and $v^{0}>0$ for a positive energy state. Since we are interested in momentum scales in the infrared region $p^{\mu} \simeq \Lambda_{\mathrm{QCD}}$, where $\Lambda_{\mathrm{QCD}}$ is the typical low-energy scale of QCD of a few hundred MeV~\cite{Agashe:2014kda}, we are able to take $p^{\mu}$ much smaller than $m_{Q}v^{\mu}$ for sufficiently large $m_{Q}$ ($p^{\mu} \ll m_{Q}$). We treat that the on-mass-shell motion of a heavy quark is not affected by the interaction with light particles (light quarks and gluons). We call the reference frame in which the heavy quark is moving at the velocity $v^\mu$ the {\it $v$-frame}. In such a frame, if we extract the on-mass-shell motion, the dynamics of the heavy quark is only described by the residual momentum $p^\mu$. Therefore, it is convenient to redefine the heavy quark field $\Psi$ so that the on-mass-shell motion is made manifest~\cite{Manohar:2000dt,Neubert:1993mb}: we introduce an effective field $\Psi_{v}$ by $\Psi_{v} = \frac{1+v\hspace{-0.5em}/}{2} e^{im_{Q}\cdot v} \Psi$ where $\frac{1+v\hspace{-0.5em}/}{2}$ is a projection operator on the positive-energy component of the heavy quark field, and $e^{im_{Q}v\cdot x}$ represents the on-mass-shell motion. We notice that, after eliminating the on-mass-shell dynamics, $\Psi_{v}$ depends only on the residual momentum $p^{\mu}$ as a dynamical variable. It should be also noticed that the heavy quark effective theory is originally invented to describe the dynamics of a single heavy quark in a heavy-light bound state. If there are several heavy quarks and they are moving at different velocities $v^\mu_i$, analysis becomes quite complicated. Thus, we assume that all the heavy quarks are moving at the same velocity and in the same direction,  $v^\mu_i=v^\mu$. This is, however, a natural situation for heavy quarks because it implies that we can take the rest frame where all the heavy quarks are at rest. 

In the following, we work in the frame where all the heavy quarks are at rest ({\it static frame}), or equivalently, we simply take  $v^{\mu}=(1,\vec{0}\,)$. Then, the projection operator $\frac{1+v\hspace{-0.5em}/}{2}=\frac{1+\gamma^0}{2}$ is, in the standard representation of the Dirac matrices\footnote{Throughout the paper, we use the Dirac representation for $\gamma$ matrices: $\gamma^0=\left(\begin{array}{cc}1&0\\ 0&-1\end{array}\right)$, $\gamma^i=\left(\begin{array}{cc}0&\sigma^i\\ -\sigma^i&0\end{array}\right)$, $\gamma_5=\left(\begin{array}{cc}0&1\\ 1&0\end{array}\right)$.}, nothing but the projection on the upper two components, as known in the nonrelativistic limit of a Dirac fermion. In the present paper, we will work in the standard representation for the Dirac matrices, but we express the upper two components of the heavy quark as $\Psi_v$ too, to avoid introducing too many notations. 
It is important that the heavy quark spin ``up" and ``down" components are not mixed in the heavy-quark mass limit, so that the heavy quark spin is always the conserved quantity.
This is called the {\it heavy quark (spin) symmetry} (HQS), $\mathrm{SU}(2)_{\mathrm{HQS}}$, which is the symmetry for the invariance under interchange of the spin up and down components.

We construct the model so that the Lagrangian has the chiral symmetry for (massless) $N_f$ light quarks $\mathrm{U}(N_{f})_{\mathrm{V}}\times\mathrm{U}(N_{f})_{\mathrm{A}}$ and the heavy quark symmetry $\mathrm{SU}(2)_{\mathrm{HQS}}$ for one flavor heavy quark. Since we are interested in the Kondo effect, we only focus on the interaction between the light and heavy quarks.\footnote{We are able to study the effects of interactions between light quarks which induce the chiral symmetry breaking, which is reported elsewhere.
} Notice that the color currents $j^a_\mu=\bar \psi \gamma_\mu T^a \psi$ and $J^a_\mu =\bar \Psi \gamma_\mu T^a \Psi= \bar \Psi_v \gamma_\mu T^a \Psi_v$ are invariant under the chiral transformation and the spin rotation, respectively. Here, $T^{a}=\lambda^{a}/2$ ($a=1,\dots,N_{c}^{2}-1$) with the Gell-Mann matrices $\lambda^{a}$ are the generators of color $\mathrm{SU}(N_{c})$ symmetry. Thus we introduce in the Lagrangian the current-current interaction $G_cj^a_\mu J^{a\mu}$ between the light quark and the heavy quark~\cite{Yasui:2013xr,Yasui:2016svc,Yasui:2016yet}:
\begin{eqnarray}
{\cal L}
&=&
\bar{\psi} ( i\partial\hspace{-0.52em}/
 + \mu \gamma_{0} ) \psi
+\bar{\Psi}_{v} v \!\cdot\!  
i\partial \Psi_{v}
\nonumber \\
&&
- G_{c}
\sum_{a =1}^{N_{c}^{2}-1} (\bar{\psi}\gamma^{\mu}T^{a}\psi) (\bar{\Psi}_{v} \gamma_{\mu} T^{a} \Psi_{v}).
\label{eq:Lagrangian_current}
\end{eqnarray}
The first term is the kinetic term for massless light quarks and indices for $N_{f}$ light flavors $\psi=(\psi_{1},\dots,\psi_{N_{f}})^{t}$ are implicit. We assume that light quarks have a common chemical potential $\mu$.
The second term is the (off-mass-shell) kinetic term for a heavy quark, and the spatial derivative is meant to be associated with the off-mass-shell momentum $p^\mu$. 
The third term is the four-point interaction, i.e. the color-current color-current interaction representing exchange of a color between a light quark and a heavy quark as mimicking the one-gluon exchange in QCD~\footnote{When the color-current color-current interaction in Eq.~(\ref{eq:Lagrangian_current}) is replaced by the interaction between only light quarks, it is nothing but the Nambu--Jona-Lasinio interaction used for description of the dynamical breaking of chiral symmetry in vacuum~\cite{Klimt:1989pm,Vogl:1989ea,Klevansky:1992qe,Hatsuda:1994pi,Buballa:2003qv}.}.
The non-Abelian property of the color exchange interaction is important in the Kondo effect, as shown in our previous works. As is common to fermionic effective theories with contact-type interactions, we need to introduce a cutoff $\Lambda$ to remedy possible ultraviolet divergences. 
As for the numerical values for the coupling constant $G_c$ and the cutoff $\Lambda$, we use $G_{c}\Lambda^{2}=(9/2)4.0$ with the three-dimensional momentum cutoff $\Lambda=0.65 \, \mathrm{GeV}$ in our numerical calculations as estimated in Appendix~\ref{sec:coupling_constant}.

Notice that the Lagrangian (\ref{eq:Lagrangian_current}) contains chemical potentials $\mu$ only for the light quarks, and lacks the information on the distribution of heavy quarks. We are going to analyze the situation where the heavy quark impurities are randomly distributed in a light quark matter and thus are not treated as a Fermi gas~\cite{Yasui:2016svc}, which is not simply expressed by the introduction of a chemical potential. We assume that the information of randomly distributed heavy quarks can be specified by a static function  $n_Q(\vec x)$ for the number density. On the other hand, since the operator for the number density of heavy quarks is given by $\Psi^\dagger_v \Psi_v=\bar \Psi_v \Psi_v$ (notice $\bar \Psi_v=\Psi_v^\dagger$), we impose 
\begin{eqnarray}
 \bar{\Psi}_{v}(\vec{x}) \Psi_{v}(\vec{x}) = n_{Q}(\vec{x}).
 \label{eq:constraint}
\end{eqnarray}
This condition can be easily included in the Lagrangian (\ref{eq:Lagrangian_current}) by the introduction of a Lagrange multiplier $\lambda$ so that $\delta[f(x)]=\int {\mathcal D}\lambda\, e^{-i\lambda f}$ with $f(x)=\bar{\Psi}_{v}(\vec{x}) \Psi_{v}(\vec{x}) - n_{Q}(\vec{x})$. Together with this constraint term, the Lagrangian is now redefined as (in the momentum space) 
\begin{eqnarray}
{\cal L}_{\mathrm{eff}}
&=&
\bar{\psi} ( p\hspace{-0.45em}/ + \mu \gamma_{0} ) \psi
+\bar{\Psi}_{v} v \!\cdot\!  p \Psi_{v}
\nonumber \\
&&
- G_{c}
\sum_{a=1}^{N_{c}^{2}-1} (\bar{\psi}\gamma^{\mu}T^{a}\psi) (\bar{\Psi}_{v} \gamma_{\mu} T^{a} \Psi_{v})
\nonumber \\
&&
-\lambda(\bar{\Psi}_{v} \Psi_{v} - n_{Q}).
\label{eq:Lagrangian_current_eff}
\end{eqnarray}
The value of $\lambda$ will be determined by the stationary condition $\partial \Omega/\partial \lambda=0$ for the thermodynamic potential $\Omega$ with a given function $n_{Q}$. 

The number density  $n_{Q}(\vec{x})$ of $N$ heavy quark impurities that are located at positions $\vec x=\vec x_i$ is given by 
\begin{eqnarray}
 n_{Q}(\vec{x}) = \sum_{i=1}^N \delta^{(3)}(\vec{x}-\vec{x}_{i}),
\end{eqnarray}
where $\delta^{(3)}(\vec{x})$ is a three-dimensional $\delta$-function.  Since we assume all the heavy quarks are at rest, $n_Q(\vec x)$ does not depend on time. 
In general, positions of heavy quarks are random, and it is convenient to treat the number density averaged over the positions. Namely, we introduce a number $n_Q$ given by  
\begin{eqnarray}
n_{Q} \equiv \left\langle n_{Q}(\vec{x}) 
\right\rangle_Q , \label{nQ_const}
\end{eqnarray}
where $\langle \cdots \rangle_Q $
denotes the average over random configuration $\{\vec x_i\}$. Notice that, after averaging over the random configuration, $n_Q$ is no longer a function of $\vec x$ and can be treated as a constant number. In the present paper, we perform the average over the random configuration first, and treat the number distribution $n_Q(\vec x)$ as if it is just a constant $n_Q(\vec x)=n_Q$ as given by Eq.~(\ref{nQ_const}). This is technically convenient because the translational invariance is kept with a constant number density. Besides, we can regard such a replacement as a good approximation for the dynamics of light quarks having wave lengths longer than a typical coherence length of the Kondo state or a typical distance between heavy impurities. 
% !TEX root = 0_main.tex
\section{Kondo phase: ground state}
\label{sec:ground_state}

In this section, we present a detailed investigation on the properties of the ground state within the mean-field approximation. After solving the gap equation at zero and finite temperatures, which reproduces the previous work \cite{Yasui:2016svc}, we further discuss the symmetry breaking pattern and topological properties of the ground state with the Kondo condensate.

\subsection{Mean-field approximation}
\label{sec:mean_field}

We determine the ground state of the Lagrangian (\ref{eq:Lagrangian_current_eff}) by the mean-field approximation to the color-current interaction. Since we are now interested in enhanced correlations between a light quark and a heavy quark impurity that are represented as operators of the type $\bar\psi {\cal O}\Psi_v$ or $\bar \Psi_v{\cal O}\psi$,  we rearrange the interaction by using the Fierz transformation so that the form of $\bar{\psi}{\cal O}\psi \bar{\Psi}_{v}{\cal O}\Psi_{v}$ in the original interaction is transformed into the form of $\bar{\psi}{\cal O}'\Psi_{v}\bar{\Psi}_{v}{\cal O}'\psi$. Among several different Fierz transformations as summarized in Appendix B, 
we adopt Eq.~(\ref{eq:Fierz_Dirac_1}) for the Dirac matrices, and Eq.~(\ref{eq:Fierz_SUN_2}) for the Gell-Mann matrices. The Fierz transformations are performed for each flavor $i=1,\dots,N_{f}$. 
Then, we divide the interaction term in Eq.~(\ref{eq:Lagrangian_current_eff}) into color-singlet and color-octet parts:
\begin{eqnarray}
{\cal L}_{\mathrm{int}} = {\cal L}_{\mathrm{int}}^{\mathrm{sing}} + {\cal L}_{\mathrm{int}}^{\mathrm{oct}},
\end{eqnarray}
with
\begin{widetext}
\begin{eqnarray}
{\cal L}_{\mathrm{int}}^{\mathrm{sing}}
&=&
 \frac{N_{c}^{2}-1}{2N_{c}^{2}}G_{c}
\sum_{i=1}^{N_{f}}
\Bigl\{
(\bar{\psi}_{i}\Psi_{v}) (\bar{\Psi}_{v}\psi_{i}) + (\bar{\psi}_{i}i\gamma_{5}\Psi_{v}) (\bar{\Psi}_{v}i\gamma_{5}\psi_{i})
 - \frac{1}{2} (\bar{\psi}_{i}\gamma^{\mu}\Psi_{v}) (\bar{\Psi}_{v}\gamma_{\mu}\psi_{i})
-\frac{1}{2} (\bar{\psi}_{i}\gamma^{\mu}\gamma_{5}\Psi_{v}) (\bar{\Psi}_{v}\gamma_{\mu}\gamma_{5}\psi_{i})
\Bigr\},
\nonumber \\
\label{eq:int_L_singlet} \\
{\cal L}_{\mathrm{int}}^{\mathrm{oct}}
&=&
-
\frac{1}{4N_{c}}G_{c} \sum_{i=1}^{N_{f}} \sum_{a=1}^{N_{c}^{2}-1}
\Bigl\{
(\bar{\psi}_{i}\lambda^{a}\Psi_{v}) (\bar{\Psi}_{v}\lambda^{a}\psi_{i}) + (\bar{\psi}_{i}i\gamma_{5}\lambda^{a}\Psi_{v}) (\bar{\Psi}_{v}i\gamma_{5}\lambda^{a}\psi_{i}) - \frac{1}{2} (\bar{\psi}_{i}\gamma^{\mu}\lambda^{a}\Psi_{v}) (\bar{\Psi}_{v}\gamma_{\mu}\lambda^{a}\psi_{i}) 
\nonumber \\
&& \hspace{8em}
-\frac{1}{2} (\bar{\psi}_{i}\gamma^{\mu}\gamma_{5}\lambda^{a}\Psi_{v}) (\bar{\Psi}_{v}\gamma_{\mu}\gamma_{5}\lambda^{a}\psi_{i}) 
\Bigr\},
\label{eq:int_L_octet}
\end{eqnarray}
\end{widetext}
where the light flavor indices $i=1,\dots,N_{f}$ are explicitly shown. The minus sign due to the exchange of the fermion fields is included. We notice that the coupling constant of the color-octet part ${\cal L}_{\mathrm{int}}^{\mathrm{oct}}$ is suppressed by a factor of $1/N_{c}$, while that of the color-singlet part ${\cal L}_{\mathrm{int}}^{\mathrm{sing}}$ is not. Therefore, in the following, we consider only the color-singlet part as the dominant interaction~\footnote{\label{fn:Fierz}Instead of Eq.~(\ref{eq:Fierz_SUN_2}), we applied a different type of the Fierz transformation in the previous works~\cite{Yasui:2016svc,Yasui:2016yet}:
\begin{eqnarray}
 \sum_{a=1}^{N_{c}^{2}-1} (\lambda^{a})_{ij} (\lambda^{a})_{kl} = 2 \delta_{il} \delta_{kj} - \frac{2}{N_{c}} \delta_{ij} \delta_{kl}.
 \label{eq:Fierz_color_0}
\end{eqnarray}
We ignored the second term and performed the mean-field approximation for the interaction coming from the first term. In this case, however, the color singlet and octet parts are not completely separated, and thus the strength of the dominant interaction is different from that of the color-singlet interaction in the present article.}.

By using the relation $\gamma^{0}\Psi_{v}=\Psi_{v}$ for the heavy quark in the rest frame, one can rewrite the singlet interaction~(\ref{eq:int_L_singlet}) as
\begin{widetext}
\begin{eqnarray}
{\cal L}_{\mathrm{int}}^{\mathrm{sing}}
=
\frac{N_{c}^{2}-1}{4N_{c}^{2}} G_{c}
\sum_{i=1}^{N_{f}}
\Bigl\{
(\bar{\psi}_{i}\Psi_{v}) (\bar{\Psi}_{v}\psi_{i}) + (\bar{\psi}_{i}i\gamma_{5}\Psi_{v}) (\bar{\Psi}_{v}i\gamma_{5}\psi_{i}) 
 + (\bar{\psi}_{i}\vec{\gamma}\Psi_{v}) (\bar{\Psi}_{v}\vec{\gamma}\psi_{i})
 + (\bar{\psi}_{i}\vec{\gamma}\gamma_{5}\Psi_{v}) (\bar{\Psi}_{v}\vec{\gamma}\gamma_{5}\psi_{i})
\Bigr\},
 \label{eq:Lint_RL_Nf_singlet}
\end{eqnarray}
\end{widetext}
which is composed of the scalar, pseudoscalar, vector and axial vector terms.
Before we perform the mean-field approximation, let us rewrite each term 
\begin{widetext}
\begin{eqnarray}
\frac{N_{c}^{2}-1}{4N_{c}^{2}} G_{c} (\bar{\psi}_{i}\Psi_{v}) (\bar{\Psi}_{v}\psi_{i})
&=& 
(\bar{\Psi}_{v}\psi_{i}) \Phi_{i} + \Phi_{i}^{\dag} (\bar{\psi}_{i}\Psi_{v}) - \frac{4N_{c}^{2}}{(N_{c}^{2}-1)G_{c}} |\Phi_{i}|^{2}, \\
\frac{N_{c}^{2}-1}{4N_{c}^{2}} G_{c} (\bar{\psi}_{i} i\gamma_{5} \Psi_{v}) (\bar{\Psi}_{v} i\gamma_{5} \psi_{i})
&=& 
(\bar{\Psi}_{v} i\gamma_{5} \psi_{i}) \Phi_{i5} + \Phi_{i5}^{\dag} (\bar{\psi}_{i} i\gamma_{5} \Psi_{v}) - \frac{4N_{c}^{2}}{(N_{c}^{2}-1)G_{c}} |\Phi_{i5}|^{2}, \\
\frac{N_{c}^{2}-1}{4N_{c}^{2}} G_{c} (\bar{\psi}_{i} \vec{\gamma} \Psi_{v}) (\bar{\Psi}_{v} \vec{\gamma} \psi_{i})
&=& 
(\bar{\Psi}_{v} \vec{\gamma} \psi_{i}) \vec{\Phi}_{i} + \vec{\Phi}_{i}^{\dag} (\bar{\psi}_{i}\vec{\gamma} \Psi_{v}) - \frac{4N_{c}^{2}}{(N_{c}^{2}-1)G_{c}} |\vec{\Phi}_{i}|^{2}, \\
\frac{N_{c}^{2}-1}{4N_{c}^{2}} G_{c} (\bar{\psi}_{i} \vec{\gamma}\gamma_{5} \Psi_{v}) (\bar{\Psi}_{v} \vec{\gamma}\gamma_{5} \psi_{i})
&=& 
(\bar{\Psi}_{v} \vec{\gamma}\gamma_{5} \psi_{i}) \vec{\Phi}_{i5} + \vec{\Phi}_{i5}^{\dag} (\bar{\psi}_{i}\vec{\gamma}\gamma_{5} \Psi_{v}) - \frac{4N_{c}^{2}}{(N_{c}^{2}-1)G_{c}} |\vec{\Phi}_{i5}|^{2},
\end{eqnarray}
\end{widetext}
with the bosonic fields defined by 
\begin{eqnarray}
\Phi_{i}&=&\frac{(N_{c}^{2}-1)G_{c}}{4N_{c}^{2}}\bar{\psi}_{i}\Psi_{v}, \label{Phi}\\
\Phi_{i5}&=&\frac{(N_{c}^{2}-1)G_{c}}{4N_{c}^{2}}\bar{\psi}_{i}i\gamma_{5}\Psi_{v}, \\
\vec{\Phi}_{i}&=&\frac{(N_{c}^{2}-1)G_{c}}{4N_{c}^{2}}\bar{\psi}_{i}\vec{\gamma}\Psi_{v}, \\
\vec{\Phi}_{i5}&=&\frac{(N_{c}^{2}-1)G_{c}}{4N_{c}^{2}}\bar{\psi}_{i}\vec{\gamma}\gamma_{5}\Psi_{v},\label{Phi_vec5}
\end{eqnarray}
for the scalar, pseudoscalar, vector and axial vector terms, respectively. The expectation values of these bosonic fields are the ``order parameters" for the QCD Kondo effect, corresponding to the correlations between the hole of the light quark $h$ and the heavy quark $Q$. Then, the singlet interaction  (\ref{eq:Lint_RL_Nf_singlet}) can be rewritten as 
\begin{eqnarray}
 {\cal L}_{\mathrm{int}}^{\mathrm{sing}}
&=& 
\sum_{i=1}^{N_{f}}
\biggl[
\bar{\Psi}_{v} \bigl( \Phi_{i} + \vec{\gamma} \!\cdot\! \vec{\Phi}_{i} + i\gamma_{5} \Phi _{i5}+ \vec{\gamma}\gamma_{5} \!\cdot\! \vec{\Phi}_{i5} \bigr) \psi_{i}
\nonumber \\
&&
\hspace{0em} +
\bar{\psi}_{i} \bigl( \Phi_{i}^{\dag} + \vec{\gamma} \!\cdot\! \vec{\Phi}_{i}^{\dag} + i\gamma_{5} \Phi_{i5}^{\dag} + \vec{\gamma}\gamma_{5} \!\cdot\! \vec{\Phi}_{i5}^{\dag} \bigr) \Psi_{v}
\nonumber \\
&&
\hspace{0em}
- \frac{4N_{c}^{2}}{(N_{c}^{2}-1)G_{c}}
\Bigl( |\Phi_{i}|^{2} + |\vec{\Phi}_{i}|^{2} + |\Phi_{i5}|^{2} + |\vec{\Phi}_{i5}|^{2} \Bigr)
\biggr]. \nonumber \\
\label{singlet_int}
\end{eqnarray}

We assume that the order parameters have the following structure in the ground state: 
\begin{eqnarray}
\langle \Phi_{i} \rangle=\Delta \delta_{i1}, \quad 
\langle \Phi_{i5} \rangle =0,\label{orderparameter1}\\
\langle \vec{\Phi}_{i} \rangle=\vec{\Delta}\delta_{i1}, \quad 
\langle \vec{\Phi}_{i5} \rangle=0,\label{orderparameter2}
\end{eqnarray} 
where we have introduced a complex number $\Delta$ and a complex vector $\vec{\Delta}$ and chosen  the light flavor direction $i=1$ which could be arbitrary reflecting the flavor symmetry. We call these condensates the {\it Kondo condensates}. We further introduce the hedgehog ansatz in momentum space:
\begin{eqnarray}
 \vec{\Delta} = \Delta \, \hat{p},\quad \hat{p}=\frac{\vec{p}}{|\vec{p}|},
\label{hedgehog}
\end{eqnarray}
for the three-dimensional momentum $\vec{p}$ of the light quark which can be also identified with the off-mass-shell momentum of the heavy quark. In the previous works~\cite{Yasui:2016svc,Yasui:2016yet}, the explicit form of the Kondo condensate was specified for a heavy-light operator $\langle \bar\psi_\alpha \Psi_{v\delta}\rangle$ with $\alpha,\delta$ being the Lorentz indices as $\langle \bar\psi_\alpha \Psi_{v\delta}\rangle\sim \Delta \left( \frac{1+\gamma_0}{2}(1-\hat p\cdot \vec \gamma) \right)_{\delta\alpha}$. This structure with the $\hat p\cdot \vec \gamma$ term was found from the completeness relation for heavy quark fields. By contracting the operator $\langle \bar\psi_\alpha \Psi_{v\delta}\rangle$ with ${\bf 1}_{\alpha\delta}$, $\vec \gamma_{\alpha\delta}$ and so on, one can easily confirm that this structure is equivalent to the hedgehog ansatz (\ref{hedgehog}) with Eqs.~(\ref{orderparameter1}) and (\ref{orderparameter2}). Although our choice of the Kondo condensate is still an ansatz and it is not easy to check that it gives the lowest energy among all the possible forms of condensates, there are a few reasons to expect that it indeed gives the lowest energy. First of all, as we will see below, our choice of course gives lower energy than the state with vanishing condensates. Next, our choice of condensate allows for a mixing between a light quark and a heavy quark in a clear manner while the other types of condensates induce more entangled mixing patterns. Third, if there was another lower energy state, our choice of the ground state would be unstable against fluctuations. As we will discuss later, however, there is no instability in the fluctuations around the Kondo condensates with the hedgehog ansatz, and thus our choice will not fall into other states.  Last,  an analysis in a simpler model supports our choice: As we discuss in Appendix C, a model with a Weyl fermion and a heavy quark allows us to directly compare two cases: (i) only a scalar (or vector) condensate $\langle \Phi_i\rangle=\Delta \delta_{i1}$ or $\langle \vec \Phi_i\rangle=\Delta \hat p \delta_{i1}$ (and the others are vanishing) and (ii) nonzero scalar and vector condensates coexisting with the hedgehog ansatz $\langle \Phi_i\rangle=\Delta \delta_{i1},\langle \vec \Phi_i\rangle=\Delta \hat p\, \delta_{i1}$. The latter case gives lower energy than the former case.

If we simply replace the bosonic operators (\ref{Phi})-(\ref{Phi_vec5}) in the Lagrangian (with only the singlet interaction (\ref{singlet_int})) by their expectation values (mean-fields) given in Eqs.~(\ref{orderparameter1}) and (\ref{orderparameter2}), we obtain the mean-field Lagrangian in momentum space:
\begin{eqnarray}
{\cal L}_{\mathrm{MF}}^{\mathrm{sing}}
&=&
\sum_{i=1}^{N_{f}}
\bar{\psi}_{i} ( p \hspace{-0.4em}/ + \mu \gamma_{0} ) \psi_{i}
+ \bar{\Psi}_{v} v \!\cdot\!  p \Psi_{v}
+
\Delta \bar{\Psi}_{v} \bigl( 1 + \vec{\gamma} \!\cdot\! \hat{p} \bigr) \psi_{1}
\nonumber \\
&&
+
\Delta^{\ast} \bar{\psi}_{1} \bigl( 1 + \vec{\gamma} \!\cdot\! \hat{p} \bigr) \Psi_{v}
- \frac{8N_{c}^{2}}{(N_{c}^{2}-1)G_{c}} |\Delta|^{2}
\nonumber \\
&&
-\lambda \! \left( \bar{\Psi}_{v} \Psi_{v} - n_{Q} \right).
\label{eq:L_RL_Nf_MF}
\end{eqnarray}
We notice that the $\vec{\gamma} \!\cdot\! \hat{p}$ terms appear due to the hedgehog ansatz.
We emphasize that only the light quark field $\psi_{1}$ ($i=1$) couples to the heavy quark due to our choice of the light flavor axis of the condensate (see Eqs.~(\ref{orderparameter1}) and (\ref{orderparameter2})). The mean-field Lagrangian will be used to determine the ground state in the presence of the Kondo condensate. Stability of the ground state and excitation modes will be discussed in the next section by using the Lagrangian that includes fluctuations around the mean fields.

Notice that the mean-field Lagrangian (\ref{eq:L_RL_Nf_MF}) is bilinear with respect to fermion fields $\psi$ and $\Psi_v$. Thus it is convenient to introduce the following shorthand notation: 
\begin{eqnarray}
\phi \equiv 
\left( \!\!\!
\begin{array}{c}
 \psi_{1} \\
 \psi_{2} \\
 \vdots \\
 \Psi_{v}  
\end{array}
\!\!\! \right),
\hspace{2em}
 \bar{\phi} \equiv (\bar{\psi}_{1}, \bar{\psi}_{2}, \dots, \bar{\Psi}_{v}).
 \label{eq:phi_def}
\end{eqnarray}
Then, the mean-field Lagrangian (\ref{eq:L_RL_Nf_MF}) can be rewritten in a compact form:
\begin{eqnarray}
\hspace{-1em}
{\cal L}_{\mathrm{MF}}^{\mathrm{sing}}
=
\bar{\phi} \,
 G(p_{0},\vec{p}\,)^{-1}
\phi
-\frac{8N_{c}^{2}}{(N_{c}^{2}-1)G_{c}} |\Delta|^{2} + \lambda n_{Q},
\label{eq:L_RL_Nf_MF2}
\end{eqnarray}
with the inverse of the propagator given by
\begin{widetext}
\begin{eqnarray}
 G(p_{0},\vec{p}\,)^{-1}
\equiv
\left(
\begin{array}{ccccc}
 p \hspace{-0.4em}/ + \mu \gamma_{0} &  0 & \hdots & 0 & \Delta^{\ast}(1+\vec{\gamma}\!\cdot\!\hat{p}\,)\\
 0  & p \hspace{-0.4em}/ + \mu \gamma_{0}  & \hdots & 0 & 0 \\
 \vdots & \vdots  &  \ddots  & \vdots & \vdots \\ 
 0 & 0 & \hdots & 0 & 0 \\
 \Delta (1+\vec{\gamma}\!\cdot\!\hat{p}\,) & 0 & \hdots & 0 & p_{0}-\lambda 
\end{array}
\right),
\label{eq:Ginverse_RL_Nf}
\end{eqnarray}
\end{widetext}
as the $(N_{f}+1)\times(N_{f}+1)$ dimensional matrix in flavor space including the light quarks and the heavy quark.

\begin{figure}[tbp]
  \begin{center}
 \begin{minipage}{0.49\linewidth}
  \hspace*{-8em}
     \centering
   \includegraphics[keepaspectratio,scale=0.3]{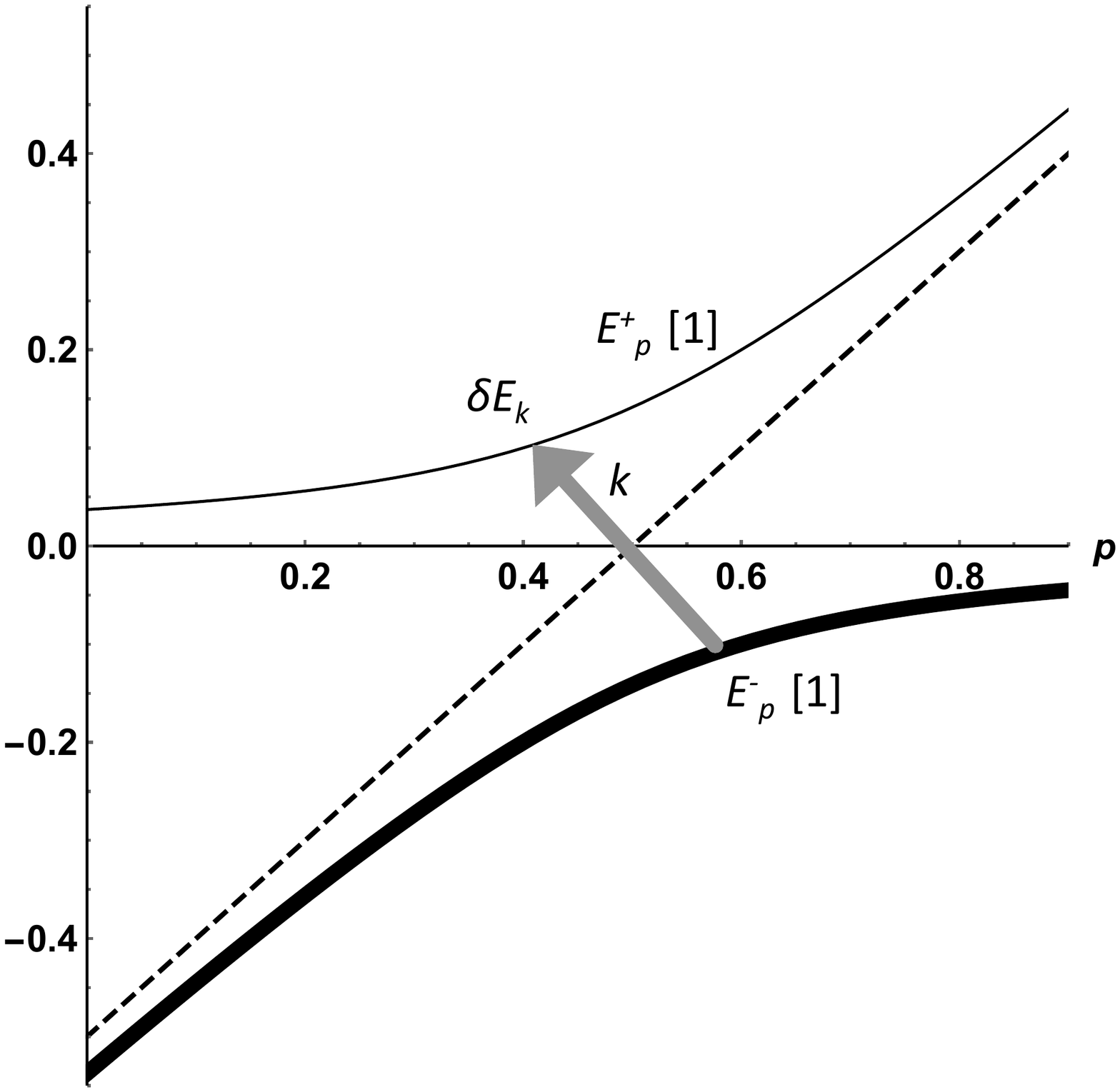}
 \end{minipage} \\
 \begin{minipage}{0.49\linewidth}
  \hspace*{-8em}
     \centering
   \includegraphics[keepaspectratio,scale=0.3]{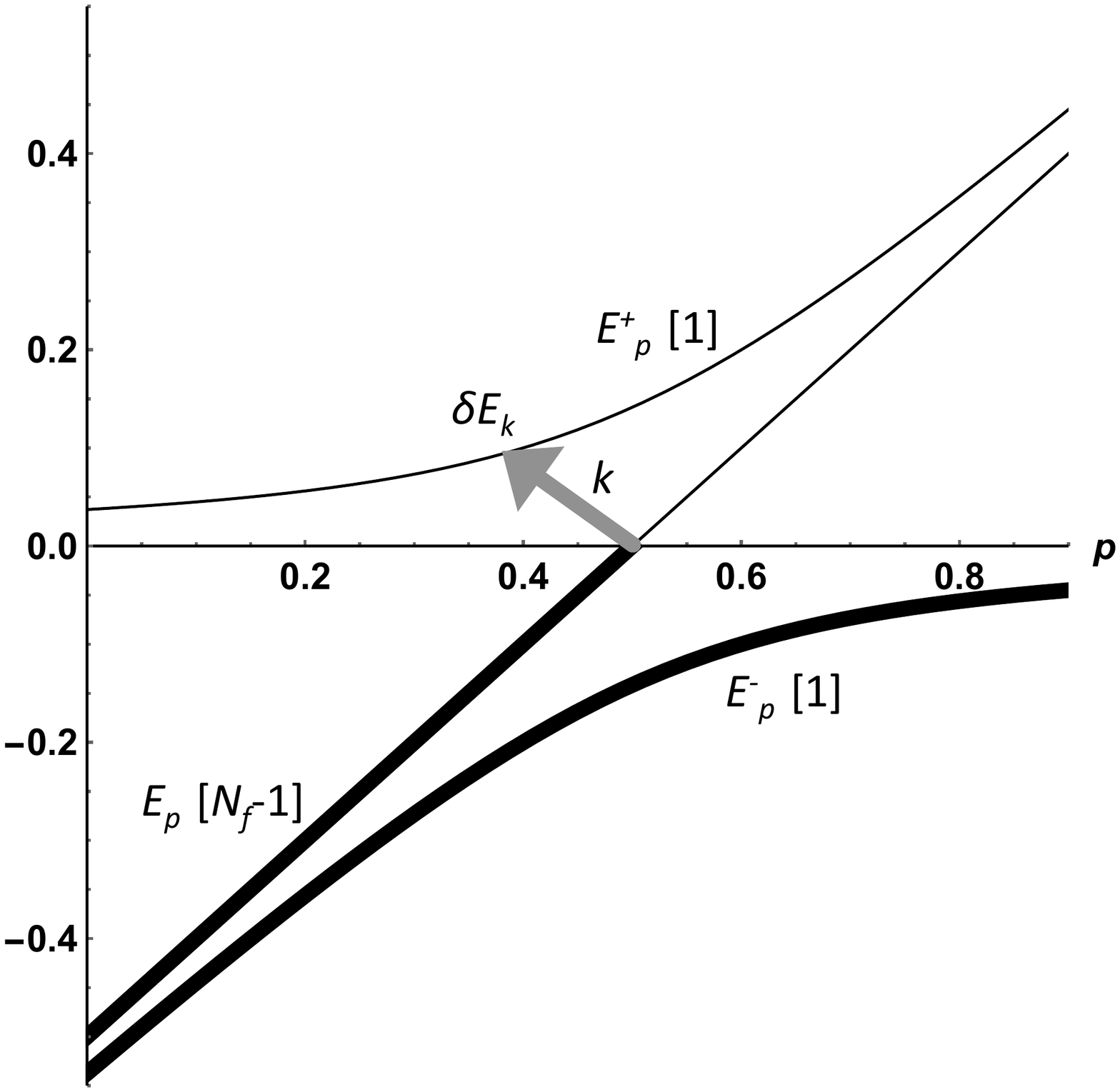}
 \end{minipage}
 \vspace*{-1em}
   \end{center}
 \caption{
Schematic figures of the energy-momentum dispersion relations in the presence of the Kondo condensate: $N_{f}=1$ (top panel) and $N_{f} \ge 2$ (bottom panel). The units are arbitrary, but we here fixed $\mu=0.5$ and $\lambda=0$. The origin of the vertical axis corresponds to the Fermi surface. The thick solid lines are the occupied states below the Fermi sphere. The dashed line for $N_{f}=1$ means the dispersion of a free light quark at finite density. The numbers in the square brackets indicate the number of degeneracy for each of spin up and down. The arrows stand for the excitations from the quark in the ground state (cf.~Sec.~\ref{sec:hole_Q_modes}). 
 }
  \label{fig:dispersion_MF}
\end{figure}

The poles of $G(p_{0},\vec{p}\,)$ in the energy $p_0$ plane correspond to physical modes in the presence of the Kondo condensate. Reflecting the original $2N_f+1$ degrees of freedom (for spin ``up") including $N_f$ light quarks, $N_f$ light antiquarks, and one heavy quark (without a heavy antiquark), there should appear $2N_f+1$ modes. Indeed, one can easily confirm that there are $2N_f+1$ dispersion relations as the poles of $G(p_{0},\vec{p}\,)$. As we already commented before, the Kondo condensate of the hedgehog type induces a mixing between the particle mode of $\psi_1$ and $\Psi_v$ (that has the particle (or positive energy) mode alone) leaving the other particle modes and all the antiparticle modes unchanged. Namely, the mixing between the particle mode of $\psi_1$ and $\Psi_v$ gives the following two modes:
\begin{eqnarray}
E_{p}^{\pm} = \frac{1}{2} \Bigl( p-\mu+\lambda \pm \sqrt{(p-\mu-\lambda)^{2}+8|\Delta|^{2}} \Bigr),
\label{eq:eigen_energy1}
\end{eqnarray}
while all the other modes are unchanged: the particle modes of the other light quarks $\psi_i$ ($i\neq 1$) 
\begin{eqnarray}
E_{p} = p-\mu, 
\label{eq:eigen_energy2}
\end{eqnarray}
and all the antiparticle modes of light quarks 
\begin{eqnarray}
\widetilde{E}_{p} = -p-\mu.
\label{eq:eigen_energy3}
\end{eqnarray}
Here $p$ is the three dimensional momentum $p=|\vec{p}\,|$. 
These dispersion relations are common for spin ``up" and ``down" components. In Sec. \ref{sec:hedgehog}, we will see that the spin ``up" and ``down" correspond to the chirality $\pm1$ of the light quark and the helicity $\pm1$ of the heavy quark. We display schematic figures of the energy-momentum dispersion relations for $N_{f}=1$ and $N_{f} \ge 2$ in Fig.\ref{fig:dispersion_MF}.
At zero temperature, all the states below the Fermi surface ($E=0$) are occupied as shown by the thick solid lines.
In the case of $N_{f}=1$, the branch of Eq.~(\ref{eq:eigen_energy2}) is absent, as indicated by the dashed line. As is evident from the figure, due to the presence of a nonzero $\Delta$, two original dispersions ($E_q=p-\mu$ for a light quark and $E_Q=\lambda$ for a heavy quark) show level crossing at $p=\mu+\lambda$. The magnitude of $\Delta$ controls the region of mixing: a larger $\Delta$ induces mixing in a wider momentum region. After the mixing, the upper (lower) mode with dispersion $E_p^+$ ($E_p^-$) is more like a heavy (light) quark at small $p \ll \mu+\lambda$ and more like a light (heavy) quark at large $p\gg \mu+\lambda$.

For later convenience, let us show the explicit form of the propagator $G(p_0, \vec p)$ which is an inverse of $G(p_0,\vec p)^{-1}$ given in Eq.~(\ref{eq:Ginverse_RL_Nf}). Here we show the propagator in the case for $N_f=1$. The propagator for $N_f>1$ is easily obtained since the mixing occurs only between $\psi_1$ and $\Psi_v$. The propagator in the space of $\psi_1$ and (the upper component of) $\Psi_v$ is given as 
\begin{widetext}
\begin{eqnarray}
G(p_0,\vec p)&=&\frac{1}{(p_0-\tilde E_p)(p_0-E_p^+)(p_0-E_p^-)}\nonumber\\
&&\times \left(
\begin{array}{ccc}
(p_0-\lambda)(p_0+\mu)-|\Delta|^2 & -\{p(p_0-\lambda)+|\Delta|^2\}\hat p \cdot \vec \sigma & -\Delta^* (p_0+p+\mu)\\
\{p(p_0-\lambda)+|\Delta|^2\}\hat p \cdot \vec \sigma & - (p_0-\lambda)(p_0+\mu)+|\Delta|^2 & -\Delta^* (p_0+p+\mu)\hat p \cdot \vec \sigma\\
-\Delta (p_0+p+\mu)& \Delta (p_0+p+\mu)\hat p \cdot \vec \sigma& (p_0-p+\mu)(p_0+p+\mu)
\end{array}
\right), \label{propagator_Nf=1}
\end{eqnarray}
\end{widetext}
with $p=|\vec p|$. For $N_f=1$, there are only three poles $p_0=\tilde E_p, E_p^\pm$ as mentioned before (see the upper panel of Fig.~\ref{fig:dispersion_MF} where we show two dispersions $p_0=E_p^\pm$).

Analysis in the simpler model with a Weyl fermion (see Appendix~\ref{sec:Weyl_right}) is helpful to understand how the Kondo condensate occurs. As mentioned before, this model also leads to the Kondo condensates in the scalar and vector channels with the hedgehog ansatz. If the Weyl fermion $\chi$ is right-handed, the Kondo condensate is constructed by the linear combination of $\chi$ and the spin ``up" component of the heavy quark $\Psi_{v \uparrow}$. In contrast, the spin ``down" component $\Psi_{v \downarrow}$ would couple to a left-handed Weyl fermion $\varphi$ which is absent in the simple model, and thus does not form the Kondo condensate. Coming back to our present study, the Dirac field $\psi$ contains both the right-handed and left-handed Weyl fermions. Therefore, the two types of condensates, the mixing between $\chi$ and $\Psi_{v\uparrow}$ and the mixing between $\varphi$ and $\Psi_{v\downarrow}$, exist simultaneously.

%%%%%%%%%%%%%%%%%%%%%%%%%%%%%%%%%%%
%%%%%%%%%%%%%%%%%%%%%%%%%%%%%%%%%%%

\subsection{Thermodynamic potential and gap equation}
\label{sec:gap_equation}

We compute the thermodynamic potential $\Omega$ from the mean-field Lagrangian (\ref{eq:L_RL_Nf_MF}) or (\ref{eq:L_RL_Nf_MF2}), then derive the gap equation for $\Delta$ as the stationary condition for $\Omega$ with respect to $\Delta$. By using the dispersion relations (\ref{eq:eigen_energy1})-(\ref{eq:eigen_energy3}), the thermodynamic potential is given by
\begin{eqnarray}
 \Omega(T,\mu,\lambda;\Delta)
 &=& 2N_{c}\! \int_{0}^{\Lambda}  \frac{p^{2}\mathrm{d}p}{2\pi^{2}} \, 
f(T,\mu,\lambda;p)
 \nonumber \\
&& +\, \frac{8N_{c}^{2}}{(N_{c}^{2}-1)G_{c}} |\Delta|^{2} - \lambda n_{Q}, \label{eq:thermo_potential}
\end{eqnarray}
where $\beta$ is the inverse temperature $\beta=1/T$ and 
\begin{eqnarray}
f(T,\mu,\lambda;p)
&=& -\frac{1}{\beta} 
\left[ \ln  (1+e^{-\beta E_{p}^{+}}) + \ln (1+e^{-\beta E_{p}^{-}}) \right. 
\nonumber \\
&& \left. + \ln (1+e^{-\beta E_{p}})^{N_{f}\!-\!1} + \ln (1+e^{-\beta \widetilde{E}_{p}})^{N_{f}} \right].
\nonumber
\end{eqnarray}
The factor two in front of the integral comes from the spin degeneracy, and $\Lambda$ is the three-momentum cutoff to regularize the ultraviolet divergence. In addition to an external parameter $T$, the thermodynamic potential depends on three parameters $\mu$, $\lambda$ and $\Delta$. All of them are dynamically determined by the stationary conditions with respect to each parameter. In particular, the value of $|\Delta|$ is obtained from the ``gap equation", namely
\begin{eqnarray}
 \frac{\partial }{\partial \Delta^{\ast}} \Omega(T,\mu,\lambda;\Delta) = 0.
 \label{eq:gap_equation}
\end{eqnarray}
Similarly, the chemical potential $\mu$ and the Lagrange multiplier $\lambda$ are determined for fixed values of the light quark number density $n_{q}$ and the heavy quark number density $n_{Q}$ by 
\begin{eqnarray}
 \frac{\partial }{\partial \mu} \Omega(T,\mu,\lambda;\Delta) &=& -n_{q}, \label{eq:nq_stationary} \\
 \frac{\partial }{\partial \lambda} \Omega(T,\mu,\lambda;\Delta) &=& 0. \label{eq:nQ_stationary}
\end{eqnarray}
To find solutions to Eqs.~(\ref{eq:gap_equation})-(\ref{eq:nQ_stationary}) simultaneously corresponds to determining the ground state.

Analytic evaluation of the thermodynamic potential and the gap equation is possible for $T=0$ ($\beta\rightarrow \infty$) and $\lambda=0$. Notice that, in the limit $\beta\to \infty$, the factor $-\frac{1}{\beta}\ln (1+e^{-\beta E})$ in $f(T;\mu,\lambda;\Delta)$ simplifies to $E$ (or $0$) for $E\le 0$ (or $E>0$). Thus, in the zero temperature limit, only the modes below the Fermi energy ($E_p^-<0$, $E_p<0$ and $\widetilde E_p<0$) survive. Thus we find 
\begin{eqnarray}
&& \hspace{-5mm}\Omega(T=0,\mu,\lambda=0;\Delta)\nonumber\\
&&=
2N_{c} \! \int_{0}^{\Lambda}\! \frac{p^{2}\mathrm{d}p}{2\pi^{2}} \, E_{p}^{-}
\, +\, 
2N_{c}(N_{f}\!-\!1) \! \int_{0}^{\mu} \! \frac{p^{2}\mathrm{d}p}{2\pi^{2}} \, E_{p}
\nonumber \\
&&
+\, 
2N_{c}N_{f} \! \int_{0}^{\Lambda} \!\frac{p^{2}\mathrm{d}p}{2\pi^{2}} \, \widetilde{E}_{p} \, 
+\, \frac{8N_{c}^{2}}{(N_{c}^{2}-1)G_{c}} |\Delta|^{2}.
\label{eq:RL_thermo_0}
\end{eqnarray}
From the stationary condition (\ref{eq:gap_equation}), we obtain the gap equation 
\begin{eqnarray}
 \Delta = \frac{N_{c}^{2}-1}{4N_{c}}\, G_{c}\!\! \int_{0}^{\Lambda}\! \frac{p^{2}\mathrm{d}p}{2\pi^{2}}\, \frac{\Delta}{\frac{1}{2}\sqrt{(p-\mu)^{2}+8|\Delta|^{2}}}.
  \label{eq:gap_equation_0}
\end{eqnarray}
Notice that the gap equation is independent of $N_f$ because the $N_f$ dependent terms in $\Omega$ do not depend on the gap $\Delta$. This is natural because the Kondo condensate is formed by a single light quark and a heavy quark impurity. All the other ($N_f-1$) light quarks do not participate in the Kondo effect. In addition to a trivial solution $|\Delta|=0$, we find a nonzero solution to the gap equation: 
\begin{eqnarray}
|\Delta|
\simeq
\alpha \sqrt{\frac{(\Lambda-\mu)\mu}{2}}  \exp \left\{ - \frac{2\pi^{2}}{(N_{c}-1/N_{c})G_{c}\mu^{2}} \right\},
 \label{gap_solution}
\end{eqnarray}
with $\alpha=\exp\left\{(\Lambda^{2}+2\Lambda\mu-6\mu^{2})/(4\mu^{2}) \right\}$. This approximate solution was obtained under the assumption that $|\Delta|$ is much smaller than $\mu$ and $\Lambda$. Some comments are in order: First of all, as is evident from the dependence on $G_c$ which is very similar to the superconductivity gap, the gap (\ref{gap_solution}) becomes larger as the coupling $G_c$ becomes stronger. This implies that the Kondo effect becomes stronger with increasing value of $G_c$, which is intuitively acceptable. Second, the gap increases also with increasing $\mu$. Since the momentum cutoff $\Lambda$ must be taken larger than $\mu$, we may write it as $\Lambda = \kappa \mu$ with a parameter $\kappa>1$, which immediately implies that the gap (\ref{gap_solution}) increases with $\mu$. Last, by drawing the thermodynamical potential (\ref{eq:RL_thermo_0}) as a function of $|\Delta |$, one finds that the nontrivial value of the gap (\ref{gap_solution}) corresponds to the minimum of the potential and thus gives lower energy than the trivial one $|\Delta |=0$ (see Ref.~\cite{Yasui:2016svc}). If we take the parameter set $G_{c} = 2\times \frac{9}{2} \times 2.0 / \Lambda^{2}$, $\Lambda = 0.65$ GeV, $N_{c}=3$, $\mu=0.5$ GeV, we obtain $|\Delta|=0.085$ GeV.
We note that we can numerically investigate the finite temperature $T$ and $\lambda$ dependences of the thermodynamic potential and the gap, and can draw a phase diagram where the Kondo condensate is nonzero (the Kondo phase). For example, on the $\mu-\lambda$ plane at $T=0$, the Kondo phase appears at large $\mu$ and relatively small $\lambda$. See Ref.~\cite{Yasui:2016svc} for more figures.

%%%%%%%%%%%%%%%%%%%%%%%%%%%%%%%%%%%
\subsection{Another form of gap equation}
\label{sec:another_gap}

We defined in Eq.~(\ref{eq:phi_def}) a shorthand notation $\phi$ for the light and heavy quarks $\psi$ and $\Psi_v$. Correspondingly, let us also define the vertices $\Gamma_{\ell}^{(i)}$ and $\bar{\Gamma}_{\ell}^{(i)}$ which are $(N_{f}+1)\times(N_{f}+1)$ dimensional matrices in flavor space. The lower index $\ell$ distinguishes the vertex type: $ \ell =\{\mathrm{S}, \mathrm{V}, \mathrm{P}, \mathrm{A}\}$ (V and A further have spatial indices $k=1,2,3$) and the upper index $i$ is for flavors: $i=1,\dots,N_{f}$. 
Their explicit forms are defined by
\begin{eqnarray}
&& \Gamma_{\mathrm{S}}^{(1)}
\!\!=\!\!
\left(\!
\begin{array}{cccc}
 0 &  0 & \!\!\!\hdots\!\!\!  & 0 \\
 0  & 0 & \!\!\!\hdots\!\!\! & 0 \\
 \vdots & \vdots  &  \!\!\!\ddots\!\!\!  & \vdots \\ 
 1 & 0 & \!\!\!\hdots\!\!\!  & 0
\end{array}\!
\right)\!,
\hspace{0.1em}
 \Gamma_{\mathrm{S}}^{(2)}
\!\!=\!\!
\left(\!
\begin{array}{cccc}
 0 &  0 & \!\!\!\hdots\!\!\!  & 0 \\
 0  & 0 & \!\!\!\hdots\!\!\! & 0 \\
 \vdots & \vdots  &  \!\!\!\ddots\!\!\!  & \vdots \\ 
 0 & 1 & \!\!\!\hdots\!\!\!  & 0
\end{array}\!
\right)\!,
\dots,
\nonumber \\
&&
 \bar{\Gamma}_{\mathrm{S}}^{(1)}
\!\!=\!\!
\left(\!
\begin{array}{cccc}
 0 &  0 & \!\!\!\hdots\!\!\!  & 1 \\
 0  & 0 & \!\!\!\hdots\!\!\! & 0 \\
 \vdots & \vdots  &  \!\!\!\ddots\!\!\!  & \vdots \\ 
 0 & 0 & \!\!\!\hdots\!\!\!  & 0
\end{array}\!
\right)\!,
\hspace{0.1em}
 \bar{\Gamma}_{\mathrm{S}}^{(2)}
\!\!=\!\!
\left(\!
\begin{array}{cccc}
 0 &  0 & \!\!\!\hdots\!\!\!  & 0 \\
 0  & 0 & \!\!\!\hdots\!\!\! & 1 \\
 \vdots & \vdots  &  \!\!\!\ddots\!\!\!  & \vdots \\ 
 0 & 0 & \!\!\!\hdots\!\!\!  & 0
\end{array}\!
\right)\!,  \dots,
\label{eq:scalar_matrices}
\end{eqnarray}
for the scalar channel, 
\begin{eqnarray}
 && \Gamma_{\mathrm{V},k}^{(1)}
\!\!=\!\!
\left(\!
\begin{array}{cccc}
 0 &  0 & \!\!\!\hdots\!\!\!  & 0 \\
 0  & 0 & \!\!\!\hdots\!\!\! & 0 \\
 \vdots & \vdots  &  \!\!\!\ddots\!\!\!  & \vdots \\ 
 \gamma^{k} & 0 & \!\!\!\hdots\!\!\!  & 0
\end{array}\!
\right)\!,
\hspace{0.1em}
 \Gamma_{\mathrm{V},k}^{(2)}
\!\!=\!\!
\left(\!
\begin{array}{cccc}
 0 &  0 & \!\!\!\hdots\!\!\!  & 0 \\
 0  & 0 & \!\!\!\hdots\!\!\! & 0 \\
 \vdots & \vdots  &  \!\!\!\ddots\!\!\!  & \vdots \\ 
 0 & \gamma^{k} & \!\!\!\hdots\!\!\!  & 0
\end{array}\!
\right)\!,
\dots,
\nonumber \\
&&
 \bar{\Gamma}_{\mathrm{V},k}^{(1)}
\!\!=\!\!
\left(\!
\begin{array}{cccc}
 0 &  0 & \!\!\!\hdots\!\!\!  & \gamma^{k} \\
 0  & 0 & \!\!\!\hdots\!\!\! & 0 \\
 \vdots & \vdots  &  \!\!\!\ddots\!\!\!  & \vdots \\ 
 0 & 0 & \!\!\!\hdots\!\!\!  & 0
\end{array}\!
\right)\!,
\hspace{0.1em}
 \bar{\Gamma}_{\mathrm{V},k}^{(2)}
\!\!=\!\!
\left(\!
\begin{array}{cccc}
 0 &  0 & \!\!\!\hdots\!\!\!  & 0 \\
 0  & 0 & \!\!\!\hdots\!\!\! & \gamma^{k} \\
 \vdots & \vdots  &  \!\!\!\ddots\!\!\!  & \vdots \\ 
 0 & 0 & \!\!\!\hdots\!\!\!  & 0
\end{array}\!
\right)\!,  \dots,
\label{eq:vector_matrices}
\end{eqnarray}
with $k=1,2,3$ for the vector channel,
\begin{eqnarray}
&& \Gamma_{\mathrm{P}}^{(1)}
\!\!=\!\!
\left(\!
\begin{array}{cccc}
 0 &  0 & \!\!\!\hdots\!\!\!  & 0 \\
 0  & 0 & \!\!\!\hdots\!\!\! & 0 \\
 \vdots & \vdots  &  \!\!\!\ddots\!\!\!  & \vdots \\ 
 i\gamma_{5} & 0 & \!\!\!\hdots\!\!\!  & 0
\end{array}\!
\right)\!,
\hspace{0.1em}
 \Gamma_{\mathrm{P}}^{(2)}
\!\!=\!\!
\left(\!
\begin{array}{cccc}
 0 &  0 & \!\!\!\hdots\!\!\!  & 0 \\
 0  & 0 & \!\!\!\hdots\!\!\! & 0 \\
 \vdots & \vdots  &  \!\!\!\ddots\!\!\!  & \vdots \\ 
 0 & i\gamma_{5} & \!\!\!\hdots\!\!\!  & 0
\end{array}\!
\right)\!,
\dots,
\nonumber \\
&&
 \bar{\Gamma}_{\mathrm{P}}^{(1)}
\!\!=\!\!
\left(\!
\begin{array}{cccc}
 0 &  0 & \!\!\!\hdots\!\!\!  & i\gamma_{5} \\
 0  & 0 & \!\!\!\hdots\!\!\! & 0 \\
 \vdots & \vdots  &  \!\!\!\ddots\!\!\!  & \vdots \\ 
 0 & 0 & \!\!\!\hdots\!\!\!  & 0
\end{array}\!
\right)\!,
\hspace{0.1em}
 \bar{\Gamma}_{\mathrm{P}}^{(2)}
\!\!=\!\!
\left(\!
\begin{array}{cccc}
 0 &  0 & \!\!\!\hdots\!\!\!  & 0 \\
 0  & 0 & \!\!\!\hdots\!\!\! & i\gamma_{5} \\
 \vdots & \vdots  &  \!\!\!\ddots\!\!\!  & \vdots \\ 
 0 & 0 & \!\!\!\hdots\!\!\!  & 0
\end{array}\!
\right)\!,  \dots,
\label{eq:pseudoscalar_matrices}
\end{eqnarray}
for the pseudoscalar channel, and
\begin{eqnarray}
&&  \Gamma_{\mathrm{A},k}^{(1)}
\!\!=\!\!
\left(\!\!
\begin{array}{cccc}
 0 &  0 & \!\!\!\hdots\!\!\!  & 0 \\
 0  & 0 & \!\!\!\hdots\!\!\! & 0 \\
 \vdots & \vdots  &  \!\!\!\ddots\!\!\!  & \vdots \\ 
 \gamma^{k}\gamma_{5} & 0 & \!\!\!\hdots\!\!\!  & 0
\end{array}\!\!
\right)\!,
\hspace{0.1em}
 \Gamma_{\mathrm{A},k}^{(2)}
\!\!=\!\!
\left(\!\!
\begin{array}{cccc}
 0 &  0 & \!\!\!\hdots\!\!\!  & 0 \\
 0  & 0 & \!\!\!\hdots\!\!\! & 0 \\
 \vdots & \vdots  &  \!\!\!\ddots\!\!\!  & \vdots \\ 
 0 & \gamma^{k}\gamma_{5} & \!\!\!\hdots\!\!\!  & 0
\end{array}\!\!
\right)\!,
\dots,
\nonumber \\
&&
 \bar{\Gamma}_{\mathrm{A},k}^{(1)}
\!\!=\!\!
\left(\!\!
\begin{array}{cccc}
 0 &  0 & \!\!\!\hdots\!\!\!  & \gamma^{k}\gamma_{5} \\
 0  & 0 & \!\!\!\hdots\!\!\! & 0 \\
 \vdots & \vdots  &  \!\!\!\ddots\!\!\!  & \vdots \\ 
 0 & 0 & \!\!\!\hdots\!\!\!  & 0
\end{array}\!\!
\right)\!,
\hspace{0.1em}
 \bar{\Gamma}_{\mathrm{A},k}^{(2)}
\!\!=\!\!
\left(\!\!
\begin{array}{cccc}
 0 &  0 & \!\!\!\hdots\!\!\!  & 0 \\
 0  & 0 & \!\!\!\hdots\!\!\! & \gamma^{k}\gamma_{5} \\
 \vdots & \vdots  &  \!\!\!\ddots\!\!\!  & \vdots \\ 
 0 & 0 & \!\!\!\hdots\!\!\!  & 0
\end{array}\!\!
\right)\!,  \dots,
\label{eq:axialvector_matrices}
\end{eqnarray}
with $k=1,2,3$ for the axial vector channel.
Using these matrices, we are able to express the interaction Lagrangian (\ref{eq:Lint_RL_Nf_singlet}) in a compact form:
\begin{eqnarray}
 {\cal L}_{\mathrm{int}}^{\mathrm{sing}}
=
 \frac{N_{c}^{2}-1}{4N_{c}^{2}} G_{c}
\sum_{\ell}\sum_{i=1}^{N_{f}}
\bar{\phi} \, \Gamma_{\ell}^{(i)} \phi \, \bar{\phi} \, \bar{\Gamma}_{\ell}^{(i)} \phi,
\label{eq:Lint_RL_Nf_singlet_phi}
\end{eqnarray}
with $\ell=\{\mathrm{S}, \mathrm{V}, \mathrm{P}, \mathrm{A}\}$.
If we consider the mean fields only for $\ell=\mathrm{S}$ and $\mathrm{V}$ as we have done, the gap equation is expressed as
\begin{widetext}
\begin{eqnarray}
 iG(p_{0},\vec{p}\,)^{-1}
&=&
iG_{0}(p_{0},\vec{p}\,)^{-1} 
+ (-1)i \frac{N_{c}^{2}-1}{4N_{c}^{2}} G_{c}
\sum_{i=1}^{N_{f}}
\sum_{\ell=\mathrm{S}, \mathrm{V} } 
\Biggl[
\Gamma_{\ell}^{(i)} \mathrm{Tr}\!\!\int\!\!\frac{\mathrm{d}^{4}q}{(2\pi)^{2}} iG(q_{0},\vec{q}\,) \bar{\Gamma}_{\ell}^{(i)}
+
\bar{\Gamma}_{\ell}^{(i)} \mathrm{Tr}\!\!\int\!\!\frac{\mathrm{d}^{4}q}{(2\pi)^{2}} iG(q_{0},\vec{q}\,) \Gamma_{\ell}^{(i)}
\Biggr],
\nonumber \\
\label{eq:gap_diagram}
\end{eqnarray}
\end{widetext}
where $\mathrm{Tr}$ is the trace over the Dirac and color matrices, and 
 $G_{0}(p_{0},\vec{p}\,)^{-1}$ is the inverse of the free propagator
\begin{eqnarray}
G_{0}(p_{0},\vec{p}\,)^{-1}
=
\left(
\begin{array}{cccc}
 p \hspace{-0.4em}/ + \mu \gamma_{0} &  0 & \hdots  & 0 \\
 0  & p \hspace{-0.4em}/ + \mu \gamma_{0}  &  \hdots & 0 \\
 \vdots & \vdots  &  \ddots  & \vdots \\ 
 0 & 0 & \hdots  & (p_{0}-\lambda) \frac{1+\gamma_{0}}{2} 
\end{array}
\right).
\nonumber \\
\label{eq:propagator}
\end{eqnarray}
The sum over $\ell=\mathrm{S}$ and $\mathrm{V}$ is taken, where $k=1,2,3$ in Eq.~(\ref{eq:vector_matrices}) are also summed for $\ell=\mathrm{V}$. This is the most general form of the gap equation at $T=0$. Of course, Eq.~(\ref{eq:gap_diagram}) reproduces Eq.~(\ref{eq:gap_equation_0}) for $\lambda=0$. The gap equation at finite temperature can be easily obtained by the replacement of the integral over $q_0$ in Eq.~(\ref{eq:gap_diagram}) by the Matsubara summation, which should be equivalent to the gap equation (\ref{eq:gap_equation}).

%%%%%%%%%%%%%%%%%%%%%%%%%%%%%%%%%%%
\subsection{Symmetry of the ground state}
\label{sec:symmetry}

\subsubsection{Pattern of symmetry breaking}

Formation of nonzero Kondo condensates breaks the original symmetry in a nontrivial way. Recall that we have constructed the model so that it possesses the chiral symmetry for the light quarks and the heavy-quark spin symmetry for the heavy quark impurities. Since the current-current interaction is invariant under each transformation, nonzero Kondo condensates connecting the light and heavy quarks must break these symmetries. Here we discuss the breaking pattern of the symmetry in the Kondo phase. Let us first show the whole symmetry of the system and its breaking pattern: 
\begin{eqnarray}
 G \rightarrow H,
 \label{eq:symmetry_breaking}
\end{eqnarray}
with the group
\begin{eqnarray}
 G
 &=&
 \mathrm{SO}(3)_{\mathrm{space}} \times \mathrm{SU}(2)_{\mathrm{HQS}} \times \mathrm{U}(1)_{\mathrm{Q}} \times \mathrm{U}(1)_{\mathrm{V}} \times \mathrm{U}(1)_{\mathrm{A}} \nonumber \\
 && \times \mathrm{SU}(N_{f})_{\mathrm{V}} \times \mathrm{SU}(N_{f})_{\mathrm{A}},
\end{eqnarray}
and the subgroup
\begin{eqnarray}
 H
 &=&
 \mathrm{SO}(3)_{\mathrm{space}} \times \mathrm{U}(1)_{\mathrm{Q}+\mathrm{V}} \times \mathrm{U}(1)_{\mathrm{A}+\mathrm{HQS}_{h}}
 \nonumber \\
 && \times \mathrm{SU}(N_{f}-1)_{\mathrm{V}} \times \mathrm{SU}(N_{f}-1)_{\mathrm{A}}.
\end{eqnarray}
In the group $G$, $\mathrm{SO}(3)_{\mathrm{space}}$ is the rotational symmetry in the three-dimensional space.
$\mathrm{SU}(2)_{\mathrm{HQS}}$ and  $\mathrm{U}(1)_{\mathrm{Q}}$ are the HQS of $\Psi_{v}$ and the vector symmetry for the overall phase of $\Psi_{v}$, respectively.
$\mathrm{U}(1)_{\mathrm{V}} \times \mathrm{U}(1)_{\mathrm{A}} \times \mathrm{SU}(N_{f})_{\mathrm{V}} \times \mathrm{SU}(N_{f})_{\mathrm{A}}$ is the usual chiral symmetry of light flavor $\psi$.
Notice that the rotational symmetry holds as long as $n_{Q}$ is constant.
As more general situations, when $n_{Q}(\vec{x})$ depends on the position $\vec{x}$, this symmetry does not necessarily hold. 
The breaking pattern (\ref{eq:symmetry_breaking}) can be separated into four different pieces of pattern: (i) $\mathrm{SO}(3)_{\mathrm{space}} \times \mathrm{SU}(2)_{\mathrm{HQS}} \rightarrow \mathrm{SO}(3)_{\mathrm{space}}$, (ii) $\mathrm{U}(1)_{\mathrm{Q}} \times \mathrm{U}(1)_{\mathrm{V}} \rightarrow \mathrm{U}(1)_{\mathrm{Q}+\mathrm{V}}$, (iii) $\mathrm{U}(1)_{\mathrm{A}} \times \mathrm{SU}(2)_{\mathrm{HQS}} \rightarrow \mathrm{U}(1)_{\mathrm{A}+\mathrm{HQS}_{h}}$, and (iv) $\mathrm{SU}(N_{f})_{\mathrm{V},\mathrm{A}} \rightarrow \mathrm{SU}(N_{f}-1)_{\mathrm{V},\mathrm{A}}$.
Below we explain each of them.

\begin{description}

 \item[(i) $\mathrm{SO}(3)_{\mathrm{space}} \times \mathrm{SU}(2)_{\mathrm{HQS}} \rightarrow \mathrm{SO}(3)_{\mathrm{space}}$]\mbox{}\\
The  Lagrangian (\ref{eq:Lagrangian_current_eff}) possesses a global rotation $\mathrm{SO}(3)_{\mathrm{space}}$ for $\psi$ and $\Psi_{v}$. In the heavy quark limit, however, we can perform the rotation of $\Psi_{v}$ independently from the $\mathrm{SO}(3)_{\mathrm{space}}$ rotation, and can regard the rotational symmetry for $\Psi_v$ as an internal symmetry independent of $\mathrm{SO}(3)_{\mathrm{space}}$. This is the HQS and we represent it as $\mathrm{SU}(2)_{\mathrm{HQS}}$. This is one of the important properties of the heavy quark systems~\cite{Manohar:2000dt,Neubert:1993mb}. In the presence of the Kondo condensate with nonzero $\langle \Phi_{1}\rangle$ and $\langle \vec{\Phi}_{1}\rangle$, the heavy quark spin cannot be transformed independently. In other words, the mixing between a light quark and a heavy quark induces disappearance of the HQS.\footnote{
Recalling that the existence of HQS leads to the degeneracy of spin up and down states, the ``HQS doublet"~\cite{Isgur:1991wq} (see also Refs.~\cite{Manohar:2000dt,Neubert:1993mb}), one may wonder if the disappearance of HQS could induce mass splitting of the doublet states. However, since the origin of mass splitting should come from the next-leading contribution (the effects of the order of $1/m_Q$ such as the color magnetic spin interaction) to the heavy mass limit, the mixing effect due to the Kondo condensate which occurs in the leading order does not lead to mass splitting. Indeed, the dispersion relations~(\ref{eq:eigen_energy1})-(\ref{eq:eigen_energy3}) have the same form for spin up and down components.
}
 Thus, the original symmetry $\mathrm{SO}(3)_{\mathrm{space}} \times \mathrm{SU}(2)_{\mathrm{HQS}}$ just goes back to $\mathrm{SO}(3)_{\mathrm{space}}$. However, we notice that the U(1) part of $\mathrm{SU}(2)_{\mathrm{HQS}}$ survives in a nontrivial way. We will discuss this in (iii) below.

\item[(ii) $\mathrm{U}(1)_{\mathrm{Q}} \times \mathrm{U}(1)_{\mathrm{V}} \rightarrow \mathrm{U}(1)_{\mathrm{Q}+\mathrm{V}}$]\mbox{}\
\\
The $\mathrm{U}(1)_{\mathrm{Q}}$ and $\mathrm{U}(1)_{\mathrm{V}}$ symmetries are global $\mathrm{U}(1)$ symmetries 
for $\Psi_{v}$ and for $\psi=(\psi_{1},\dots,\psi_{N_{f}})$, respectively.
Since the Kondo condensates $\langle \Phi_{1}\rangle$ and $\langle \vec{\Phi}_{1}\rangle$ induce a mixing between the light quark $\psi_1$ and the heavy quark $\Psi_v$, the $\mathrm{U}(1)_{\mathrm{Q}}$ symmetry for $\Psi_{v}$ and the $\mathrm{U}(1)$ symmetry for $\psi_{1}$ are no longer independent symmetries. Indeed, the mean-field Lagrangian shows that it is invariant only if we rotate $\psi_1$ and $\Psi_v$ with the same angle. Therefore, after the Kondo condensation, we have only one $\mathrm{U}(1)$ transformation, which we symbolically denote $\mathrm{U}(1)_{\mathrm{Q}+\mathrm{V}}$.
Here we wrote $\mathrm{U(1)_{Q+V}}$ with V because the U(1) transformation for $\psi_1$ must be performed simultaneously with the other U(1) transformations for the light quarks $\psi_i\ (i=2,\dots, N_f)$ to form the $\mathrm{U(1)_{V}}$ transformation. In other words, in the presence of the Kondo condensate, we rotate the heavy quark field $\Psi_v$ with the same angle as that of the light quarks $\psi_i\ (i=1,\dots, N_f)$.

\item[(iii) $\mathrm{U}(1)_{\mathrm{A}} \times \mathrm{U}(1)_{\mathrm{HQS}} \rightarrow \mathrm{U}(1)_{\mathrm{A}+\mathrm{HQS}_{h}}$]\mbox{} 
\\ 
In order to see what kind of transformation will leave the Kondo condensates unchanged, let us rewrite $\psi_{1}$ by using the right-handed field $\chi_{1}$ and the left-handed field $\varphi_{1}$:
\begin{eqnarray}
 \psi_{1}
= \frac{1}{\sqrt{2}}
\left(
\begin{array}{c}
\chi_{1}+\varphi_{1} \\
\chi_{1}-\varphi_{1}
\end{array}
\right).
\end{eqnarray}
In the mean-field Lagrangian (\ref{eq:L_RL_Nf_MF}), the Kondo condensates give the mixing term $\Delta \bar \Psi_v (1+\vec \gamma \cdot \hat p)\psi_1=\bar\Psi_v \langle \Phi_1 + \vec \gamma \cdot \vec \Phi_1\rangle \psi_1$. The Kondo condensate part is rewritten as 
\begin{eqnarray}
&& \hspace{-1.5em}\langle \Phi_{1}+\vec{\gamma}\!\cdot\!\vec{\Phi}_{1} \rangle
\nonumber\\
&=& \langle \bar{\psi}_{1} (1+\hat{p}\!\cdot\!\vec{\gamma}) 
\Psi_{v} \rangle\nonumber \\
&=&
\frac{1}{\sqrt{2}}
\left\langle \left\{ (\chi_{1}^{\dag}+\varphi_{1}^{\dag})+(\chi_{1}^{\dag}-\varphi_{1}^{\dag})\hat{p}\!\cdot\!\vec{\sigma} \right\} 
\Psi_{v}
\right\rangle. 
\label{eq:axial_chiral_transformation}
\end{eqnarray}
We notice that this combination of the Kondo condensates is invariant under the $\mathrm{U}(1)_{\mathrm{A}}$ transformation $\psi_{1} \rightarrow e^{i\alpha\gamma_{5}} \psi_{1}$ for $\psi_1$, combined with $\mathrm{U}(1)$ transformation  $\Psi_{v} \rightarrow e^{i\alpha \hat{p}\cdot\vec{\sigma}} \Psi_{v}$ for $\Psi_v$. Indeed, for infinitesimal parameter $\alpha$, the $\mathrm{U}(1)_{\mathrm{A}}$ transformation generates $\chi_{1} \rightarrow \chi_{1} + i\alpha \chi_{1}$ and $\varphi_{1} \rightarrow \varphi_{1} - i\alpha \varphi_{1}$, which is compensated by the U(1) transformation $\Psi_{v} \rightarrow \Psi_{v} + i\alpha \hat{p}\cdot\vec{\sigma} \Psi_{v}$ (to show the invariance, use the relation $(\hat{p}\cdot\vec{\sigma})^{2}=1$). Note also that the U(1) transformation for $\Psi_v$ is a subgroup of $\mathrm{SU}(2)_{\mathrm{HQS}}$ ($e^{i\vec\theta\cdot \vec \sigma}\!\in$ $\mathrm{SU}(2)$ with a parameter $\vec\theta=\alpha \hat p$), and conserves the helicity since it is the rotation around the axis along the direction $\hat{p}$ in momentum space. We denote it $\mathrm{U}(1)_{\mathrm{HQS}_{h}}$. This is reminiscent of the Color-Flavor-Locked state in color-superconductivity, and we call the remaining symmetry $\mathrm{U}(1)_{\mathrm{A}+\mathrm{HQS}_{h}}$ the {\it chiral-HQS locked ($\chi$HQSL) symmetry}.
Here we wrote　$\mathrm{U(1)_{A+HQS_h}}$ with A due to the same reason mentioned in case (ii). The axial U(1) transformation for $\psi_1$ must be performed simultaneously with the other axial U(1) transformations.
Lastly, we emphasize that it is the factor of $\hat{p} \cdot \vec{\sigma}$ coming from the hedgehog ansatz, $\vec{\Delta}=\Delta \,\hat{p}$, that makes the $\chi$HQSL symmetry possible.

\item[(iv) $\mathrm{SU}(N_{f})_{\mathrm{V},\mathrm{A}} \rightarrow \mathrm{SU}(N_{f}-1)_{\mathrm{V},\mathrm{A}}$]\mbox{}\\
We assumed that only the flavor $i=1$ of the light quark forms the Kondo condensate with the heavy quark,  while the other flavors $i=2,\dots,N_{f}$ do not. 
Thus, the original $\mathrm{SU}(N_{f})_{\mathrm{V}} \times \mathrm{SU}(N_{f})_{\mathrm{A}}$ symmetry is broken to the $\mathrm{SU}(N_{f}-1)_{\mathrm{V}} \times \mathrm{SU}(N_{f}-1)_{\mathrm{A}}$ symmetry.

\end{description}

\subsubsection{Spin-coherent state}

The $\chi$HQSL symmetry introduced in (iii) provides an intuitive picture of the heavy quark state. We notice that the helicity conservation for the heavy quark spin leads to the spin-coherent state of the heavy quark, as known in atomic physics~\cite{PhysRevA.6.2211}. 
For the direction $\hat{p}=\vec{p}/|\vec{p}\,|=(\sin\theta \cos\varphi, \sin\theta\sin\varphi,\cos\theta)$ in three-dimensional momentum space, the spin-coherent state $|\hat{p}\rangle$ is defined as follows.
We first introduce the ``vacuum" $| 0 \rangle$ as the state whose eigenvalue for the third component of the spin operator, $\vec{S}=(S_{1},S_{2},S_{3})=\vec{\sigma}/2$, is the largest: $S_{3} |0\rangle=S |0\rangle$.
Then, by using the vacuum state, we define the spin-coherent state as 
\begin{eqnarray}
 |\hat{p}\rangle = 
 \left( \cos \frac{\theta}{2} \right)^{2S} 
   e^{\tan \frac{\theta}{2} e^{i\varphi} S_{-}} | 0 \rangle,
 \label{eq:spin_coherent}
\end{eqnarray}
where $S=1/2$ and the lowering operator is defined by $S_{-}=S_{1}-iS_{2}$.
This state satisfies the normalization $\langle \hat{p} | \hat{p} \rangle=1$. 
The expectation value of the spin operator $\vec{S}$ with respect to the spin-coherent state is given by $\langle\hat{p}|S_{i}|\hat{p}\rangle=S\,\hat{p}_{i}$ with $i=1,2,3$. Because the directions of the heavy quark spins are aligned along $\hat{p}$, it implies the hedgehog configuration for the heavy quark spin.
It is easy to confirm that $|\hat{p}\rangle$ is the eigenstate of $\hat{p}\!\cdot\!\vec{\sigma}$, namely $\hat{p}\!\cdot\!\vec{\sigma}|\hat{p}\rangle =  |\hat{p}\rangle$. From this relation, we notice that the $\mathrm{U}(1)_{\mathrm{HQS}_{h}}$ transformation, $|\hat{p}\rangle \rightarrow e^{i\alpha \hat{p}\cdot\vec{\sigma}} |\hat{p}\rangle$, induces only the  phase rotation, $|\hat{p}\rangle \rightarrow e^{i\alpha} |\hat{p}\rangle$, and does not produce essentially a new state which is different from the state before the transformation. Therefore, the $|\hat{p}\rangle$ state is mapped to a two-dimensional sphere $\mathrm{SU}(2)_{\mathrm{HQS}}/\mathrm{U}(1)_{\mathrm{HQS}_{h}} \simeq S^{2}$. The mapping from a unit vector $\hat{p}$ to $S^{2}$ implies the existence of a topological invariant characterized by the winding number. Thus, the spin-coherent state is important for the topological properties of the Kondo condensate as discussed in detail in the next subsection.\footnote{See, for example, Ref.~\cite{Volovik:2003fe} for applications of topological concepts to a variety of systems including condensed matter and quark matter.}

\subsection{Hedgehog configuration of heavy quark spin}
\label{sec:hedgehog}

Let us investigate in more detail the properties of the wavefunctions in the mean-field approximation. Here we focus on the light quark $\psi_1$ and the heavy quark $\Psi_v$ since only these fields mix with each other in a nontrivial way as we saw in the dispersion relations (\ref{eq:eigen_energy1})-(\ref{eq:eigen_energy3}). The other light quark fields $\psi_i$ $(i=2,\dots,N_f)$ do not participate in formation of the Kondo condensates and thus are not important in the level of the mean-field approximation (as we will see later in Sec.~\ref{sec:Kondo_excitations}, however, they play interesting roles in quantum fluctuations around the mean-field).

We solve the eigenvalue equation in a restricted Hilbert space of $(\psi_{1},\Psi_{v})$. The canonically obtained Hamiltonian density has the form $\bar \phi h \phi + \cdots$ with $\phi=(\psi_1,\Psi_v)$, and the eigenvalue equation reads $h\phi=\varepsilon \gamma^0 \phi$ (note that we discard the contributions represented by $``\cdots"$ which only give a trivial shift of the energy). Thus, we define ${\cal H}\equiv \gamma^0 h$ so that the eigenvalue equation can be written as 
\begin{eqnarray}
 {\cal H} u = \varepsilon u,
\end{eqnarray}
where the eigenvector $u$ has six components with the first four components being the light quark $\psi_{1}$ and the second two components being the positive energy part of the heavy quark $\Psi_{v}$. The mean-field ``Hamiltonian" ${\cal H}$ in this restricted space $(\psi_{1},\Psi_{v})$ is given by (see Eq.~(\ref{eq:L_RL_Nf_MF})) 
\begin{widetext}
\begin{eqnarray}
 {\cal H}
=
\left(
\begin{array}{cccccc}
 -\mu & 0 & p \cos \theta & p e^{-i\varphi} \sin \theta & - \Delta^{\ast} & 0 \\
 0 & -\mu & p e^{i\varphi} \sin \theta & - p \cos \theta & 0 & -\Delta^{\ast} \\
 p \cos \theta & p e^{-i\varphi} \sin \theta & -\mu & 0 & -\Delta^{\ast} \cos \theta & - \Delta^{\ast} e^{-i\varphi} \sin\theta \\
 p e^{i\varphi} \sin \theta & - p\cos \theta & 0 & -\mu & -\Delta^{\ast} e^{i\varphi} \sin \theta & \Delta^{\ast} \cos \theta \\
 -\Delta & 0 & -\Delta \cos \theta & -\Delta e^{-i\varphi} \sin \theta & \lambda & 0 \\
 0 & -\Delta & -\Delta e^{i\varphi} \sin \theta & \Delta \cos \theta & 0 & \lambda
\end{array}
\right),
\label{eq:Hamiltonian_MF}
\end{eqnarray}
\end{widetext}
where the three-dimensional momentum $\vec p$ is parametrized as $\vec{p}=(p\sin\theta \cos\varphi, p\sin\theta\sin\varphi, p\cos\theta)$. The left-upper ($4\times 4$) part and right-lower ($2\times 2$) part of the Hamiltonian correspond to the `kinetic' parts for $\psi_1$ and $\Psi_v$, respectively, and the other (non-diagonal) parts depending on $\Delta$ or $\Delta^*$ correspond to the mixing between $\psi_1$ and $\Psi_v$.

In the basis of $(\psi_{1},\Psi_{v})$, $\gamma_{5}$ matrix for $\psi_{1}$ and spin operator matrices $\vec{S}=\vec{\sigma}/2$ for $\Psi_{v}$ are extended to $\Gamma_{5}$ and $\vec{\cal S}=({\cal S}_{1},{\cal S}_{2},{\cal S}_{3})$, which are defined by
\begin{eqnarray}
 \Gamma_{5}
=
\left(
\begin{array}{cccccc}
 0 & 0 & 1 & 0 & 0 & 0 \\
 0 & 0 & 0 & 1 & 0 & 0 \\
 1 & 0 & 0 & 0 & 0 & 0 \\
 0 & 1 & 0 & 0 & 0 & 0 \\
 0 & 0 & 0 & 0 & 0 & 0 \\
 0 & 0 & 0 & 0 & 0 & 0
\end{array}
\right),
\end{eqnarray}
and
\begin{eqnarray}
 {\cal S}_{1}
&=&
\frac{1}{2}\!
\left(
\begin{array}{cccccc}
 0 &  0 &  0 & 0 & 0 & 0 \\
 0 &  0 &  0 &  0 & 0 & 0 \\
 0 &  0 &  0 & 0 & 0 & 0 \\
 0 &  0 &  0 &  0 & 0 & 0 \\
 0 &  0 &  0 & 0 & 0 & 1 \\
 0 &  0 &  0 & 0 & 1 & 0 
\end{array}
\right)\!,
\\
 {\cal S}_{2}
&=&
\frac{1}{2}\!
\left(
\begin{array}{cccccc}
 0 &  0 &  0 & 0 & 0 & 0 \\
 0 &  0 &  0 &  0 & 0 & 0 \\
 0 &  0 &  0 & 0 & 0 & 0 \\
 0 &  0 &  0 &  0 & 0 & 0 \\
 0 &  0 &  0 & 0 & 0 & -i \\
 0 &  0 &  0 & 0 & i & 0 
\end{array}
\right)\!,
\\
 {\cal S}_{3}
 &=&
\frac{1}{2}\!
\left(
\begin{array}{cccccc}
 0 &  0 &  0 & 0 & 0 & 0 \\
 0 &  0 &  0 &  0 & 0 & 0 \\
 0 &  0 &  0 & 0 & 0 & 0 \\
 0 &  0 &  0 &  0 & 0 & 0 \\
 0 &  0 &  0 & 0 & 1 & 0 \\
 0 &  0 &  0 & 0 & 0 & -1 
\end{array}
\right)\!.
\end{eqnarray}
We note that $\Gamma_{5}$ and ${\cal S}_{i}$ ($i=1,2,3$) are not commutative with ${\cal H}$: $[\Gamma_{5},{\cal H}]\neq0$ and $[{\cal S}_{i},{\cal H}]\neq0$.
Hence the chirality for the light quark $\psi_{1}$ and the helicity for the heavy quark $\Psi_{v}$ are not good quantum numbers in the Kondo condensate.
Instead, the 
 operator defined by
\begin{eqnarray}
\Sigma_{5}(\hat{p})
&=& \Gamma_{5}+2\hat{p} \!\cdot\! \vec{\cal S}
\nonumber \\
&=&
\left(
\begin{array}{cccccc}
 0 & 0 & 1 & 0 & 0 & 0 \\
 0 & 0 & 0 & 1 & 0 & 0 \\
 1 & 0 & 0 & 0 & 0 & 0 \\
 0 & 1 & 0 & 0 & 0 & 0 \\
 0 & 0 & 0 & 0 & \cos \theta & e^{-i\varphi} \sin\theta \\
 0 & 0 & 0 & 0 & e^{i\varphi} \sin \theta & - \cos \theta
\end{array}
\right)\!,
\label{eq:Sigma_5}
\end{eqnarray}
with $\hat{p}=\vec{p}/|\vec{p}\,|=(\sin\theta \cos\varphi, \sin\theta\sin\varphi,\cos\theta)$ 
is commutative with ${\cal H}$: $[\Sigma_{5},{\cal H}]=0$. Therefore, we can use the eigenvalues of $\Sigma_{5}(\hat{p})$ to specify the wavefunctions. In fact, $\Sigma_5$ has eigenvalues $\pm 1$ because it satisfies $(\Sigma_5)^2=1$. Notice also that $\Sigma_{5}(\hat{p})$ is the generator of the $\chi$HQSL symmetry $\mathrm{U}(1)_{\mathrm{A}+\mathrm{HQS}_{h}}$.

We already know the eigenvalues of the mean-field Hamiltonian (\ref{eq:Hamiltonian_MF}): they are given by the dispersions $\widetilde{E}_{p}$ in Eq.~(\ref{eq:eigen_energy3}) and $E_{p}^{\pm}$ in Eq.~(\ref{eq:eigen_energy1}), corresponding to the antiparticle mode of the light fermion $\psi_1$ and two mixed states, respectively. In the following, we give the wavefunctions for each mode and consider the properties of them.

\noindent{\it a) Wavefunctions of antiparticle modes: $\widetilde{E}_{p}$\\}
The eigenstates of $\widetilde{E}_{p}$ are given by
\begin{eqnarray}
 u_{\widetilde{E}_{p}}^{(+)}(\vec{p}\,)
&=&
\frac{1}{\sqrt{2}}
\left(\!\!
\begin{array}{c}
 \sin \frac{\theta}{2} \\
 -e^{i\varphi}\cos\frac{\theta}{2} \\
 \sin \frac{\theta}{2}  \\
 -e^{i\varphi}\cos\frac{\theta}{2} \\
 0 \\
 0
\end{array}\!\!
\right),
 \\ 
 u_{\widetilde{E}_{p}}^{(-)}(\vec{p}\,)
&=&
\frac{1}{\sqrt{2}}
\left(\!\!
\begin{array}{c}
 \cos \frac{\theta}{2} \\
 e^{i\varphi}\sin\frac{\theta}{2} \\
 -\cos \frac{\theta}{2}  \\
 -e^{i\varphi}\sin\frac{\theta}{2} \\
 0 \\
 0
\end{array}\!\!
\right),
\end{eqnarray}
where the upper indices $(\pm)$ are eigenvalue of the $\Sigma_{5}(\hat{p})$ operator: $\Sigma_{5}(\hat{p}) u_{\widetilde{E}_{p}}^{(\gamma)}=\gamma u_{\widetilde{E}_{p}}^{(\gamma)}$. Since the antiparticle modes have dispersions of free antiquarks without coupling to the heavy quark $\Psi_v$ (thus the wavefunction has only upper four components), they are also eigenstates of the chirality $\Gamma_{5}$ operator: $\Gamma_{5} u_{\widetilde{E}_{p}}^{(\gamma)}=\gamma u_{\widetilde{E}_{p}}^{(\gamma)}$. In other words, the operator $\Sigma_{5}(\hat{p})=\Gamma_{5}+2\hat{p} \!\cdot\! \vec{\cal S} $ reduces to the chirality operator $\Gamma_5$ for the antiquark dispersion, and thus these two operators have the same eigenvalues.\\

\noindent{\it b) Wavefunctions of mixed states: $E_{p}^{\pm}$}\\
The eigenstates of $E_{p}^{+}$ are given by
\begin{eqnarray}
 u_{E_{p}^{+}}^{(+)}(\vec{p}\,)
&=&
{\cal N}_{+}
\left(\!\!
\begin{array}{c}
 -\frac{E_{p}^{+}-\lambda}{2\Delta} e^{-i\varphi} \cos \frac{\theta}{2} \\
 -\frac{E_{p}^{+}-\lambda}{2\Delta}  \sin \frac{\theta}{2} \\
 -\frac{E_{p}^{+}-\lambda}{2\Delta} e^{-i\varphi} \cos \frac{\theta}{2} \\
 -\frac{E_{p}^{+}-\lambda}{2\Delta}  \sin \frac{\theta}{2} \\
 e^{-i\varphi} \cos \frac{\theta}{2} \\
  \sin \frac{\theta}{2}
\end{array}\!\!
\right),
\label{eq:u_E+1} \\ 
 u_{E_{p}^{+}}^{(-)}(\vec{p}\,)
&=&
{\cal N}_{+}
\left(\!\!
\begin{array}{c}
 \frac{E_{p}^{+}-\lambda}{2\Delta} e^{-i\varphi} \sin \frac{\theta}{2} \\
 -\frac{E_{p}^{+}-\lambda}{2\Delta}  \cos \frac{\theta}{2} \\
 -\frac{E_{p}^{+}-\lambda}{2\Delta} e^{-i\varphi} \sin \frac{\theta}{2} \\
 \frac{E_{p}^{+}-\lambda}{2\Delta} \cos \frac{\theta}{2} \\
 - e^{-i\varphi} \sin \frac{\theta}{2} \\
  \cos \frac{\theta}{2}
\end{array}\!\!
\right),
\label{eq:u_E+2}
\end{eqnarray}
with $ {\cal N}_{+}^{-1} =\sqrt{1+(E_{p}^{+}-\lambda)^{2}/(2|\Delta|^{2})}
$ being the normalization constant. This time, all the six components are nonzero reflecting that this state corresponds to one of the mixed dispersions. Again, the upper indices $(\pm)$ are the eigenvalues of the operator $\Sigma_5(\hat p)$. However, it is not the eigenstate of $\Gamma_{5}$ and ${\cal S}_{3}$. Similarly, the eigenstates of $E_{p}^{-}$ are given by replacement of $E_p^+$ in the above by $E_p^-$: 
\begin{eqnarray}
 u_{E_{p}^{-}}^{(+)}(\vec{p}\,)
&=&
{\cal N}_{-}
\left(\!\!
\begin{array}{c}
 -\frac{E_{p}^{-}-\lambda}{2\Delta} e^{-i\varphi} \cos \frac{\theta}{2} \\
 -\frac{E_{p}^{-}-\lambda}{2\Delta}  \sin \frac{\theta}{2} \\
 -\frac{E_{p}^{-}-\lambda}{2\Delta} e^{-i\varphi} \cos \frac{\theta}{2} \\
 -\frac{E_{p}^{-}-\lambda}{2\Delta} \sin \frac{\theta}{2} \\
 e^{-i\varphi} \cos \frac{\theta}{2} \\
  \sin \frac{\theta}{2}
\end{array}\!\!
\right),
\label{eq:u_E-1} \\ 
 u_{E_{p}^{-}}^{(-)}(\vec{p}\,)
&=&
{\cal N}_{-}
\left(\!\!
\begin{array}{c}
 \frac{E_{p}^{-}-\lambda}{2\Delta} e^{-i\varphi} \sin \frac{\theta}{2} \\
 -\frac{E_{p}^{-}-\lambda}{2\Delta}  \cos \frac{\theta}{2} \\
 -\frac{E_{p}^{-}-\lambda}{2\Delta} e^{-i\varphi} \sin \frac{\theta}{2} \\
 \frac{E_{p}^{-}-\lambda}{2\Delta}  \cos \frac{\theta}{2} \\
 - e^{-i\varphi}\sin \frac{\theta}{2} \\
  \cos \frac{\theta}{2}
\end{array}\!\!
\right),
\label{eq:u_E-2}
\end{eqnarray}
with $ {\cal N}_{-}^{-1} = \sqrt{1+(E_{p}^{-}-\lambda)^{2}/(2|\Delta|^{2})}
$ for the normalization constant.

\begin{figure}[tbp]
  \begin{center}
 \begin{minipage}[b]{0.49\linewidth}
   \centering
   \includegraphics[keepaspectratio,scale=0.13]{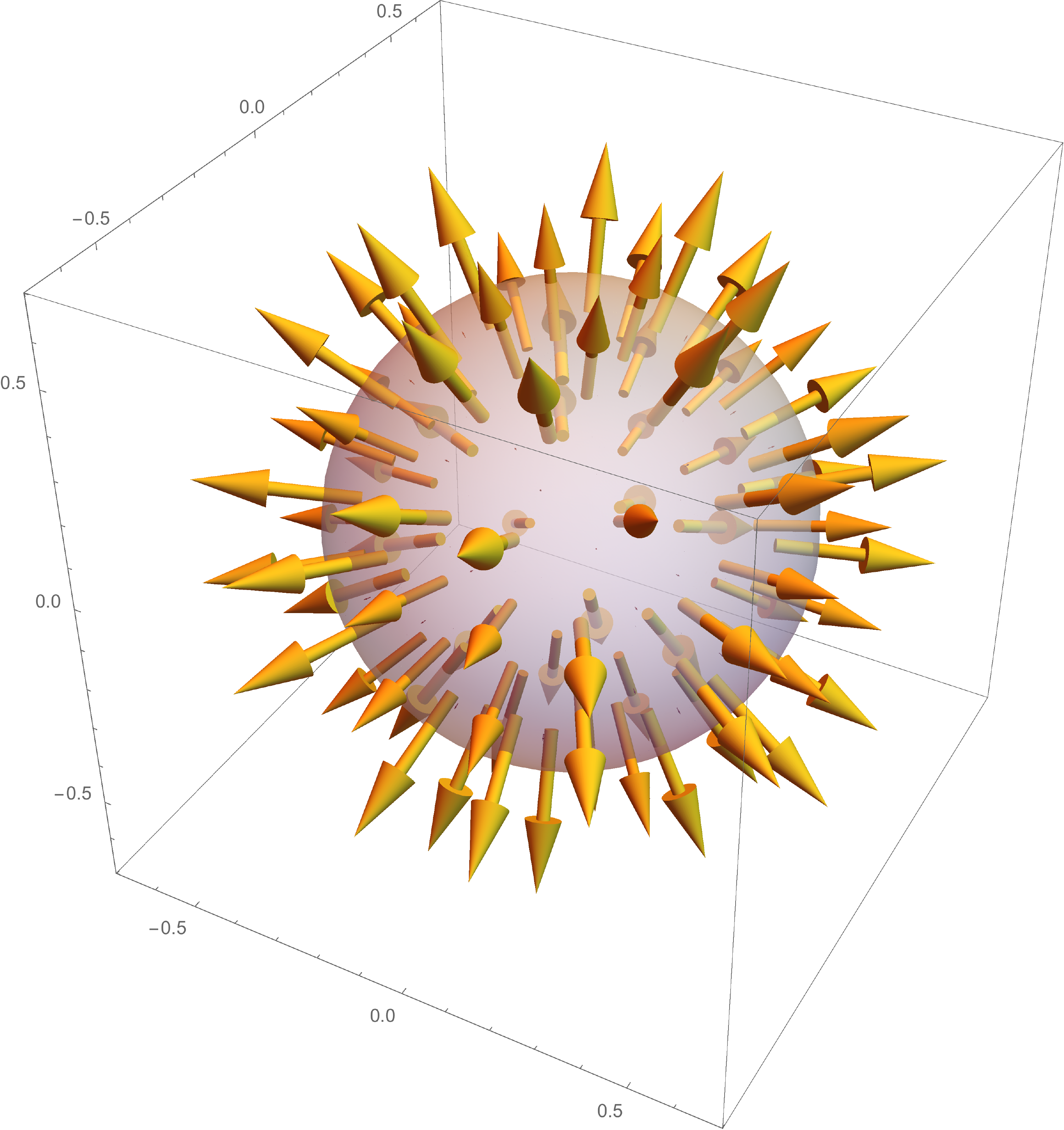}
 \end{minipage} 
 \begin{minipage}[b]{0.49\linewidth}
    \centering
   \includegraphics[keepaspectratio,scale=0.13]{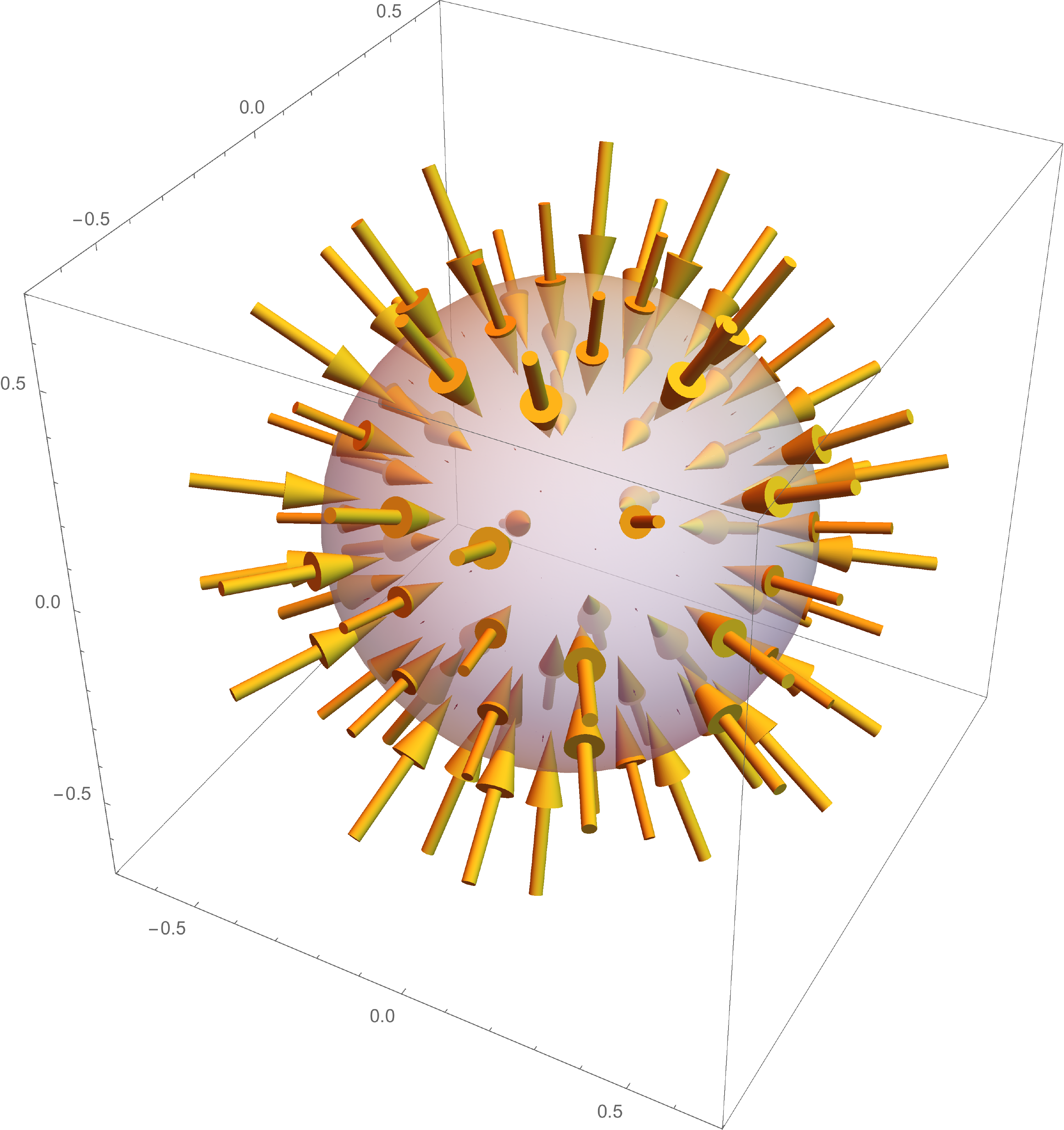}
 \end{minipage}
   \end{center}
 \caption{Plots of $\vec{S}_{E_{p}^{-}}^{(\gamma)}$ (left: $\gamma=+1$ and right: $\gamma=-1$) in momentum space (Eq.~(\ref{eq:hedgehog})).
 }
  \label{fig:hedgehog}
\end{figure}

Let us consider the expectation value of the heavy quark spin operator $\vec{\cal S}$ for the wave functions $u_{\widetilde{E}_{p}}^{(\gamma)}$ and $u_{E_{p}^{\pm}}^{(\gamma)}$.
When $\vec{\cal S}$ is sandwiched by $u_{\widetilde{E}_{p}}^{(\gamma)}$, it leads to the absence of the heavy quark spin:
$\langle u_{\widetilde{E}_{p}}^{(\gamma)} | \vec{\cal S} | u_{\widetilde{E}_{p}}^{(\gamma)} \rangle = \vec{0}$. 
This is natural because $u_{\widetilde{E}_{p}}^{(\gamma)}$ has no component of the heavy quark.
On the other hand, when $\vec{\cal S}$ is sandwiched by $u_{E_{p}^{\pm}}^{(\gamma)}$, one finds 
\begin{eqnarray}
\vec{S}_{E_{p}^{\pm}}^{(\gamma)}(\vec{p}\,)
&=&
 \langle u_{E_{p}^{\pm}}^{(\gamma)}(\vec{p}\,) | \vec{\cal S} | u_{E_{p}^{\pm}}^{(\gamma)}(\vec{p}\,) \rangle
\nonumber \\
 &=&
   \gamma S_{\pm}(p) \hat{p},
  \label{eq:hedgehog}
\end{eqnarray}
where the magnitude of heavy quark spin is given by
\begin{eqnarray}
S_\pm(p)=\frac12\cdot \frac{1}{1+\frac{(E_p^\pm-\lambda )^2}{2|\Delta|^2}}.
   \label{eq:hedgehog_spin_S}
\end{eqnarray}
We call $S_{\pm}(p)$ the ``effective" magnitude of the heavy quark spin, because it is a function of the momentum $p$ and it can deviate from $S=1/2$.
We notice that $\vec{S}_{\pm}^{(\gamma)}(\vec{p}\,)$ is proportional to $\hat p$ and $\gamma$ can be $\pm 1$. Thus, it is directed outwardly for $\gamma=+1$ or inwardly for $\gamma=-1$ along the direction of $\hat{p}$ in momentum space.
Therefore, the heavy quark spin forms a hedgehog configuration, which may be called the {\it HQS hedgehog}, as shown in Fig.~\ref{fig:hedgehog}.
This configuration is essentially the same as the spin-coherent state in Eq.~(\ref{eq:spin_coherent}).
The slight difference is that Eq.~(\ref{eq:spin_coherent}) has only the information on the direction of the heavy quark spin, while Eq.~(\ref{eq:hedgehog}) has both the information about the direction and magnitude.

%%%%%%%%%%%%%%%%%%%%%%%%%%
%%%%%%%%%%%%%%%%%%%%%%%%%%

\begin{figure}[tbp]
  \centering
  \includegraphics[keepaspectratio,scale=0.35]{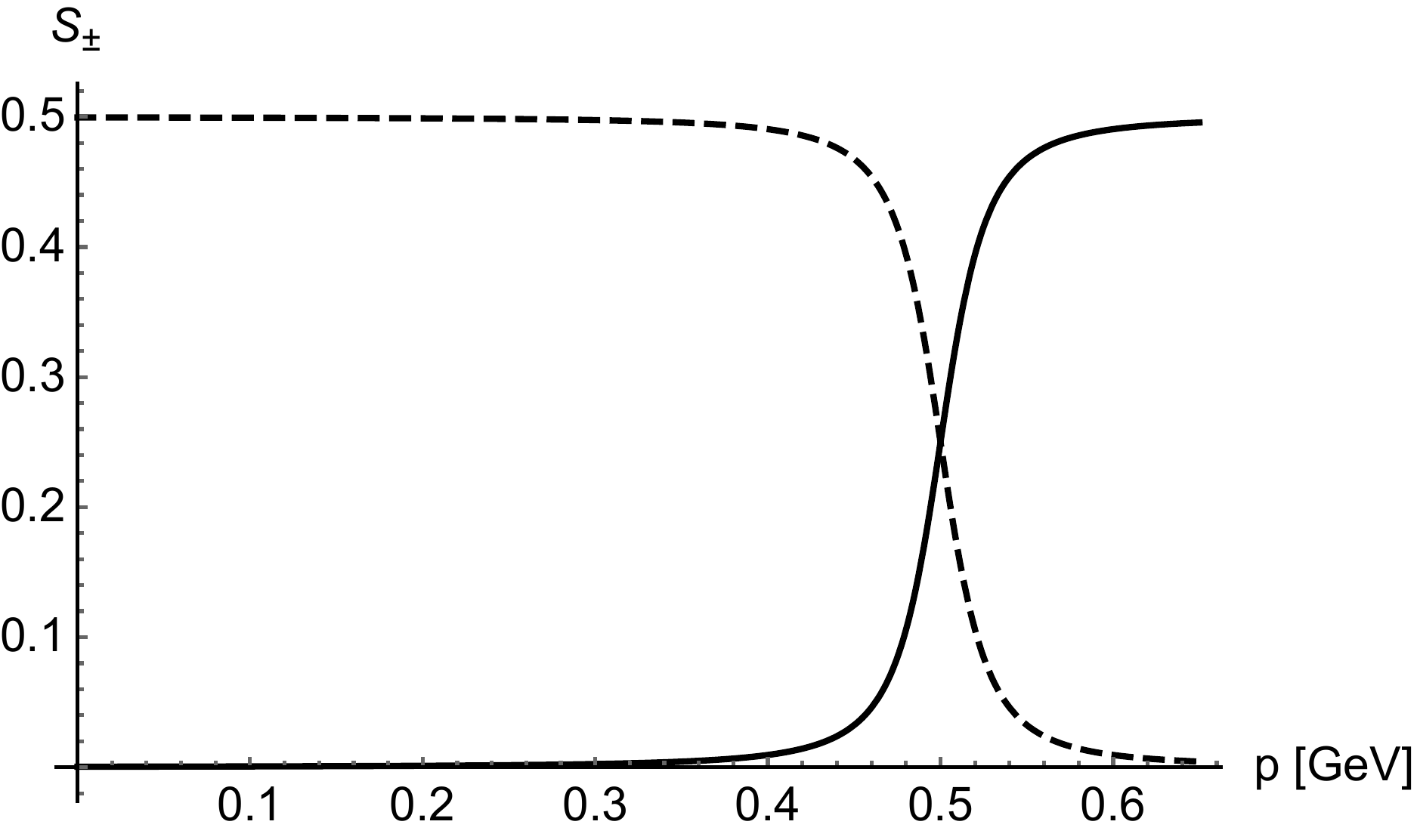}
\caption{
Effective heavy quark spins $S_{\pm}(p)$ as functions of $p$ (Eq.~(\ref{eq:hedgehog_spin_S})). The solid (dashed) line corresponds to $S_{-}(p)$ ($S_{+}(p)$). We used typical values of parameters: $\mu=0.5$ GeV, $\lambda=0$, $|\Delta|=0.01$ GeV.
}
  \label{fig:hedgehog2}
\end{figure}

In Fig.~\ref{fig:hedgehog2}, we plot $S_{\pm}(p)$ as the functions of momentum $p$ for typical values of parameters: $\mu=0.5$~GeV, $\lambda=0,$ and $|\Delta|=0.01$~GeV. As is evident from the figure, $S_+(p)$ and $S_-(p)$ show complementary behaviors. For the $E_{p}^{-}$ mode, we find that $S_{-}(p)$ starts from $S_-\simeq0$ at $p=0$ and reaches $S_-\simeq 1/2$ for large $p$. In contrast, for the $E_{p}^{+}$ mode,  we find that $S_{+}(p)$ starts from $S_+\simeq 1/2$ at $p=0$, and reaches $S_+ \simeq0$ for large $p$. This behavior is reasonable because $u_{E_{p}^{-}}^{(\gamma)}(\vec{p}\,)$ is dominated by the light quark component for $p \ll \mu$, while that is dominated by the heavy quark component for $p \gg \mu$ (cf.~Eqs.~(\ref{eq:u_E-1}), (\ref{eq:u_E-2})). This is also seen in the behavior that $E_{p}^{-}\simeq p-\mu$ for $p \ll \mu$ and $E_{p}^{-}\simeq \lambda$ for $p \gg \mu$ in Fig.~\ref{fig:dispersion_MF}. As for the $E_{p}^{+}$ mode, $u_{E_{p}^{+}}^{(\gamma)}(\vec{p}\,)$ is dominated by the heavy quark component for $p \ll \mu$, while that is dominated by the light quark component for $p \gg \mu$ (cf.~Fig.~\ref{fig:dispersion_MF} and Eqs.~(\ref{eq:u_E+1}), (\ref{eq:u_E+2})).

%%%%%%%%%%%%%%%%%%%%%%%%%%
%%%%%%%%%%%%%%%%%%%%%%%%%%

As is always the case with hedgehog-type configurations in quantum field theories, we are able to assign topological interpretation to the HQS hedgehog configuration (\ref{eq:hedgehog}). Indeed, by using a unit vector along the heavy quark spin, 
\begin{eqnarray}
 \vec{m}_{\varepsilon}^{(\gamma)}(\vec{p}\,) \equiv 
 \frac{\vec{S}_{\varepsilon}^{(\gamma)}(\vec{p}\,)}{S_{\pm}(p)}
 =\gamma \, \hat{p},
 \label{eq:m_vector}
\end{eqnarray}
with $\varepsilon=E_{p}^{\pm}$, we can define the winding number by
\begin{eqnarray}
w_{\varepsilon}^{(\gamma)} 
= \frac{1}{8\pi} \epsilon_{ijk}\!\! \int_{\Sigma} \!
  \vec{m}_{\varepsilon}^{(\gamma)} \!\cdot\! \left( \! \frac{\partial \vec{m}_{\varepsilon}^{(\gamma)}}{\partial p_{i}} \! \times \! \frac{\partial \vec{m}_{\varepsilon}^{(\gamma)}}{\partial p_{j}} \!\right) \mathrm{d}S^{k},
 \label{eq:winding_number}
\end{eqnarray}
where $\Sigma$ is a surface surrounding the origin in the momentum space, $\mathrm{d}S^{k}$ is the area element on $\Sigma$, $\epsilon_{ijk}$ ($i,j,k=1,2,3$) is an epsilon tensor with $\epsilon_{123}=1$, and the sum over $i,j,k$ is taken (cf.~Ref.~\cite{Volovik:2003fe}). Inserting Eq.~(\ref{eq:m_vector}) into Eq.~(\ref{eq:winding_number}), we obtain $w _{\varepsilon}^{(\gamma)}=\gamma$. Therefore, the hedgehog configuration with the outward (inward) direction of the heavy quark spin has a winding number $\gamma=+1$ ($\gamma=-1$). Notice that the winding number thus defined does not distinguish the $E^+_p$ and $E^-_p$ modes.

We can also define the ``fraction of light-quark component" in the wavefunctions of the mixed states $E_p^\pm$. Recall that the operator $\Sigma_5(p)$ is conserved and gives eigenvalues $\gamma=\pm 1$ for the wavefunctions $u_\epsilon^{(\gamma)}(p)$. This can be interpreted as a kind of spin in the mixed states. Equation (\ref{eq:Sigma_5}) defines the decomposition of the total ``spin" into the light and heavy quark components. Namely, $\langle \Sigma_5 \rangle_\gamma = \gamma=\langle \Gamma_5 \rangle_\gamma + 2\hat p \cdot \langle \vec {\cal S} \rangle_\gamma $ where $\langle \cdots \rangle_\gamma $ is the expectation value with respect to the eigenstate specified by $\gamma$. In order to make the decomposition for positive quantity, we multiply $\gamma$, so that we find $1=\langle \gamma \Gamma_5 \rangle_\gamma + 2\gamma \hat p \cdot \langle \vec {\cal S} \rangle_\gamma $. Therefore, we can regard the quantity $\langle \gamma \Gamma_5 \rangle_\gamma$ as the fraction of light-quark component, which we denote $n_\pm(p)$. Explicitly, 
\begin{eqnarray}
 n_{\pm}(p) \equiv \langle u_{E_{p}^{\pm}}^{(\gamma)}(\vec{p}\,) | \gamma \Gamma_{5} | u_{E_{p}^{\pm}}^{(\gamma)}(\vec{p}\,) \rangle.
 \label{eq:light_quark_fraction}
\end{eqnarray}
Notice that this quantity does not depend on $\gamma$ by construction ($\gamma^2=1$).

\begin{figure}[tbp]
  \begin{center}
 \begin{minipage}[b]{1.0\linewidth}
 \centering
   \includegraphics[keepaspectratio,scale=0.35]{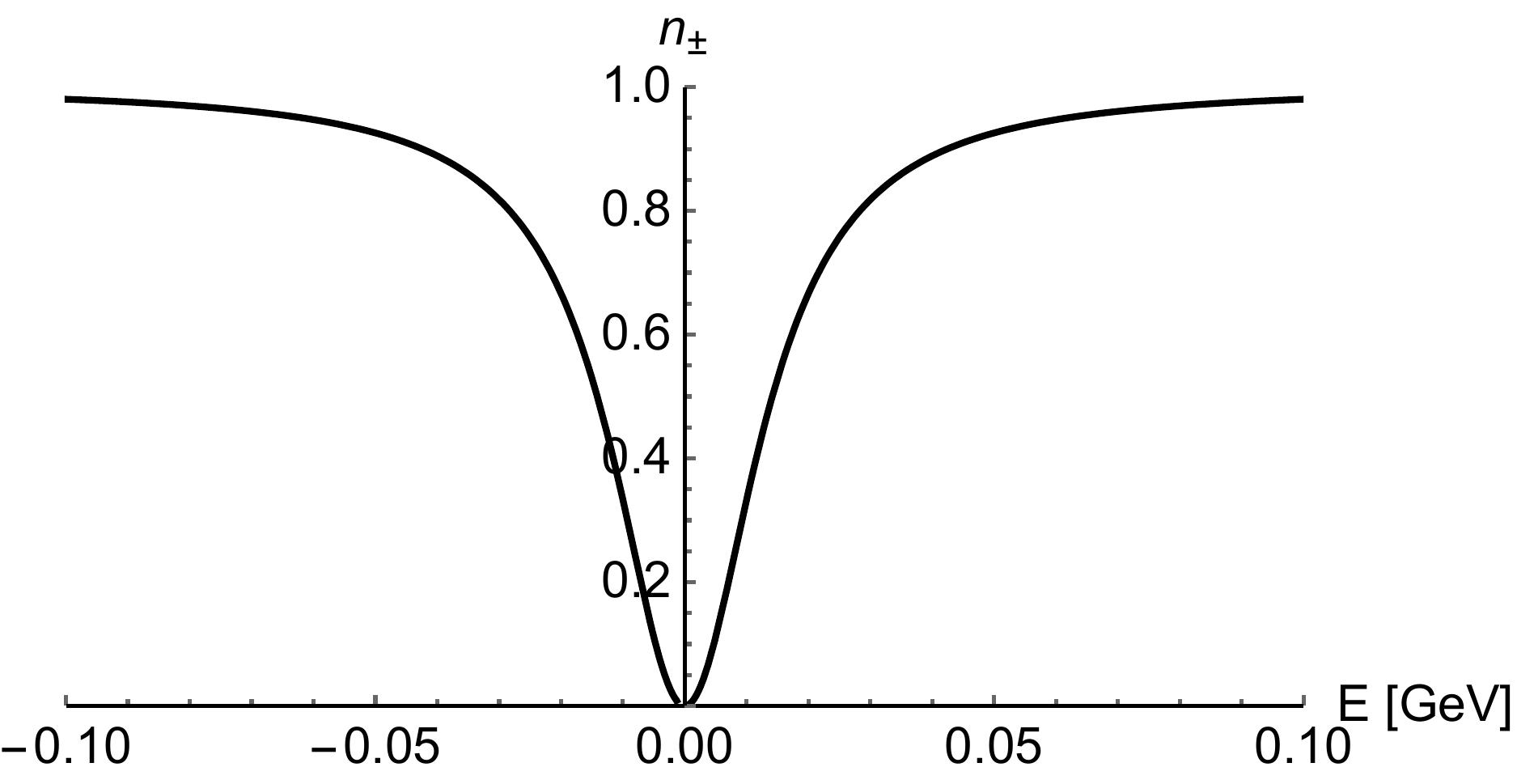}
 \end{minipage}
\\
 \vspace{1em}
 \begin{minipage}[b]{1.0\linewidth}
 \centering
   \includegraphics[keepaspectratio,scale=0.35]{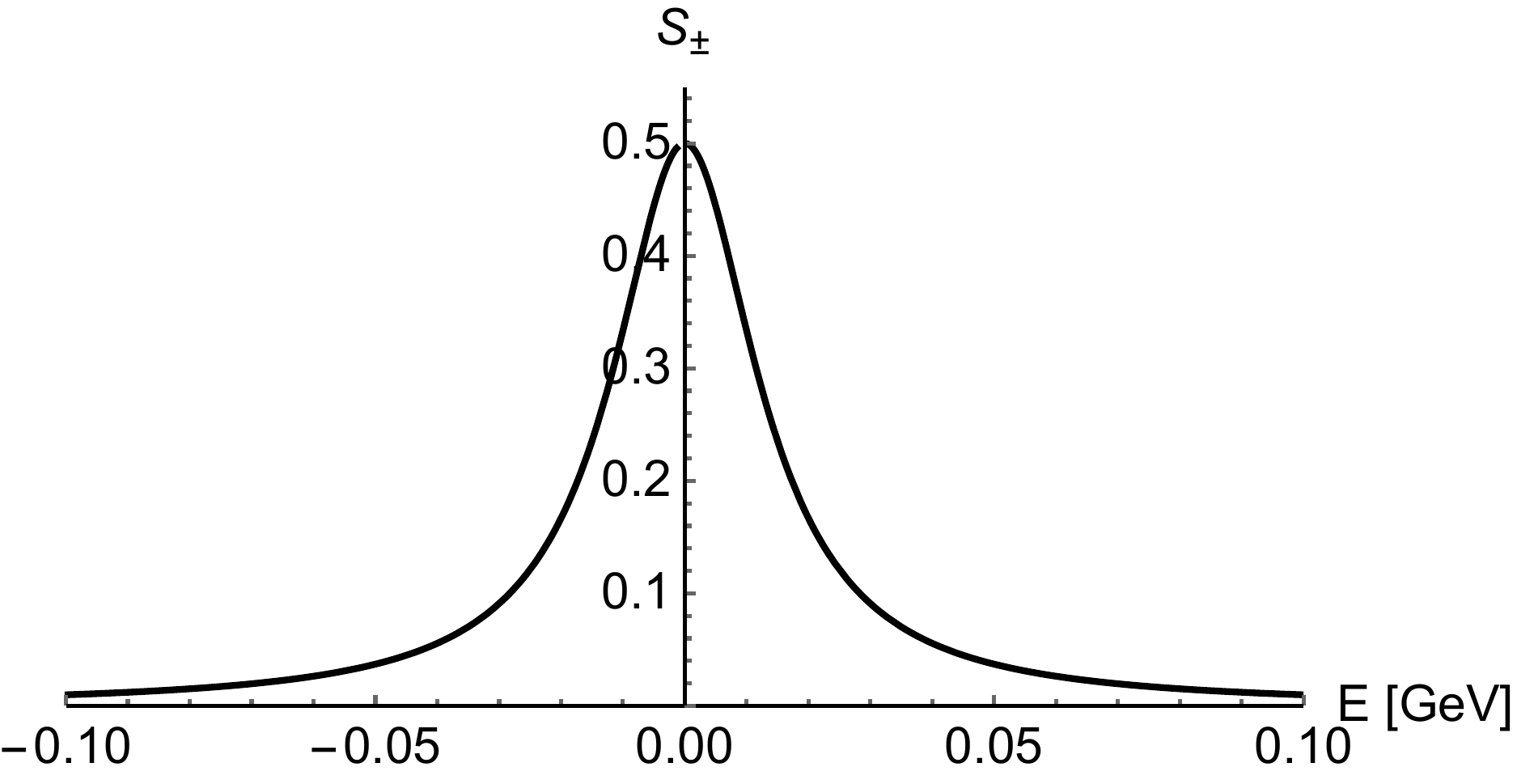}
 \end{minipage}
   \end{center}
 \caption{
The fraction of the light quark component $n_{\pm}(p)$ (Eq.~(\ref{eq:light_quark_fraction})) (top panel) and the effective heavy quark spin $S_{\pm}(p)$ (Eq.~(\ref{eq:hedgehog_spin_S})) (bottom panel) as functions of energy $E$ for $\lambda=0$.
}
  \label{fig:hedgehog3}
\end{figure}

It is quite instructive to plot $n_{\pm}$ and $S_{\pm}$ as functions of energy $E$ in place of momentum $p$. Consider the case with $\lambda=0$ where the location of the Fermi surface corresponds to $E=0$. As is evident from Fig.~\ref{fig:dispersion_MF}, the Fermi sea is purely occupied by the $E_p^-$ mode which has negative energies, while the $E_p^+$ mode with positive energies is not occupied at all. Figure~\ref{fig:hedgehog3} is the plots of $n_{\pm}(E)$ and $S_{\pm}(E)$.  The negative energy region (in the Fermi sea) corresponds to $E=E_{p}^{-}$, while the positive, $E=E_{p}^{+}$. Remarkably, $n_{\pm}(E)$ and $S_\pm(E)$ show opposite behaviors. Around the Fermi surface $E=0$, the light quark fraction is the smallest while the effective heavy quark spin is the maximum. Therefore, the physics at the Fermi surface (the ground state) and around the Fermi surface (excitations around the ground state) is dominantly given by the heavy quark dynamics. This suggests that the transport and thermodynamic properties of the Kondo phase will be predominantly determined by the heavy quark dynamics, which is counter intuitive. In reality, however, since heavy quarks do not propagate in matter, they will not be able to contribute to transport phenomena. Therefore, suppression of the light quark degree of freedom and inability of heavy quarks around the Fermi surface imply that the transport and thermodynamic properties are generally much suppressed in the Kondo phase, compared to the free light quark gas.\footnote{This expectation will be true when the number of light flavors $N_f$ is small. When $N_f$ is large, both the transport and thermodynamic properties will be dominantly determined by the unmixed $N_f-1$ flavor light quarks.} 

When $\lambda\neq 0$, we need to modify the above discussion. For $\lambda>0$, the crossing point of the two dispersions $E_q=p-\mu$ and $E_Q=\lambda$ moves towards positive energy side, while for $\lambda<0$, it moves towards negative energy side. If we define the Fermi surface as $E=0$, then either $E_p^+$ or $E_p^-$ will cut across the $E=0$ line, at which both the light and heavy quark components are non-negligible. This means that when $\lambda\neq 0$, the ground state and excitations around it will be affected by both components.  This can be seen in the plots of $n_\pm(E)$ and $S_\pm(E)$ for $\lambda\neq 0$. The maximum or minimum point will move depending on the value of $\lambda$, and at the Fermi surface ($E=0$), both components will be in general finite.

%%%%%%%%%%%%%%%%%%%%%%%%%%
%%%%%%%%%%%%%%%%%%%%%%%%%%

\subsection{Berry phase and monopole}

We have seen that the wavefunctions $u_{\varepsilon}^{(\gamma)}(p)$ ($\varepsilon=E_{p}^{\pm}$) of the ground state in the Kondo phase give hedgehog configurations for the effective heavy quark spin which have nonzero winding numbers. This  fact implies the necessity of topological interpretation of the states. Here we provide another property related to topology, the Berry phase~\cite{Berry45}. We define the Berry connection (or gauge field) as follows:
\begin{eqnarray}
 \vec{A}_{\varepsilon}^{(\gamma)}(\vec{p}\,) = \langle u_{\varepsilon}^{(\gamma)}(\vec{p}\,) | (-i) \vec{\nabla}_{p} | u_{\varepsilon}^{(\gamma)}(\vec{p}\,) \rangle,
\end{eqnarray}
where the derivative is defined  with respect to momentum $\vec{\nabla}_{p} = \left( \frac{\partial}{\partial p_{x}}, \frac{\partial}{\partial p_{y}}, \frac{\partial}{\partial p_{z}} \right)$. 
By using the wavefunctions~(\ref{eq:u_E+1})--(\ref{eq:u_E-2}), the Berry connection is explicitly given 
in the polar coordinate $\vec{A}_{\varepsilon}^{\,(\gamma)}(\vec{p}\,) = \left( A_{\varepsilon,p}^{(\gamma)}(\vec{p}\,), A_{\varepsilon,\theta}^{(\gamma)}(\vec{p}\,),  A_{\varepsilon,\varphi}^{(\gamma)}(\vec{p}\,) \right)$ as 
\begin{eqnarray}
 A_{\varepsilon,p}^{(\gamma)}(\vec{p}\,) &=& 0, \label{eq:Berry_gauge_1} \\
 A_{\varepsilon,\theta}^{(\gamma)}(\vec{p}\,) &=& 0, \label{eq:Berry_gauge_2} \\
 A_{\varepsilon,\varphi}^{(\gamma)}(\vec{p}\,)
&=&
\left\{
\begin{array}{l}
-\frac{1}{2p} \cot \! \frac{\theta}{2} 
 \hspace{0.6em}\mathrm{for}\hspace{0.5em}\gamma=+1,\\
-\frac{1}{2p} \tan \! \frac{\theta}{2} 
 \hspace{0.5em}\mathrm{for}\hspace{0.5em}\gamma=-1.
\end{array}
\right. \label{eq:Berry_gauge_3}
\end{eqnarray}
We notice that $A_{\varepsilon,\varphi}^{(\gamma)}(\vec{p}\,)$ has the singularities at $\theta=0$ ($\gamma=+1$) and $\theta=\pi$ ($\gamma=-1$), which is the so-called Dirac string~\cite{Wu:1976ge}.
As we will see later, the singularity lines can be moved by gauge transformation to other places in the momentum space, but cannot be eliminated in all over the space.
Next we define the Berry curvature (or magnetic field) as
\begin{eqnarray}
  \vec{B}_{\varepsilon}^{(\gamma)}(\vec{p}\,) = \vec{\nabla}_{p} \times \vec{A}_{\varepsilon}^{(\gamma)}(\vec{p}\,).
\end{eqnarray}
From Eqs.~(\ref{eq:Berry_gauge_1})--(\ref{eq:Berry_gauge_3}), the components in the polar coordinate, $\vec{B}_{\varepsilon}^{(\gamma)}=\left( B_{\varepsilon,p}^{(\gamma)}(\vec{p}\,), B_{\varepsilon,\theta}^{(\gamma)}(\vec{p}\,), B_{\varepsilon,\varphi}^{(\gamma)}(\vec{p}\,) \right)$, are given by
\begin{eqnarray}
 B_{\varepsilon,p}^{(\gamma)}(\vec{p}\,) &=& \frac{\gamma}{2p^{2}}, \label{eq:B_r} \\
 B_{\varepsilon,\theta}^{(\gamma)}(\vec{p}\,) &=& 0, \label{eq:B_theta} \\
 B_{\varepsilon,\varphi}^{(\gamma)}(\vec{p}\,) &=& 0. \label{eq:B_varphi}
\end{eqnarray}
We notice that only the radial component has a nonzero magnetic field whose magnitude is $1/(2p^{2})$.
This result indicates the existence of a monopole at the origin in the momentum space.

We define the monopole charge by the surface integral of $\vec{B}_{\varepsilon}^{(\gamma)}(\vec{p}\,)$ on the surface $\Sigma$ surrounding the origin in the momentum space: 
\begin{eqnarray}
 Q_{\varepsilon}^{(\gamma)} \equiv \frac{1}{2\pi} \int_{\Sigma} \vec{B}_{\varepsilon}^{(\gamma)}(\vec{p}\,) \cdot \mathrm{d}\vec{S}.
 \label{eq:charge_monopole}
\end{eqnarray}
Substituting Eqs.~(\ref{eq:B_r})-(\ref{eq:B_varphi}), we obtain $Q_{\varepsilon}^{(\gamma)} = \gamma$. This coincides with the winding number $w_{\varepsilon}^{(\gamma)}=\gamma$ obtained in Sec.~\ref{sec:hedgehog}.
Therefore, we conclude that the winding number associated with the effective heavy quark spin is equivalent to the monopole charge in the Berry phase. We emphasize that $Q_{\varepsilon}^{(\gamma)}$ and $w_{\varepsilon}^{(\gamma)}$ are topologically conserved quantities for adiabatic and continuous deformation of the wavefunctions.

Finally, we consider the gauge transformation of the Berry connection. Recall that the monopole charge $Q_\epsilon^{(\gamma)}=\gamma$ is the eigenvalue of the operator $\Sigma_5(\vec p)$. This implies that the Berry connection is associated with the ${\mathrm U(1)}$ transformation generated by the charge operator $\Sigma_5(\vec p)$. Thus, we define the gauge (phase) transformation for the wavefunction $u_{\varepsilon}^{(\gamma)}(\vec{p}\,)$ as
\begin{eqnarray}
 u_{\varepsilon}^{(\gamma)}(\vec{p}\,) &\rightarrow& e^{i\xi(\vec{p}\,) \, \Sigma_{5}(\hat{p})} u_{\varepsilon}^{(\gamma)}(\vec{p}\,)
\nonumber \\
& &=
 e^{i\gamma \xi(\vec{p}\,)} u_{\varepsilon}^{(\gamma)}(\vec{p}\,),
\end{eqnarray}
with an arbitrary real smooth function $\xi(\vec{p}\,)$.
In the last line, we have used the relation $\Sigma_{5}(\hat{p}) u_{\varepsilon}^{(\gamma)}(\vec{p}\,) = \gamma \, u_{\varepsilon}^{(\gamma)}(\vec{p}\,)$. 
For the Berry connection, we define the gauge transformation
\begin{eqnarray}
 \vec{A}_{\varepsilon}^{(\gamma)}(\vec{p}\,) \rightarrow \vec{A}_{\varepsilon}^{(\gamma)}(\vec{p}\,)+ \gamma \vec{\nabla}_{p} \xi (\vec{p}\,).
\end{eqnarray}
We can easily check that $\vec{B}_{\varepsilon}^{(\gamma)}(\vec{p}\,)$ is invariant for any smooth $\xi(\vec{p}\,)$.
Therefore, $\vec{B}_{\varepsilon}^{(\gamma)}$ and $Q_{\varepsilon}^{(\gamma)}$ are gauge-invariant. This also implies that we can move the Dirac string by the gauge transformation while keeping the monopole charge unchanged. \\
% !TEX root = 0_main.tex
\section{Kondo excitons: collective excitations}
\label{sec:Kondo_excitations}

In the previous section, we determined the ground state within the mean-field approximation. In this section, we go beyond the mean-field approximation by including the effects of fluctuations around the mean-fields. Then we are able to discuss stability of the mean-field ground state and collective excitations around the mean-fields. We consider the ``$hQ$ modes" as excitation modes which are made of a light quark {\it hole} $h$ and a heavy quark $Q$ (in a small momentum region). The analysis will be performed within the random-phase approximation (RPA) where we use quark propagators determined by the mean-field Lagrangian and include the effects of interactions by summing up the bubble diagrams each of which represents the light quark and heavy quark excitation. As we will see later the resulting excitation modes consist of superposition of quark fields with different momenta and we call these collective excitations the {\it Kondo excitons}. Indeed, they appear as bound states when the interaction in the $hQ$ channel is strongly attractive. Recall that we have selected the mean-fields so that they have nonzero values only for those containing the first flavor of light quark $q_1$. Thus the effects of mixing is expected to appear only in the excitation modes containing the $q_1$ field. It is important to distinguish the excitations involving $q_{1}$ and $Q$ from those involving $q_{i}$ ($i=2,\dots,N_{f}$) and $Q$.

\subsection{RPA equation}
\label{sec:hole_Q_modes}

In the following calculation, we use the quark propagators obtained in the mean-field approximation (see Eq.~(\ref{eq:Ginverse_RL_Nf}) for its inverse, and Eq.~(\ref{propagator_Nf=1}) for the propagator in the $N_f=1$ case), and treat the interaction Lagrangian (\ref{eq:Lint_RL_Nf_singlet_phi}).

\begin{figure}[tb]
  \centering
  \includegraphics[keepaspectratio,scale=0.15]{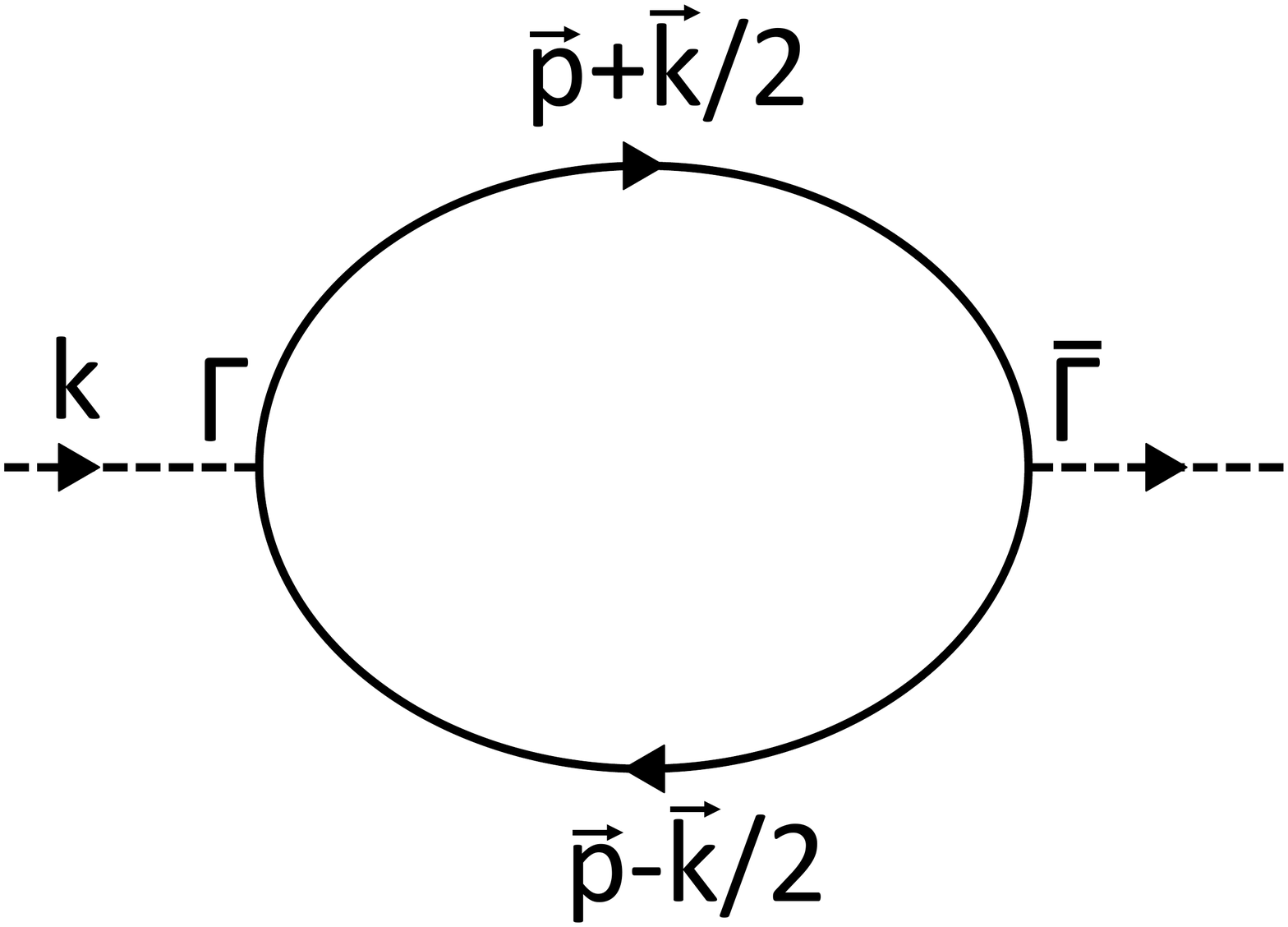}
  \caption{
A one-loop diagram of the self-energy $\frac{1}{i}\Pi(k_{0},\vec{k}\,)$. Solid lines are quark propagators $G$ in the presence of the Kondo condensate. We denote $\Gamma=\Gamma_{\ell}^{(i)}$ and $\bar{\Gamma}=\bar{\Gamma}_{\ell}^{(i)}$ for $\ell=\{\mathrm{S}, \mathrm{V}, \mathrm{P}, \mathrm{A}\}$ and $i=1,\dots,N_{f}$.
}
  \label{fig:Pi_diagram}
\end{figure}

We consider excitation modes in the channels represented by the composite fields (\ref{Phi})--(\ref{Phi_vec5}). For the channel labelled by $\ell=\{ \mathrm{S, V, P, A}\}$ and flavor $i=1,\dots,N_{f}$, we compute the ``self-energy" $\frac{1}{i} \Pi_{\ell}^{(i)}(k_{0},\vec{k}\,)$ which corresponds to a one-loop bubble diagram as shown in Fig.~\ref{fig:Pi_diagram}. Here, $k^0$ and $\vec k$ are the energy and momentum of the excitation mode: 
\begin{eqnarray}
 \frac{1}{i} \Pi_{\ell}^{(i)}(k_{0},\vec{k}\,)
&=&
- \mathrm{Tr} \! \int \!\! \frac{\mathrm{d}^{4}p}{(2\pi)^{4}}
\bar{\Gamma}_{\ell}^{(i)} iG ( p_{0}+\frac{k_{0}}{2}, \vec{p}+\frac{\vec{k}}{2} )
\nonumber \\
&&\hspace{8mm}\times \Gamma_{\ell}^{(i)}  iG ( p_{0}-\frac{k_{0}}{2}, \vec{p}-\frac{\vec{k}}{2} ), 
\label{eq:self_energy_hQ}
\end{eqnarray}
where the propagator $G(p_{0},\vec{p}\,)$ and the vertex matrices $\Gamma_{\ell}^{(i)}$, $\bar{\Gamma}_{\ell}^{(i)}$ are defined in Eq.~(\ref{eq:Ginverse_RL_Nf}), and Eqs.~(\ref{eq:scalar_matrices})-(\ref{eq:axialvector_matrices}), respectively, and $``\mathrm{Tr}"$ is the trace over the color indices, the Dirac matrices and the flavor matrices. The minus sign comes from the fermion loop. 
If we carefully look at the matrix structure for each channel, we find that the loop propagators are made of the light quark hole $h_{i}$ with light flavor $i$ and the heavy quark $Q$.

In the RPA calculation, we sum up contributions of the ring diagrams which are made of the bubble diagram, Eq.~(\ref{eq:self_energy_hQ}), to obtain 
\begin{widetext}
\begin{eqnarray}
iU_{\ell}^{(i)}(k_{0},\vec{k}\,) 
 &=&
 \bar{\Gamma}_{\ell}^{(i)}
\biggl\{
i {\cal G}_{c}
+ i {\cal G}_{c}
\left( \frac{1}{i} \Pi_{\ell}^{(i)}(k_{0},\vec{k}\,) \right) 
i {\cal G}_{c}
+ i{\cal G}_{c}
\left( \frac{1}{i} \Pi_{\ell}^{(i)}(k_{0},\vec{k}\,) \right) 
i {\cal G}_{c}
\left( \frac{1}{i} \Pi_{\ell}^{(i)}(k_{0},\vec{k}\,) \right) 
i {\cal G}_{c}
+ \dots
\biggr\}
\Gamma_{\ell}^{(i)}
\nonumber \\
&=&
\bar{\Gamma}_{\ell}^{(i)}
\frac{i{\cal G}_{c}
}{1-{\cal G}_{c}
\Pi_{\ell}^{(i)}(k_{0},\vec{k}\,)}
\Gamma_{\ell}^{(i)},
 \label{eq:vertex_qbarQ_S}
\end{eqnarray}
\end{widetext}
where we have defined ${\cal G}_{c}\equiv \frac{N_{c}^{2}-1}{4N_{c}^{2}}G_{c}$ for brevity. The pole of the last equation is called the RPA equation:
\begin{eqnarray}
 1-{\cal G}_{c}
\Pi_{\ell}^{(i)}(k_{0},\vec{k}\,) = 0.
 \label{eq:RPA_qbarQ}
\end{eqnarray}
This equation determines the relation between $k_{0}$ and $\vec{k}$, that is nothing but the dispersion relation of the excited mode.

\subsection{Dispersions of excited states}

To understand the structure of the self-energy, let us briefly discuss the case $N_f=1$ which carries the essential information on the mixing phenomena. In this case, we can use the propagator (\ref{propagator_Nf=1}) to compute the self-energy (\ref{eq:self_energy_hQ}). For example, the self-energy in the scalar mode (S) is given by 
\begin{eqnarray}
\frac{1}{i}\Pi_{\rm S}(k_0,\vec k)={\rm Tr}\int \frac{d^4p}{(2\pi)^4} G_{\Psi\Psi}(p_1)G_{\varphi\varphi}(p_2),
\end{eqnarray}
where $p_1=p+k/2$, $p_2=p-k/2$, and $G_{ij}(p)$ is an element of the propagator in the $(\varphi, \eta, \Psi_v)$ space with $\psi=(\varphi,\eta)^t$. Namely, they are explicitly given by (see Eq.~(\ref{propagator_Nf=1}))
\begin{eqnarray}
G_{\Psi\Psi}(p)&=&\frac{p_0-p+\mu}{(p_0-E_p^+)(p_0-E_p^-)}\cdot {\mathbf 1},\\
G_{\varphi\varphi}(p)&=&\frac{(p_0-\lambda)(p_0+\mu)-|\Delta|^2}{(p_0-\tilde E_p)(p_0-E_p^+)(p_0-E_p^-)} \cdot {\mathbf 1},
\end{eqnarray}
where ${\mathbf 1}$ is a $2\times 2$ identity matrix. This result clearly explains that one of the two propagators is from the light quark and the other from the heavy quark, suggesting that the excitation mode is made of a light quark hole and a heavy quark.  Notice that the heavy-quark component of the propagator $G_{\Psi\Psi}$ has only two poles because the pole at $p_0=\tilde E_p$ cancels with the same factor in the denominator (see Eq.~(\ref{propagator_Nf=1})). The same calculations for the other channels $\ell=\mathrm{V, P, A}$ lead to a surprising result: All the channels have the same structure as the scalar channel, $\Pi_{\mathrm{S}}(k) = \Pi_{\mathrm{V},k}(k) = \Pi_{\mathrm{P}}(k) = \Pi_{\mathrm{A},k}(k)$. Furthermore, this is in fact not a special result in the one-flavor case, but can be seen also in the case $N_f > 1$. Indeed, after some calculations, we find for each flavor index $i=1,\dots,N_{f}$, the self-energies in all the channels have the same structure:
\begin{eqnarray}
\Pi_{\mathrm{S}}^{(i)}(k_{0},\vec{k}\,) = \Pi_{\mathrm{V},k}^{(i)}(k_{0},\vec{k}\,) = \Pi_{\mathrm{P}}^{(i)}(k_{0},\vec{k}\,) = \Pi_{\mathrm{A},k}^{(i)}(k_{0},\vec{k}\,).
\nonumber \\
\end{eqnarray}
This means that (for each flavor) the excitation energies for scalar, vector, pseudoscalar and axial vector channels are degenerate. We notice, however, that the excitation energy for the light flavor $i=1$ is different from those of the other light flavors $i=2,\dots,N_{f}$, as we confirm in the numerical calculations.

%%%%%%%%%%%%%%%%%%%%%%%%%%
%%%%%%%%%%%%%%%%%%%%%%%%%%

\begin{figure}[tbp]
  \begin{center}
   \begin{minipage}[b]{1.0\linewidth}
   \centering
   \includegraphics[keepaspectratio,scale=0.5]{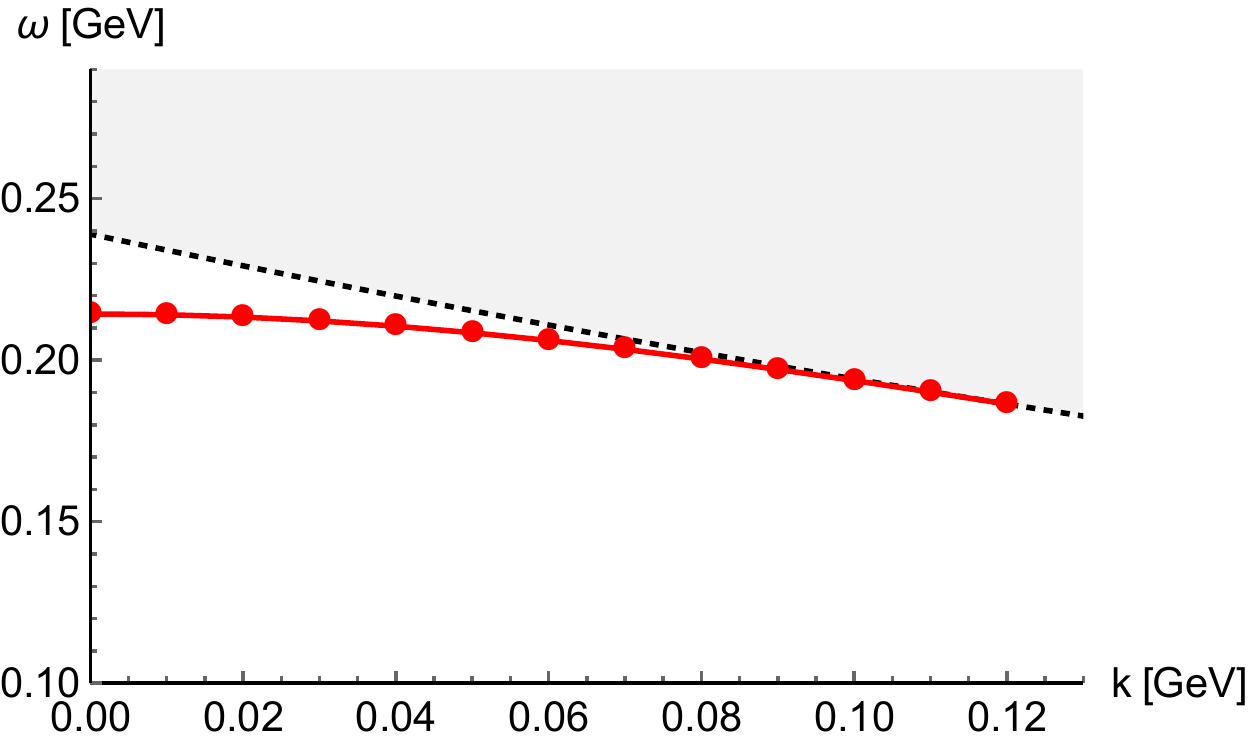}
 \end{minipage} \\
 \vspace{0.5em}
 \begin{minipage}[b]{1.0\linewidth}
    \centering
   \includegraphics[keepaspectratio,scale=0.5]{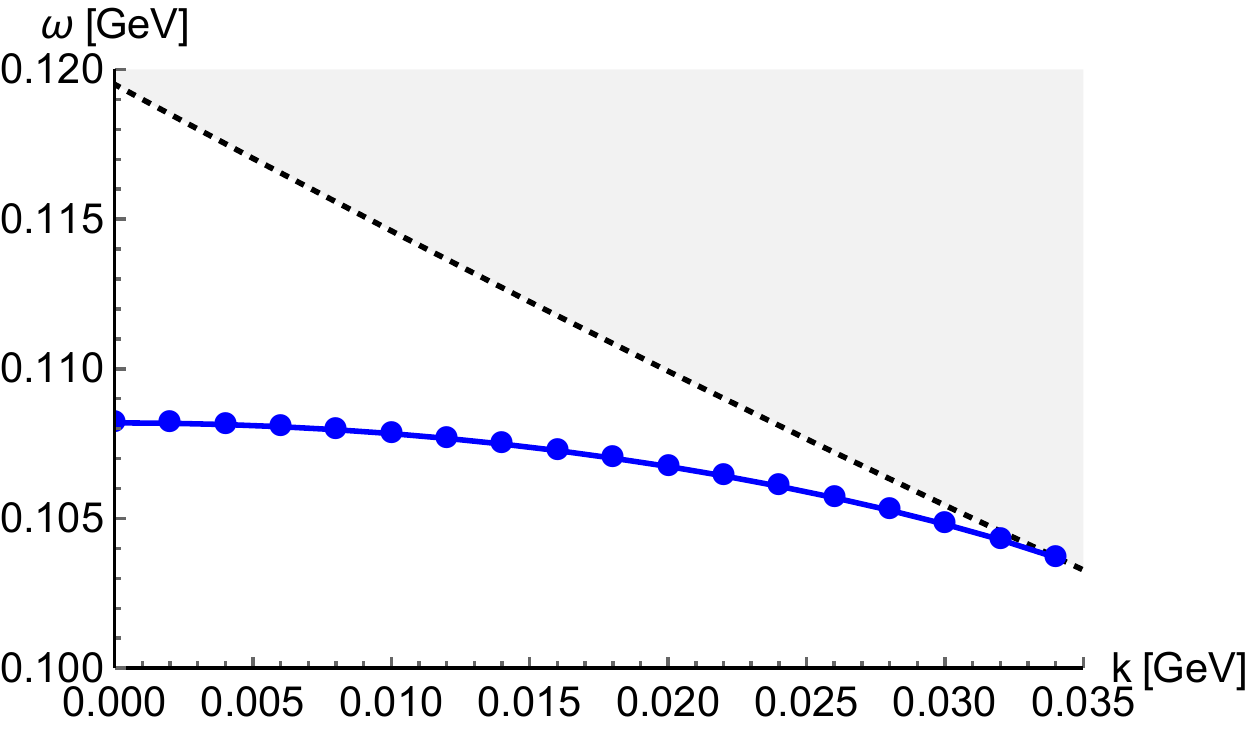}
 \end{minipage}
   \end{center}
 \caption{
Numerical results of $\omega_{k}(=k_{0})$ as functions of $k=|\vec{k}\,|$ for $h_{i}Q$ modes ($J^{P}=0^{\pm},1^{\pm}$) with $i=1$ (upper panel) and $i=2,\dots,N_{f}$ (lower panel). The dashed lines are thresholds for free $h_{i}$ and $Q$, and the upper gray regions are the scattering states. 
}
  \label{fig:RPA_solution_3}
\end{figure}

In evaluating the loop integral, we first pick up the poles which lie in the Fermi sea for the $p_0$-integral. Here, we notice that the number of poles is not necessarily equal to the number of dispersions as we have seen for the $N_f=1$ case. Then the three-momentum $|\vec{p}\,|$ integrals are performed for $0 \le p \le \Lambda$ with the three-momentum cutoff parameter $\Lambda$. For numerical evaluation, we work at $\lambda=0$ with the parameter set given in Sec.~\ref{sec:gap_equation}. In Fig.~\ref{fig:RPA_solution_3}, we show the excitation energies $\omega_{k}(=k_{0})$ for the flavor $i=1$ (upper panel) and the others $i\neq 1$ (lower panel) as functions of $k=|\vec{k}\,|$. Importantly, we find no imaginary solution for $\omega_{k}$. Therefore, we conclude that our ground state with the mean-fields introduced in Sec.~\ref{sec:mean_field} is stable against the small quantum fluctuations. Also important is the absence of zero modes. Although we discussed complicated symmetry breaking pattern associated with formation of the Kondo condensate, the RPA calculation does not produce massless modes in the channels considered here. At present, it is not clear why there is no zero modes in our RPA calculation.

We can understand these results with the help of dispersion relations shown in Fig.~\ref{fig:dispersion_MF}. First of all, consider the excitation involving the $i=1$ light quark (upper panel of Fig.~\ref{fig:RPA_solution_3}). Recall that the ground state for $\lambda=0$ is made of the occupied Fermi sea with the dispersion $p_0=E_p^-<0$. Any excitation should be constructed based on this ground state. For example, when $k=0$, the excitation should be described by a superposition of states with momenta $\vec p$ and $-\vec p$, corresponding to a particle and a hole. This excitation is drawn as a vertical transition from the (occupied) $E_p^-$ mode to the (vacant) $E_p^+$ mode.\footnote{We simply call the $hQ$ mode, but in fact the excitation appears as a superposition of states with different values of $p$ (and different content). Recall that $E_p^+$ ($E_p^-$) is more like a heavy (light) quark at small $p$ and more like a light (heavy) quark at large $p$. Thus, excitations with small $p$ are made of a hole of the light quark and a heavy quark, but  excitations with large $p$ are made of a light quark and a hole of the heavy quark.} The lowest excitation energy is given by the lowest value of $E_p^+-E_p^-=\sqrt{(p-\mu)^2+8|\Delta |^2} \ge 2\sqrt{2} |\Delta|$ which is realized at $p=\mu$. In reality, the interaction between a light quark and a heavy quark helps to form a bound state, and the excitation energy will be reduced. In the numerical result, the excitation energy is indeed close to but smaller than the threshold value $2\sqrt 2 |\Delta|\simeq 0.24~$GeV (for $|\Delta|=0.085~$GeV). When $k\neq 0$, the transition occurs from the $E_p^-$ mode with the momentum $p_1=p+k/2$ to the $E_p^+$ mode with the momentum $p_2=p-k/2$, which is represented as a gray arrow in the upper panel of Fig.~\ref{fig:dispersion_MF}. Again, the excitation mode will be made of superposition of such (single-particle) excitations. The excitation energy for this transition can be roughly evaluated for small $k$ (and at $p=\mu$) as $E_{p_2}^+-E_{p_1}^-\simeq 2\sqrt 2 |\Delta| -k/2$. The actual excitation energy $\omega(k)$ will be reduced by the interaction. The dashed line in the upper panel of Fig.~\ref{fig:RPA_solution_3} corresponds to the rough estimate of the excitation energy, and the numerical result indeed lies below this line. The negative slope of the excitation energy is also understood as a result of peculiar behavior of dispersions in the mixing phenomena. Namely, it appears due to the asymmetry that the minimum point of the energy-momentum dispersions above the Fermi surface is different from the maximum point below the Fermi surface. 

Second, consider the excitation involving the other flavor of light quarks ($i\neq 1$). Again, we can understand the numerical results with the help of dispersion relations as shown in the lower panel of Fig.~\ref{fig:dispersion_MF}. The difference from the previous case with $i=1$ is two-fold: The dispersions of the other flavor light quarks are given by $p_0=E_p=p-\mu$ and the modes with $p<\mu$ (the Fermi sea) are occupied in the ground state. Since these light quarks can interact with the heavy quarks, we are able to effectively consider the transition from the $E_p$ mode to the $E_p^+$ mode. When $k=0$, this transition should occur vertically in the figure of dispersions and the lowest excitation energy is given by $E_{p=\mu}^+-E_{p=\mu}=\sqrt 2 |\Delta|$. The actual excitation energy will be reduced due to interaction. When $k\neq 0$, we can again roughly estimate the excitation energy. In this case, we take $p_1=\mu$ for the dispersion $E_{p_1}$, then we find $E_{p_2}^+-E_{p_1}\simeq \sqrt 2 |\Delta| -k/4$. This corresponds to the dashed line  in the lower panel of Fig.~\ref{fig:RPA_solution_3} (note that $\sqrt 2 |\Delta|=0.12~$GeV). The actual excitation energy is indeed below this threshold line. Therefore, we understood that the difference between the excitation energies in the $i=1$ mode and the $i\neq 1$ mode is mainly due to the different behavior of the dispersion relations. Notice that, in both cases, the actual excitation modes appear as superpositions of excitations with different relative momenta $p$, and thus are ``collective" excitations similar to exciton modes in condensed matter physics.
It is interesting that the excitations that include the $i \neq 1$ flavor light quarks have lower energy than that for the $i=1$ mode.

\begin{figure}[tb]
  \centering
  \vspace{-3em}
  \includegraphics[keepaspectratio,scale=0.3]{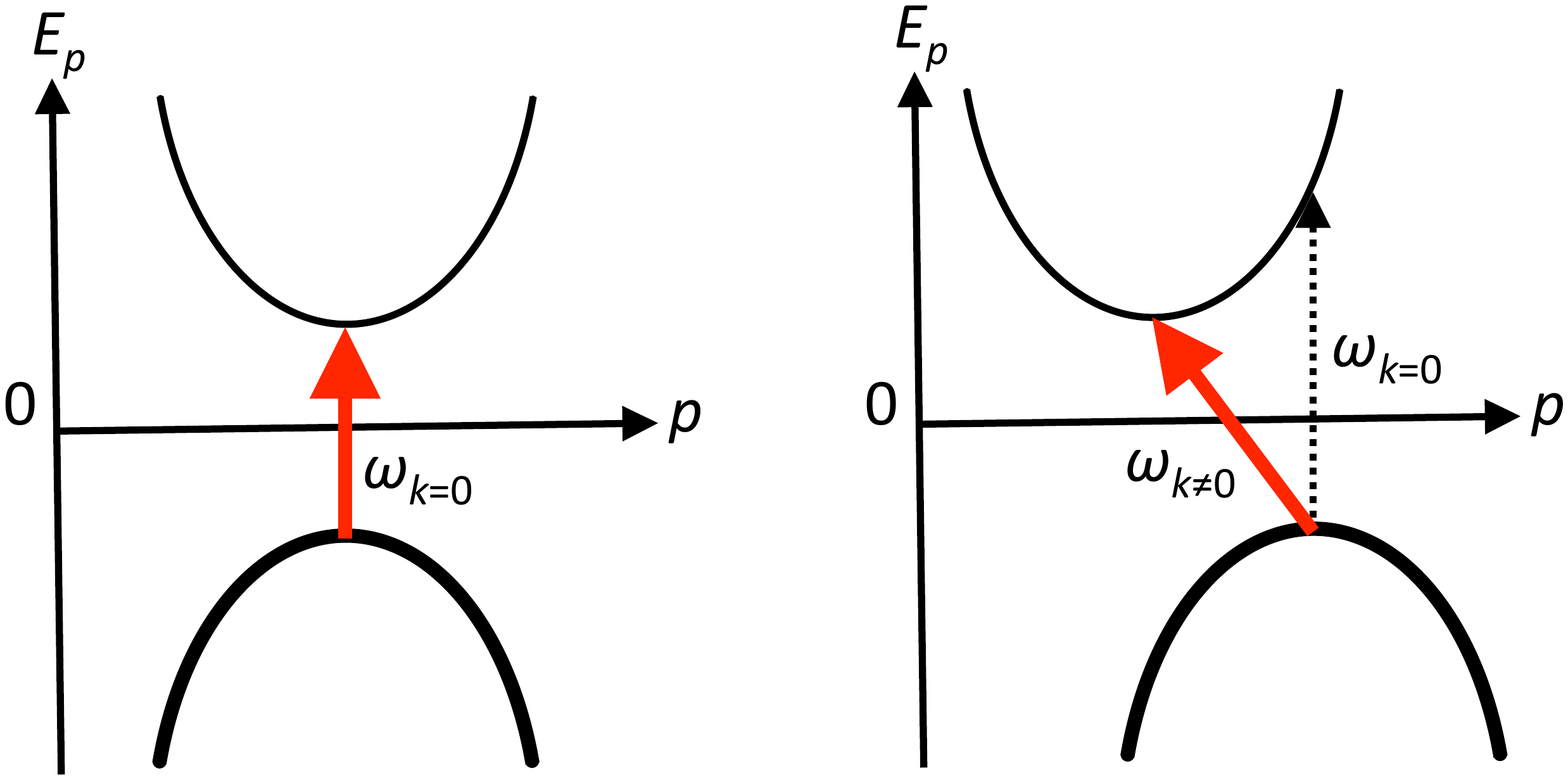}
  \vspace{-3em}
  \caption{
Schematic pictures of the particle-hole excitations from the state inside the Fermi sphere ($E_{p}<0$; thick curves) to the state outside the Fermi sphere ($E_{p}>0$; thin curves). See the text for more explanations. 
}
  \label{fig:energy_excitation}
\end{figure}

Lastly, let us again emphasize the origin of dispersions of excitations with negative slopes. In Fig.~\ref{fig:energy_excitation} we show schematic figures of two different types of excitations corresponding to the transitions from the highest state inside the Fermi sphere to the lowest state outside the Fermi sphere. Excitations with minimum excitation energy are indicated by solid arrows. The left panel presents the case where the momentum for the highest-energy state for $E_{p}<0$ and that for the lowest-energy state for $E_{p}>0$ coincide with each other. The right panel presents the case where they are different. The minimum excitation energy in the left panel is given by the zero momentum ($\omega_{k=0}$), while that in the right panel is given by the non-zero momentum ($\omega_{k\neq0}$). The dotted arrow in the right panel is a excitation with zero momentum ($\omega_{k=0}$), which is not favored. Hence, we find that, when there is asymmetry in the dispersion relations, the minimum excitation energy is given by $\omega_{k\neq0}$ with non-zero momentum. This is the origin of the decreasing behavior of the excitation energies with increasing momentum, as shown in Fig.~\ref{fig:RPA_solution_3}.

%%%%%%%%%%%%%%%%%%%%%%%%%%%%%%%%%%%
%%%%%%%%%%%%%%%%%%%%%%%%%%%%%%%%%%%

%%%%%%%%%%%%%%%%%%%%%%%%%%
%%%%%%%%%%%%%%%%%%%%%%%%%%
% !TEX root = 0_main.tex
\section{Conclusion and outlook}

We investigated the ground and excited states of a quark matter with heavy quark impurities. In order to study the Kondo effect, we worked with a model having non-Abelian current-current interactions between a light quark and a heavy quark. The ground state of this model was found to be characterized by the Kondo condensates which are the expectation values of the products of light and heavy quark operators and generate mixing between them. We nonperturbatively determined the values of the Kondo condensates in a self-consistent way within the mean-field approximation. The wavefunctions of the ground state are described by the HQS hedgehog configuration in the momentum space, which allows for topological interpretation with respect to the Berry phase. As for the excited states, we solved the RPA equations for the excitations made of a hole $h$ of the light quark and the heavy quark $Q$ and found that the ground state is stable and the excited states appear as exciton-like bound states.

The primary purpose of the present paper was to demonstrate the physical consequences of the nonzero Kondo condensates. Thus we analyzed a simple model having only the heavy-light interactions. However, in order to study the Kondo effects in more realistic quark/nuclear matter in QCD at finite densities, we need to include the effects of interactions in the light quark sector. For example, the $\mathrm{U}(1)_{\mathrm{A}}$ symmetry for light (massless) quarks should be broken by the quantum anomaly, which could affect the $\chi$HQSL symmetry in the Kondo condensate. Also it would be interesting to include the effects of interactions between two light quarks, which generate the chiral symmetry breaking at low densities, and the color superconductivity at high densities~\cite{Buballa:2003qv,Alford:2007xm,Fukushima:2010bq,Fukushima:2013rx}. In both cases, we can investigate the interplay between them and the Kondo effect (see Ref.~\cite{Kanazawa:2016ihl} as a related study).
In particular, in the color superconductivity phase with two light flavors $N_{f}=2$ (the 2SC phase), the original color symmetry $\mathrm{SU}(3)_{\mathrm{c}}$ is broken to $\mathrm{SU(2)}_{\mathrm{c}}$. If we consider the interaction between the heavy quark and the gapped light quark (which is still light compared to the heavy quark mass), we find a mismatch in the non-Abelian symmetry of the interaction: the $\mathrm{SU(2)}_{\mathrm{c}}$ symmetry of the gapped light quark and the $\mathrm{SU}(3)_{\mathrm{c}}$ symmetry of the heavy quark. Thus, the gapped light quark with a smaller symmetry cannot screen the $\mathrm{SU}(3)_{\mathrm{c}}$ charge of the heavy quark. This is a novel situation which is not seen in condensed matter, while this inability of complete screening is somewhat similar to the ``underscreening" Kondo effect. 
There are many topological structures of quark matter in real space~\cite{Eto:2013hoa}. Experimentally, to relate the present study to observables in the heavy-ion collisions, it will be also necessary to estimate the transport coefficients of quark matter taking into account the Kondo condensate. Those subjects are left for future studies.

\section*{Acknowledgements}
This work is supported partly by the Grant-in-Aid for Scientific Research (Grant No.~25247036, No.~15K17641 and No.~16K05366) from Japan Society for the Promotion of Science (JSPS).
K.S. is supported by MEXT as ``Priority Issue on Post-K computer" (Elucidation of the Fundamental Laws and Evolution of the Universe) and JICFuS.
\appendix
% !TEX root = 0_main.tex
\section{Remark on coupling constant $G_{c}$}
\label{sec:coupling_constant}

We comment how we determine the numerical value of the coupling constant $G_{c}$. Using the unified representation $\phi$ for the fields $\psi_{1}, \dots, \psi_{N_{f}}$ and $\Psi_{v}$ (see Eq.~(\ref{eq:phi_def})), we start from the following definition of the color-current interaction with the coupling constant $\bar{G}_{c}$: 
\begin{widetext}
\begin{eqnarray}
{\cal L}_{\mathrm{int}}
&=&
-\bar{G}_{c} \sum_{a} (\bar{\phi} \gamma^{\mu} T^{a} \phi) (\bar{\phi} \gamma_{\mu} T^{a} \phi)
\nonumber \\
&=&
-\bar{G}_{c} \sum_{a} 
\Bigl\{
(\bar{\psi} \gamma^{\mu} T^{a} \psi) (\bar{\psi} \gamma_{\mu} T^{a} \psi)
+2(\bar{\psi} \gamma^{\mu} T^{a} \psi) (\bar{\Psi}_{v} \gamma_{\mu} T^{a} \Psi_{v})+ \cdots 
\Bigr\}.
\end{eqnarray}
\end{widetext} 
Thus, this interaction generates the current-current interaction in the light quark sector 
\begin{eqnarray}
 {\cal L}_{\mathrm{int}}^{\mathrm{LL}}
 =
-\bar{G}_{c} \sum_{a} 
(\bar{\psi} \gamma^{\mu} T^{a} \psi) (\bar{\psi} \gamma_{\mu} T^{a} \psi),
 \label{eq:int_LL}
\end{eqnarray}
and the interaction between the light current and the heavy current: 
\begin{eqnarray}
 {\cal L}_{\mathrm{int}}^{\mathrm{LH}}
 =
-2\bar{G}_{c} \sum_{a} 
(\bar{\psi} \gamma^{\mu} T^{a} \psi) (\bar{\Psi}_{v} \gamma_{\mu} T^{a} \Psi_{v}).
 \label{eq:int_LH}
\end{eqnarray}
The latter interaction is the same as that in Eq.~(\ref{eq:Lagrangian_current}) when $G_{c}=2\bar{G}_{c}$ is satisfied.

The value of the coupling constant $G_{c}$ is given once we determine the value of $\bar{G}_{c}$ in Eq.~(\ref{eq:int_LL}). The value of $\bar{G}_{c}$ is evaluated for the $N_{f}=2$ case ($\psi=(\psi_{1},\psi_{2})^{t}$) so that the interaction (\ref{eq:int_LL}) reproduces the properties of light hadrons. We transform ${\cal L}_{\mathrm{int}}^{\mathrm{LL}}$ to the Nambu--Jona-Lasinio type interaction
\begin{eqnarray}
 {\cal L}_{\mathrm{NJL}}^{N_{f}=2}
 =
 \tilde{G}_{c}
 \bigl(
  (\bar{\psi}\psi) (\bar{\psi}\psi) + (\bar{\psi}i\gamma_{5}\vec{\tau} \psi) (\bar{\psi}i\gamma_{5}\vec{\tau}\psi)
 \bigr),
\end{eqnarray}
with $\tilde{G}_{c}=(2/9)\bar{G}_{c}$.
Here, we use the Fierz identities for SU($N_{c}$) color operators, Eq.~(\ref{eq:Fierz_SUN_2}) with $N=N_{c}=3$, where $\vec{\tau}=(\tau^{1},\tau^{2},\tau^{3})$ are the Pauli matrices for the isospin. In the literature, it is known that the low energy properties of light hadrons are well reproduced with the values of $\bar{G}_{c}$ given by $\bar{G}_{c}\Lambda^{2}=2.14$ where the three-dimensional momentum cutoff parameter is $\Lambda=0.653$ GeV~\cite{Klevansky:1992qe,Hatsuda:1994pi}. Using these parameter values, we set $\tilde{G}_{c}\Lambda^{2}=2.0$ and $\Lambda=0.65$ GeV, which leads to $\bar{G}_{c}\Lambda^{2}=(9/2)2.0$. In this way, we use $G_{c}\Lambda^{2}=(9/2)4.0$ with $\Lambda=0.65$ GeV in our numerical calculations.
% !TEX root = 0_main.tex
\section{Fierz identities}
\label{sec:Fierz}

Here we summarize the Fierz identities.
The Fierz identities for Dirac matrices are
\begin{eqnarray}
 (\gamma^{\mu})_{\alpha \beta} (\gamma_{\mu})_{\gamma \delta}
&=&
\delta_{\alpha \delta} \delta_{\gamma \beta}
+ (i\gamma_{5})_{\alpha \beta}  (i\gamma_{5})_{\gamma \delta}
\nonumber \\
&& -\frac{1}{2} (\gamma^{\mu})_{\alpha \beta}  (\gamma_{\mu})_{\gamma \delta}
-\frac{1}{2} (\gamma^{\mu}\gamma_{5})_{\alpha \beta}  (\gamma_{\mu} \gamma_{5})_{\gamma \delta},
\nonumber \\
\label{eq:Fierz_Dirac_1} \\
 (\gamma^{\mu})_{\alpha \beta}  (\gamma_{\mu})_{\gamma \delta}
 &=&
  (i\gamma_{5}C)_{\alpha \gamma}  (Ci\gamma_{5})_{\delta \beta}
 + (C)_{\alpha \gamma}  (C)_{\delta \beta}
\nonumber \\
&& -\frac{1}{2} (\gamma^{\mu}\gamma_{5}C)_{\alpha \gamma}  (C\gamma_{\mu}\gamma_{5})_{\delta \beta}
 \nonumber \\
&& -\frac{1}{2} (\gamma^{\mu}C)_{\alpha \gamma}  (C\gamma_{\mu})_{\delta \beta},
\label{eq:Fierz_Dirac_2}
\end{eqnarray}
with $C=i\gamma^{2}\gamma^{0}$, where sum over $\mu=0,1,2,3$ is taken.
The Fierz identities for the matrices of the $\mathrm{SU}(N)$ generators are
\begin{eqnarray}
 \delta_{ij}  \delta_{kl} &=& \frac{1}{N} \delta_{il}  \delta_{kj} + \frac{1}{2} (\lambda^{a})_{il}  (\lambda^{a})_{kj},
\label{eq:Fierz_SUN_1} \\
 (\lambda^{a})_{ij}  (\lambda^{a})_{kl} &=& 2\frac{N^{2}-1}{N^{2}} \delta_{il}  \delta_{kj} - \frac{1}{N} (\lambda^{a})_{il}  (\lambda^{a})_{kj},
\label{eq:Fierz_SUN_2}
\end{eqnarray}
and
\begin{eqnarray}
 (\lambda^{A})_{ij}  (\lambda^{A})_{kl} &=& \frac{N-1}{N} \delta_{ik}  \delta_{lj} - \frac{1}{2} (\lambda^{a})_{ik}  (\lambda^{a})_{lj},
\label{eq:Fierz_SUN_3} \\
 (\lambda^{S})_{ij}  (\lambda^{S})_{kl} &=& \frac{N+1}{N} \delta_{ik}  \delta_{lj} + \frac{1}{2} (\lambda^{a})_{ik}  (\lambda^{a})_{lj},
\label{eq:Fierz_SUN_4}
\end{eqnarray}
where sums over $a=1,\dots,N^{2}-1$, $S$ for symmetric matrices (including $\lambda^{0}$) and $A$ for asymmetric matrices are taken.
% !TEX root = 0_main.tex
\section{Kondo condensate of Weyl fermions}
\label{sec:Weyl_right}

In this appendix, we consider the Weyl fermion with the color-current interaction, and show that the coexistence of the scalar Kondo condensate and the vector Kondo condensate is realized in the ground state within the mean-field approximation. For simplicity, we consider the one flavor case for the light quark, $N_{f}=1$.

We determine the form of the color-current interaction for the Weyl fermion through the interaction for the Dirac fermion $\psi$. The Dirac fermion is decomposed into the right-handed Weyl fermion $\chi$ and the left-handed Weyl fermion $\varphi$ as 
\begin{eqnarray}
\psi
=
\frac{1}{\sqrt{2}}
\left(
\begin{array}{c}
 \chi + \varphi \\
 \chi -\varphi
\end{array}
\right),
\end{eqnarray}
in the standard (Dirac) representation. Applying this to the interaction term in Eq.~(\ref{eq:Lagrangian_current_eff}), picking up the terms relevant to $\chi$, and performing the Fierz transformation, we obtain the interaction Lagrangian for the right-handed Weyl fermion
\begin{eqnarray}
{\cal L}_{\mathrm{int}}^{\chi}
&=&
- \frac{G_{c}}{4} \sum_{a} \chi^{\dag} \lambda^{a} \chi \, {\Psi}_{v}^{\dag} \lambda^{a} \Psi_{v}
\nonumber \\
&=&
\frac{(N_{c}^{2}-1)G_{c}}{4N_{c}^{2}} \chi^{\dag} \Psi_{v}\, {\Psi}_{v}^{\dag} \chi
+ \frac{(N_{c}^{2}-1)G_{c}}{4N_{c}^{2}} \chi^{\dag} \vec{\sigma} \Psi_{v}\, {\Psi}_{v}^{\dag} \vec{\sigma} \chi
\nonumber \\
&&
- \frac{G_{c}}{8N_{c}} \sum_{a} \chi^{\dag} \lambda^{a} \Psi_{v}\, {\Psi}_{v}^{\dag} \lambda^{a} \chi
\nonumber \\
&&
- \frac{G_{c}}{8N_{c}} \sum_{a} \chi^{\dag} \vec{\sigma} \lambda^{a} \Psi_{v}\, {\Psi}_{v}^{\dag} \vec{\sigma} \lambda^{a} \chi.
\label{eq:int_L_Weyl_R}
\end{eqnarray}
Here the Fierz transformation is used
for SU(2) spin symmetry, Eq.~(\ref{eq:Fierz_SUN_1}) with $N=N_{f}=2$ and
for SU($N_{c}$) color symmetry, Eq.~(\ref{eq:Fierz_SUN_2}) with $N=N_{c}=3$.
In the following, we focus on the color singlet term in Eq.~(\ref{eq:int_L_Weyl_R}), namely
\begin{eqnarray}
{\cal L}_{\mathrm{int}}^{\chi,\mathrm{sing}}
=
g_{c}
\bigl(
\chi^{\dag} \Psi_{v} {\Psi}_{v}^{\dag} \chi
+ \chi^{\dag} \vec{\sigma} \Psi_{v} {\Psi}_{v}^{\dag} \vec{\sigma} \chi
\bigr),
\end{eqnarray}
with $g_{c}=(N_{c}^{2}-1)G_{c}/4N_{c}^{2}$.
In addition, including the kinetic terms and the constraint condition for the heavy quark field, we obtain the effective Lagrangian
\begin{eqnarray}
{\cal L}_{\mathrm{eff}}^{\chi,\mathrm{sing}}
&=& 
\chi^{\dag} (p_{0} + \mu - \vec{\sigma} \!\cdot\! \vec{p}\, ) \chi
+ p_{0} \Psi_{v}^{\dag} \Psi_{v}
\nonumber \\
&& 
+ g_{c}
\bigl(
\chi^{\dag} \Psi_{v} {\Psi}_{v}^{\dag} \chi
+ \chi^{\dag} \vec{\sigma} \Psi_{v} {\Psi}_{v}^{\dag} \vec{\sigma} \chi
\bigr)
\nonumber \\
&&
-\lambda \left( \Psi_{v}^{\dag}\Psi_{v} - n_{Q} \right).
\label{eq:Lagrangian_R_Nf1}
\end{eqnarray}

We proceed the same procedures as in the Dirac fermions. We first introduce the scalar filed $\Phi=g_{c}\chi^{\dag} \Psi_{v}$ and the vector field $\vec{\Phi}=g_{c}\chi^{\dag} \vec{\sigma}\, \Psi_{v}$, and rewrite the interaction terms as 
\begin{eqnarray}
\hspace*{-2em}
g_{c} \chi^{\dag} \Psi_{v} {\Psi}_{v}^{\dag} \chi
&=&
\Phi \,
 \Psi_{v}^{\dag} \chi
+\chi^{\dag} \Psi_{v}\Phi^{\dag} 
- \frac{1}{g_{c}} |\Phi|^{2}, \\
\hspace*{-2em}
g_{c} \chi^{\dag} \vec{\sigma}\, \Psi_{v} \Psi_{v}^{\dag} \vec{\sigma}\, \chi
&=&
 \vec{\Phi} \!\cdot\! \Psi_{v}^{\dag} \vec{\sigma}\, \chi
+\chi^{\dag} \vec{\sigma}\, \Psi_{v} \!\cdot\! \vec{\Phi}^{\dag} 
- \frac{1}{g_{c}} |\vec{\Phi}|^{2},
\end{eqnarray}
for the scalar channel and for the vector channel, respectively. 
Then, the Lagrangian (\ref{eq:Lagrangian_R_Nf1}) is given as
\begin{eqnarray}
{\cal L}_{\mathrm{eff}}^{\chi,\mathrm{sing}}
&=&
(\chi^{\dag},\Psi_{v}^{\dag})
\!
\left(\!\!
\begin{array}{cc}
 p_{0} + \mu - \vec{\sigma} \!\cdot\! \vec{p}\, &  \Phi^{\dag} + \hat{p} \!\cdot\! \vec{\Phi}^{\dag}  \\
 \Phi + \hat{p} \!\cdot\! \vec{\Phi} & p_{0}-\lambda
\end{array}
\!\!\right)
\!\!
\left(\!\!\!
\begin{array}{c}
 \chi \\
 \Psi_{v}
\end{array}
\!\!\!\right)
\nonumber \\
&&
-\frac{1}{g_{c}} \left( |\Phi|^{2} + |\vec{\Phi}|^{2} \right)
+\lambda n_{Q},
\label{eq:Lagrangian_R_Nf1_eff}
\end{eqnarray}
where the rest frame for the heavy quark is taken $v^{\mu}=(1,\vec{0}\,)$.
We notice that so far $\Phi$ and $\vec{\Phi}$ are treated independently, and that Eq.~(\ref{eq:Lagrangian_R_Nf1_eff}) is equivalent to Eq.~(\ref{eq:Lagrangian_R_Nf1}).

In Eq.~(\ref{eq:Lagrangian_R_Nf1_eff}), we apply the mean-field approximation by replacing $\Phi$ and $\vec{\Phi}$ by $\langle \Phi \rangle$ and $\langle \vec{\Phi} \rangle$, respectively. We consider the following two different types of condensation: (i) Either of $\langle\Phi\rangle$ or $\langle\vec{\Phi}\rangle$, but not both, has a finite expectation value, (ii) Both of $\langle\Phi\rangle$ and $\langle\vec{\Phi}\rangle$ have finite expectation values, and are related with each other. Namely, we consider two different cases:
\begin{eqnarray}
&&\mathrm{(i)}\ \ (\langle\Phi\rangle, \langle\vec{\Phi}\rangle)=(\Delta,\vec{0}\,)\ \mathrm{or} \ (0,\Delta \hat{p}),\\
&&\mathrm{(ii)}\ (\langle\Phi\rangle, \langle\vec{\Phi}\rangle)=(\Delta,\Delta \hat{p}).
\end{eqnarray}
The latter case realizes the coexistence of the scalar and vector condensates.
We can judge which type of condensate is favored by investigating the thermodynamic potentials and the gap equations. Below we discuss these two cases separately.

\noindent{\bf Case (i):}\\
Inserting the ansatz into the Lagrangian (\ref{eq:Lagrangian_R_Nf1_eff}), we find  four dispersion relations:
\begin{eqnarray}
\hspace{-2em}
 \epsilon^{\pm}_{p} &=& \frac{1}{2} \left( p-\mu+\lambda \pm \sqrt{(p-\mu-\lambda)^{2}+4|\Delta^{2}|} \right) , \\
\hspace{-2em}
\tilde{\epsilon}^{\pm}_{p} &=& \frac{1}{2} \left( -p-\mu+\lambda \pm \sqrt{(p+\mu+\lambda)^{2}+4|\Delta^{2}|} \right) .
\end{eqnarray}
Using these solutions, the thermodynamic potential is given by
\begin{eqnarray}
&&\hspace{-3mm}\Omega^{\mathrm{(i)}}(T,\mu,\lambda;\Delta)\nonumber \\
&&\hspace{-3mm}= N_{c} \int_{0}^{\Lambda} f^{\mathrm{(i)}}(T,\mu,\lambda;p) \, \frac{p^{2}\mathrm{d}p}{2\pi^{2}}
+ \frac{1}{g_{c}} |\Delta|^{2}
- \lambda n_{Q},
\label{eq:Omega_R_Nf1_1}
\end{eqnarray}
where
$f^{\mathrm{(i)}}(T,\mu,\lambda;p) = -\beta^{-1} \ln [(1+e^{-\beta \epsilon^{+}_{p}}) (1+e^{-\beta \epsilon^{-}_{p}}) (1+e^{-\beta \tilde{\epsilon}^{+}_{p}}) (1+e^{-\beta \tilde{\epsilon}^{-}_{p}})]$.
At zero temperature ($\beta\rightarrow\infty$) and $\lambda=0$, we have a simplified form
$f^{\mathrm{(i)}}(0,\mu,0;p)
=
\epsilon^{-}_{p}+\tilde{\epsilon}^{-}_{p}$,
and obtain the thermodynamic potential
\begin{eqnarray}
\Omega^{\mathrm{(i)}}(0,\mu,0;\Delta)
&=& N_{c} \int_{0}^{\Lambda} \bigl( E^{-}_{p}  + \tilde{E}^{-}_{p} \bigr) \, \frac{p^{2}\mathrm{d}p}{2\pi^{2}}
+ \frac{1}{g_{c}} |\Delta|^{2} \nonumber \\
&\simeq&
-\frac{N_{c}}{24\pi^{2}} (3\Lambda^{4}+4\Lambda^{3}\mu+\mu^{4})
\nonumber \\
&&
-\frac{N_{c}}{2\pi^{2}} (\Lambda^{2}-2\mu^{2}) |\Delta|^{2}
\nonumber \\
&&
+\frac{N_{c}}{2\pi^{2}} \mu^{2} |\Delta|^{2} \ln \frac{|\Delta|^{2}}{\Lambda^{2}-\mu^{2}}
\nonumber \\
&&
+\frac{|\Delta|^{2}}{g_{c}},
\end{eqnarray}
where, in the last line, we expanded the equation with respect to small $|\Delta|$. From the minimization condition $\partial \Omega^{\mathrm{(i)}}(0,\mu,0;\Delta)/\partial \Delta^{\ast}=0$, we obtain two solutions: 
a trivial solution $|\Delta| = 0$ and a non-trivial one
\begin{eqnarray}
|\Delta| =\alpha'\sqrt{\Lambda^{2}-\mu^{2}}  \exp \Bigl( - \frac{\pi^{2}}{N_{c}\mu^{2}g_{c}} \Bigr),
\label{eq:R_Nf1_gap1}
\end{eqnarray}
with $\alpha'=\exp\bigl( (-3 \mu^{2}+ \Lambda^{2})/(2\mu^{2}) \bigr)$.
The solution which minimizes the thermodynamic potential is given by the second solution.

%%%%%%%%%%%%%%%%%%%%%%%%%%
\begin{figure}[tb!]
  \centering
  \vspace*{-2em}
  \includegraphics[keepaspectratio,scale=0.6]{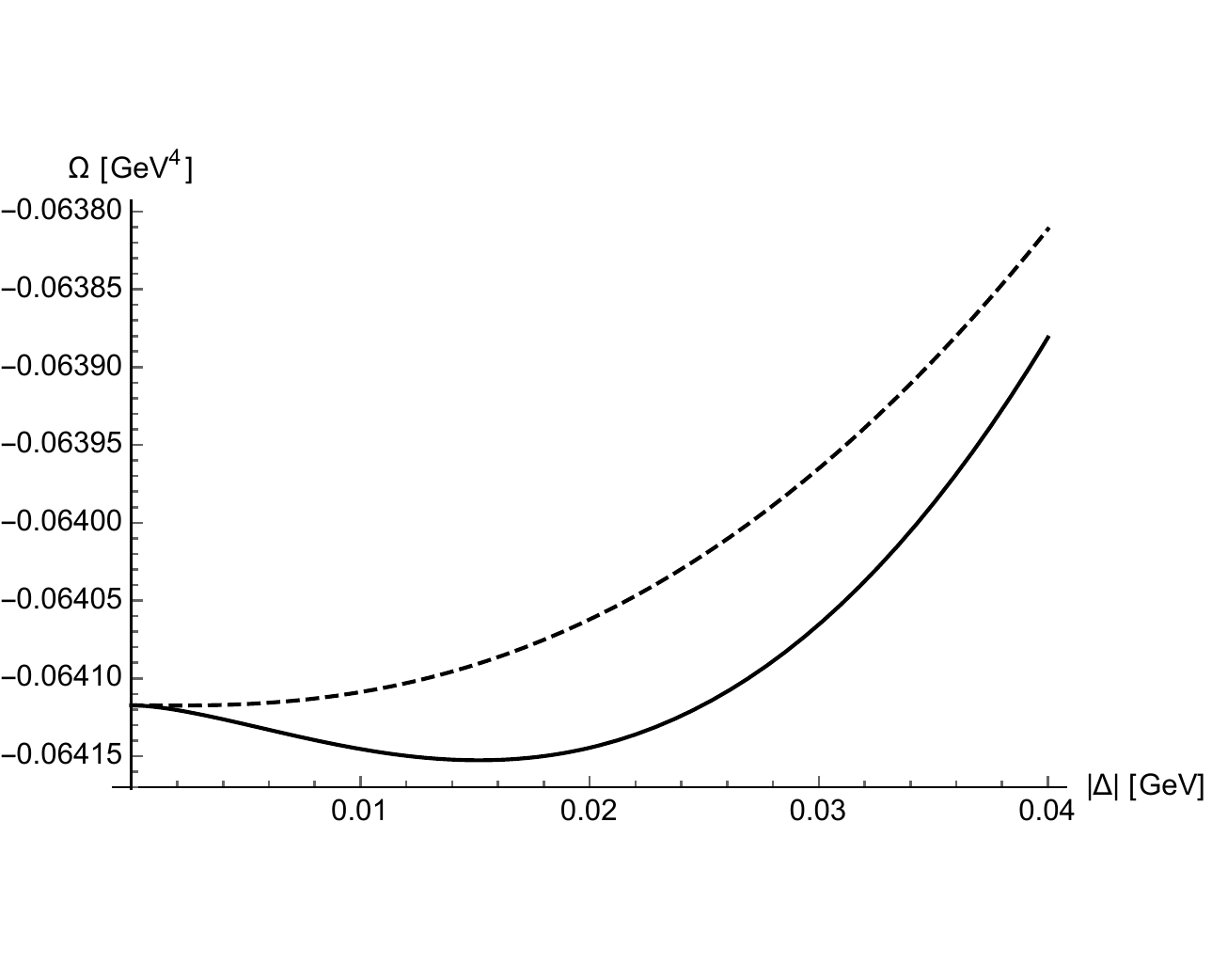}
  \vspace*{-2em}
  \caption{
Plots of the thermodynamic potentials as functions of $|\Delta|$. The dashed line is for $\Omega^{(\mathrm{i})}$, and the solid line for $\Omega^{(\mathrm{ii})}$. The parameters we used are $g_{c}\Lambda^{2}=1$, $N_{c}=3$, $\mu=0.5$~GeV, $\Lambda=1$ GeV and $T=0$.
}
  \label{fig:Omega_R_NFf1}
\end{figure}
%%%%%%%%%%%%%%%%%%%%%%%%%%

\noindent{\bf Case (ii):}\\ 
We follow the same procedures: we find four dispersions as
\begin{eqnarray}
\hspace{-2em}
\varepsilon^{\pm}_{p} &=& \frac{1}{2} \left( p-\mu+\lambda \pm \sqrt{(p-\mu-\lambda)^{2}+16|\Delta^{2}|} \right), \\
\hspace{-2em}
\tilde{\varepsilon}_{p} &=& -p-\mu, \\
\hspace{-2em}
\varepsilon_{0}  &=& \lambda.
\end{eqnarray}
Notice that the gap $|\Delta|$ appears only in two dispersions $\varepsilon^\pm_p$.
The thermodynamic potential is given by
\begin{eqnarray}
&&\hspace{-3mm}\Omega^{\mathrm{(ii)}}(T,\mu,\lambda;\Delta)\nonumber\\
&&\hspace{-3mm}= N_{c} \int_{0}^{\Lambda} \! f^{\mathrm{(ii)}}(T,\mu,\lambda;p) \, \frac{p^{2}\mathrm{d}p}{2\pi^{2}}
+ \frac{2}{g_{c}} |\Delta|^{2} 
- \lambda n_{Q},
\label{eq:Omega_R_Nf1_4}
\end{eqnarray}
with
$f^{\mathrm{(ii)}}(T,\mu,\lambda;p) = -{\beta}^{-1} \ln [(1+e^{-\beta \varepsilon^{+}_{p}}) (1+e^{-\beta \varepsilon^{-}_{p}}) (1+e^{-\beta \tilde{\varepsilon}_{p}}) (1+e^{-\beta \varepsilon_{0}})]$.
At zero temperature ($\beta\rightarrow\infty$) and $\lambda=0$, we have a simplified form
$f^{\mathrm{(ii)}}(0,\mu,0;p)
=
\varepsilon^{-}_{p}+\varepsilon^{-}_{p}$,
and obtain the thermodynamic potential
\begin{eqnarray}
\Omega^{\mathrm{(ii)}}(0,\mu,0;\Delta)
&=& N_{c} \int_{0}^{\Lambda} \bigl( \varepsilon^{-}_{p} + \tilde{\varepsilon}_{p} \bigr) \, \frac{p^{2}\mathrm{d}p}{2\pi^{2}}
+ \frac{2}{g_{c}} |\Delta|^{2} \nonumber \\
&\simeq&
-\frac{N_{c}}{24\pi^{2}} (3\Lambda^{4}+4\Lambda^{3}\mu+\mu^{4})
\nonumber \\
&&
-\frac{N_{c}}{\pi^{2}} (\Lambda^{2}+2\Lambda\mu-4\mu^{2}) |\Delta|^{2}
\nonumber \\
&&
-\frac{2N_{c}}{\pi^{2}} \mu^{2} |\Delta|^{2} \ln \frac{(\Lambda-\mu)\mu}{4|\Delta|^{2}}
\nonumber \\
&&
+\frac{2|\Delta|^{2}}{g_{c}},
\end{eqnarray}
where, in the last line, we expanded the equation with respect to small $|\Delta|$. From the minimization condition $\partial \Omega^{\mathrm{(ii)}}(0,\mu,0;\Delta)/\partial \Delta^{\ast}=0$, we find a trivial solution
$|\Delta| = 0$ and a non-trivial one
\begin{eqnarray}
|\Delta| = \frac{\alpha''}{2}\sqrt{\mu(\Lambda-\mu)}  \exp \Bigl( - \frac{\pi^{2}}{2N_{c}\mu^{2}g_{c}} \Bigr),
\label{eq:R_Nf1_gap4}
\end{eqnarray}
with $\alpha''=\exp\bigl\{ (\Lambda^{2} + 2\Lambda \mu - 6\mu^{2})/(4\mu^{2}) \bigr\}$.
The solution which minimizes the thermodynamic potential is given by the second solution.

Let us compare the gap for the case (i), Eq.~(\ref{eq:R_Nf1_gap1}), and the one for the case (ii), Eq.~(\ref{eq:R_Nf1_gap4}). We notice that the exponential factors are different; $\sim e^{-{\pi^{2}}/{(N_{c}\mu^{2}g_{c})}}$ for the case (i) and $\sim e^{-{\pi^{2}}/{(2N_{c}\mu^{2}g_{c})}}$ for the case (ii). Hence we find that the gap in (ii) is much larger than that in (i). We also find that the thermodynamic potential $\Omega^{(\mathrm{ii})}$ is indeed smaller than $\Omega^{(\mathrm{i})}$, as shown in Fig.~\ref{fig:Omega_R_NFf1}. Therefore, we conclude that the case (ii) with the coexisting scalar and vector condensates is more stable than the case (i). The same procedure is applied to the left-handed Weyl fermion, and the similar conclusion is drawn. 

\bibliography{reference}

\end{document}